\documentclass[prd,showpacs,floatfix,amsmath,amssymb,floatfix]{revtex4}
\usepackage{graphicx,dcolumn,booktabs,bm}
\usepackage{longtable,lscape}
\usepackage{txfonts}
\usepackage{overpic}
\usepackage{multirow}
\usepackage{amssymb}
\usepackage{indentfirst}
\usepackage{feynmf}   %{feynmp}
\usepackage{slashed}  %for Feynman symbols
\usepackage{cases}
\usepackage{color,ulem}
\usepackage{graphicx}
\usepackage{tikz-feynman}
\graphicspath{{Figures/}}

\begin{document}

	%%%%%%%%%%%%%%%%%%%%%%%%%%%%%%%%%%%%%%%%%%%%%%%%%%%%%%%%%%%%%%%%%%%%%%%%%%%%%%%%%%%%%%%%%%%%%%%%%%%%%%%%%
	%                                    The document begins here                                           %
	%%%%%%%%%%%%%%%%%%%%%%%%%%%%%%%%%%%%%%%%%%%%%%%%%%%%%%%%%%%%%%%%%%%%%%%%%%%%%%%%%%%%%%%%%%%%%%%%%%%%%%%%%
	
	\title{Possible $D_{1}D_{1}$, $D_{1} \bar D_{1} $, $B_{1}B_{1}$ and $B_{1} \bar B_{1} $ molecular states and the recoil corrections}
	\author{Xiao Chen$^{1}$}\email{comcn126163@bjtu.edu.cn}
	\author{Li Ma$^{1}$}\email{ma.li@bjtu.edu.cn}
	
	\affiliation{
		$^1$School of Physical Science and Engineering, Beijing Jiaotong University, Beijing 100044, China
		}
	
	\date{\today}% It is always \today, today,

	\begin{abstract}
		
	  Recoil correction appears at $O(\frac{1}{M})$, which turns out to be very essential for the hadronic molecules with heavy flavor. In the past, we always thought that the recoil corrections were unfavorable to the formation of the molecular states, but our research reveals its importance to form the di-hadron bound states. In some cases, we are unable to find the bound states without considering the recoil corrections. Under SU(2) chiral symmetry, we have studied the $D_1 D_1$, $D_1 \bar D_1$, $B_1 B_1$ and $B_1 \bar B_1$ systems in the framework of the one-boson exchange (OBE) model with the treatments of the $S$\--{}$D$ wave mixing effect and the recoil corrections. Our results indicate that both the $D_1 D_1$ system with $I(J^P)=0(1^+)$ and $I(J^P)=1(2^+)$ can form the molecular states whether with the recoil corrections or not, while the  $D_1 D_1$ system with $I(J^P)=1(0^+)$ fail to form a hadronic molecule without the recoil corrections but it turns out to be a loosely bound state after the inclusion of the recoil corrections. And we have found the deuteron-like states for the $D_1 \bar D_1$, $B_1 B_1$ and $B_1 \bar B_1$ systems, whether considering the recoil corrections or not. 
		
	\end{abstract}
	
	\pacs{14.40.Lb, 12.39.Fe, 13.60.Le} \maketitle

	%%%%%%%%%%%%%%%%%%%%%%%%%%%%%%%%%%%%%%%%%%%%%%%%
	\section{Introduction} \label{sec1}
    In the past few years, a series of $XYZ$ states such as $Y(3940)$, $Z_{c}(3900)$, $Z_{b}(10650)$ and $X(4700)$ are observed by a variety of experiments in the charmonium and bottomonium mass range\cite{ref1,ref2,ref3,ref4,ref5,ref6}.  The rapid development of the experiments makes the theoretical study of the structure and the prediction of these exotic states a hot topic. 
    Up to now, theorists have proposed several possible structures to explain the experimental observations. This interpretations encompass hadronic molecule\cite{ref7,ref8,ref9,ref10}, compact tetraquark \cite{ref11,ref12,ref13,ref14}, hybrid \cite{ref15,ref16,ref17} and glueball \cite{ref18,ref19}. Among these hypotheses, the hadronic molecule picture has made a remarkable success in describing and predicting the structure of $XYZ$ states, since many exotic states are very close to the thresholds of a pair of heavy mesons and the structure of deuteron provides robust evidence for this picture.
    \par 
    People have found many charmed mesons in the experiments, such as $D$, $D^*$, $D_{0}$, $D_{1}^{\prime}$, $D_1$ and $D_2$. Under the isospin SU(2) framework and the heavy quark symmetry \cite{ref20}, $(D,D^*)$ is the $H$ doublet, $(D_{0},D_{1}^{\prime})$ is the $S$ doublet, and $(D_1,D_2)$ is the $T$ doublet. It should be particularly noted that the $D_1$ corresponds to the $D_1(2420)$ with $I(J^{P})=\frac{1}{2}(1^{+})$ instead of the $D_1(2430)$ with a larger width, which corresponds to the $D_{1}^{\prime}$ in the $S$ doublet. We can distinguish these two states by the dynamical behavior. Both $D_1$ and $D_{1}^{\prime}$ mainly decay to $D^{*}\pi$. According to the heavy quark symmetry, the heavy degrees of freedom are hardly involved in the decay process, so the light degrees of freedom determine the decay rate. The light degrees of freedom of $D_1$ and $D_{1}^{\prime}$ are $\frac{3}{2}$ and $\frac{1}{2}$, respectively. Therefore, $D_{1}$ decays mainly by $D$ wave, which is more difficult than $D_{1}^{\prime}$ decays mainly by $S$ wave. Based on the Particle Data Group's results, $D_{1}^{\prime}$ has a very large width, which prevents it from serving as a suitable building block for the hadronic molecules. Previous research has predominantly discussed the possibility of the molecular states composed by a $H$ doublet meson and a $H$ doublet antimeson, namely $H \bar H$ states. For example, $X(3872)$ is regarded as a $D\bar D^*$ state \cite{ref21} and $Y(3940)$ is believed as a $D^* \bar D^*$ state \cite{ref22}. Recently, the $HH$ structure is also proposed to explain $T_{cc}(3875)$ \cite{ref23}. Besides, some theorists interpret $Z^+(4430)$ as a molecular state composed by a $H$ doublet meson and a $S$ doublet antimeson or a $T$ doublet antimeson, namely $H \bar S$ or $H \bar T$ state \cite{ref24,ref25}. Possibilities of $TT$ and $T \bar T$ states are also considered in 2016 and 2022, respectively \cite{ref26,ref27}.
    \par 
    However, the recoil corrections are always neglected in previous research of $T \bar T$ or $TT$ molecular states. To address this gap in the theoretical framework, we have systematically studied the $D_1D_1$ and the $D_1\bar D_1$ systems in search of the bound states. We compare the results with and without the recoil corrections. Besides, we have also discussed their bottom partners, the $\bar{B}_1 \bar{B}_1$ and the $B_1 \bar B_1$ systems. However, for the open flavor system, we actually choose the $B_{1}B_{1}$ system so that both the $D_{1}$ system and the $B_{1}$ system are composed solely of mesons. In fact, the research process and the final results of the $B_{1}B_{1}$ and the $B_1 \bar B_1$ systems are completely consistent because we can apply the G-parity operator to the $\bar{B}_{1}\bar{B}_{1}$ system to get the $B_{1}B_{1}$ system. To be more specific, the Lagrangian and the effective potential for these two systems are formally consistent. 
    \par
    The theoretical studies of the hadronic molecular states have gone through a long and arduous process. The non-perturbative nature of QCD in the low energy region requires us to seek other approaches to illuminate the dynamics of strong interactions. The attempts that have been made include Lattice QCD \cite{ref28,ref29,ref30}, QCD sum rule \cite{ref31} and the potential model \cite{ref32,ref33,ref34}. 
Among these methods, the potential model has been widely used for decades due to its obvious advantages of having lower computational complexity, providing a clear physical picture and offering a powerful predictive ability. The main idea of the potential model is to treat the interaction between baryons or mesons as the residual interaction of the quarks, and it can be equivalently represented as the exchange of light mesons between baryons or mesons. 
Theorists constantly make efforts to improve the potential model by considering various effects so that it can provide results that are closer to the experimental data. The One-Boson Exchange (OBE) model is very successful in describing the nuclear force. Much theoretical research on hadronic molecules in the framework of the OBE model has been performed in recent years\cite{ref35,ref36,ref37,ref38,ref39,ref40,ref41}. These studies show the great success of the OBE model in discussing the issues on hadronic molecules.

    \par  
    In this work, we mainly focus on the $S$ wave interactions between a pair of $D_1$ and $B_1$ mesons. We work under chiral SU(2) framework and adopt the OBE model to describe the interactions. In our calculation, the $S$\--{}$D$ wave mixing effect is considered. Furthermore, most previous works adopted the heavy quark symmetry approximation to simplify the calculation of the effective potentials at the expense of ignoring the three momentum of the external particles. However, the recoil corrections can be significant in the formation of hadronic molecules. Thus, we preserve the contributions of the recoil corrections up to $O(\frac{1}{{m_{D_1}}^2})$ and $O(\frac{1}{{m_{B_1}}^2})$ to get the more precise results. With these relativistic effective potentials, we can find the bound state solutions by solving the coupled channel Schr{\"o}dinger equation. We also compare the relativistic results with those under the heavy quark symmetry.
    \par 
    The paper is organized as follows. After the introduction, we present the formalism including the wave function, the effective Lagrangian, the effective potentials, the coupling constants, the isospin factors, and the solving of the Schr{\"o}dinger equation in Sec. \ref{sec2}. We present our numerical results in Sec. \ref{sec3}. We summarize our results and give a discussion in Sec. \ref{sec4}. We list some detailed analytical results in the Appendix.

		%%%%%%%%%%%%%%%%%%%%%%%%%%%%%%%%%%%%%%%%%%%%%%%%内容...

	\section{Formalism} \label{sec2}
	In the following section, we discuss the formalism of our study by using the systems constructed by a pair of $D_1$ mesons as an example. We can get the similar results in the systems constructed by a pair of $B_1$ mesons. The only difference is the mass of $D_1$ and $B_1$ mesons.

	\subsection{Wave functions}\label{subsecA} 
	The wave function of a system includes three parts. The first part is the isospin wave function, which produces an isospin factor in the effective potential. The second and the third parts are the orbital wave function and the spin wave function, respectively. The total wave function is the direct product of them
	\begin{eqnarray}
		\psi^{tot}=\psi^{I} \otimes \psi^{L} \otimes \psi^{S}. \label{1}
	\end{eqnarray}
	We note that the wave function should be symmetric for the $D_1 D_1$ system because it is a system of identical bosons, while there is no such restriction for the wave function of the $D_1 \bar D_1$ system.
	\par
	First, we construct the isospin wave function. The $I(J^P)$ of $D_1$ is $\frac{1}{2}(1^+)$, so the $D_1 D_1$ system can couple to isospin triplet and isospin singlet, so does the $D_1 \bar D_1$ system. We list the isospin wave functions in TABLE \ref{tab:1}. In particular, for the isospin triplet, the $\frac{1}{\sqrt{2}}\left(\left|D_{1}^{0} \bar{D}_{1}^{0}\right\rangle-\left|D_{1}^{+} D_{1}^{-}\right\rangle\right)$ state is the eigenstate of C- and G-parities with eigenvalues +1 and -1, respectively. For the isospin singlet, the $\frac{1}{\sqrt{2}}\left(\left|D_{1}^{0} \bar{D}_{1}^{0}\right\rangle+\left|D_{1}^{+} D_{1}^{-}\right\rangle\right)$ state is the eigenstate of C- and G- parities with eigenvalues -1 and +1, respectively.

	\renewcommand{\arraystretch}{1.5}
	\begin{table*}[htbp]
		\scriptsize
		\begin{center}
			\caption{\label{tab:1} The isospin wave functions for $D_1 D_1$ and $D_1 \bar D_1$systems}
			\begin{tabular}{c|c|c}\toprule[1pt]
				$\left|I, I_{3}\right\rangle$ & $D_1 D_1$ \text { system } & $D_{1} \bar D_{1}$ \text { system } \\
				\midrule[1pt]
				$|1,1\rangle$ & $\left | D^{+}_{1} D_{1}^{+}\right\rangle$ & $\left|D_{1}^{+} \bar D_{1}^{0}\right\rangle$ \\
				$|1,0\rangle$ & $\frac{1}{\sqrt{2}}\left(\left|D_{1}^{0} D_{1}^{+}\right\rangle+\left|D_{1}^{+} D_{1}^{0}\right\rangle\right)$ & $\frac{1}{\sqrt{2}}\left(\left|D_{1}^{0} \bar D_{1}^{0}\right\rangle-\left|D_{1}^{+} D_{1}^{-}\right\rangle\right)$ \\
				$|1,-1\rangle$ & $\left|D_{1}^{0} D_{1}^{0}\right\rangle$ & $\left|D_{1}^{0} D_{1}^{-}\right\rangle$ \\
				$|0,0\rangle$ & $\frac{1}{\sqrt{2}}\left(\left|D_{1}^{0} D_{1}^{+}\right\rangle-\left|D_{1}^{+} D_{1}^{0}\right\rangle\right)$ & $\frac{1}{\sqrt{2}}\left(\left|D_{1}^{0} \bar D_{1}^{0}\right\rangle+\left|D_{1}^{+} D_{1}^{-}\right\rangle\right)$ \\
				
				\bottomrule[1pt]
				
			\end{tabular}
		\end{center}
	\end{table*}
	
	\par
	Then, we consider the orbital and the spin wave functions. We mainly focus on the ground state of our systems because it is more likely to form a bound state \cite{ref42}. In general, $S$ wave is the ground state. But in our work, there exists some tensor force terms in the effective potential, which lead to $S$\--{}$D$ wave mixing. Thus, a physical ground state is the liner combination of the $S$ and the $D$ wave. According to the coupling rules of angular momentum, the total spin of our systems can be $0,1,2$, and the total angular momentum can be $0,1,2$ as well. Thus, we get three sets of the wave function.
	\begin{eqnarray}
		\mid \psi ^{J=0} \rangle &=&R_s\mid {^1}S_{0}\rangle +R_d \mid {^5}D_{0}\rangle, \label{2}\\ 
		\mid \psi ^{J=1} \rangle &=&R'_s\mid {^3}S_{1}\rangle +R'_d \mid {^3}D_{1}\rangle, \label{3}\\
		\mid \psi ^{J=2} \rangle &=&R''_s\mid {^5}S_{2}\rangle +R''_{d1} \mid {^1}D_{2}\rangle+R''_{d2} \mid {^5}D_{2}\label{4}\rangle. 
	\end{eqnarray}
	\par 
	Finally, we present all possible channels of our states. For the $D_1 D_1$ system, $I(J^P)$ can be $0(1^+)$, $1(0^+)$ or $1(2^+)$, while for the $D_1 \bar D_1$ system, $I(J^P)$ can be $0(0^+)$, $0(1^+)$, $0(2^+)$, $1(0^+)$, $1(1^+)$ or $1(2^+)$. We can get the same result in the $B_1 B_1$ and the $B_1 \bar B_1$ systems.

	\subsection{Effective Lagrangian, Isospin factors and Coupling constants}\label{subsecB}
	We construct the effective Lagrangian by following the chiral symmetry and the symmetries of the strong interaction, including parity conservation, charge parity conservation, and G-parity conservation. According to the OBE model, scalar, pseudoscalar and vector mesons may contribute to the effective potential.   \par 
	The Lagrangian for interactions among the scalar, the pseudoscalar and the vector light mesons with the heavy mesons (antimesons) reads
	\begin{eqnarray}
		\mathcal{L}_{D_1D_1\sigma }&=&-2g_\sigma m_{D_1}D_{1a}^\mu D_{1a\mu}^\dagger \sigma,\label{5}\\    
		\mathcal{L}_{D_1D_1\phi }&=&\frac{g^{\prime} }{f_\pi}\varepsilon_{\alpha \mu \lambda \nu } (D_{1b}^\mu\overleftrightarrow{\partial ^\alpha }D_{1a}^{\dagger\nu } )(\partial ^\lambda \phi_{ba}),\label{6}\\ 
		\mathcal{L}_{D_1D_1V }&=&i\beta g_V(D_{1b}^{\nu}\overleftrightarrow{\partial _\mu } D_{1a\nu }^\dagger )V_{ba}^\mu+i\lambda g_V m_{D_1}(D_{1b}^\mu D_{1a}^{\dagger\nu }-D_{1a}^{\dagger\mu } D_{1b}^{\nu })(\partial _\mu V_{\nu}-\partial _\nu V_\mu)_{ba}\label{7}
	\end{eqnarray}
	and
	\begin{eqnarray}
		\mathcal{L}_{\bar D_1\bar D_1\sigma }&=&-2g_\sigma m_{ D_1}\bar D_{1a}^\mu \bar D_{1a\mu}^\dagger \sigma,\label{8}\\   
		\mathcal{L}_{\bar D_1\bar D_1\phi }&=&\frac{g^{\prime} }{f_\pi}\varepsilon_{\alpha \mu \lambda \nu } (\bar D_{1b}^\mu\overleftrightarrow{\partial ^\alpha }\bar D_{1a}^{\dagger\nu } )(\partial ^\lambda \phi_{ba}),\label{9}\\
		\mathcal{L}_{\bar D_1\bar D_1V }&=&-i\beta g_V(\bar D_{1b}^{\nu}\overleftrightarrow{\partial _\mu } \bar D_{1a\nu }^\dagger )V_{ba}^\mu-i\lambda g_V m_{D_1}(\bar D_{1b}^\mu \bar D_{1a}^{\dagger\nu }-\bar D_{1a}^{\dagger\mu } \bar D_{1b}^{\nu })(\partial _\mu V_{\nu}-\partial _\nu V_\mu)_{ba}\label{10}.
	\end{eqnarray}
	\par
	Since we work in the isospin SU(2) framework, the heavy meson field $D_1$ represents the $(D_{1}^{0},D_{1}^{+})$ bispinor while the corresponding antimeson field $\bar D_1$ represents the $(\bar D_{1}^{0},D_{1}^{-})$ bispinor. $\phi$ and $V$ represent the matrices of the exchanged light pseudoscalar mesons and the exchanged light vector mesons, respectively, while $\sigma$ is the only exchanged light scalar meson in our framework.
	\begin{eqnarray}
		\phi=
		\begin{pmatrix}
			\frac{\pi^0}{\sqrt{2}}+\frac{\eta}{\sqrt{6}}    & \pi^+  \\
			\pi^- & -\frac{\pi^0}{\sqrt{2}}+\frac{\eta}{\sqrt{6}} 
		\end{pmatrix},
		\label{11}
		\\
	V=
	\begin{pmatrix}
		\frac{\rho^0}{\sqrt{2}}+\frac{\omega}{\sqrt{2}}    & \rho^+   \\
		\rho^- & -\frac{\rho^0}{\sqrt{2}}+\frac{\omega}{\sqrt{2}} 
	\end{pmatrix}.
	\label{12}
	\end{eqnarray}
	\par
	Expending the effective Lagrangian in Eqs.(\ref{5})-(\ref{7}) by introducing matrices Eqs.(\ref{11})-(\ref{12}) and considering the isospin wave functions given in TABLE \ref{tab:1}, we can get the isospin factors of the $D_1D_1$  effective potentials. The isospin factors of the $D_1\bar D_1$  effective potentials can be obtained in the similar way. However, in our work, we use the $G$\--{}parity rule to get it more conveniently \cite{ref43}. In certain cases, the effective potentials obtained from the two methods may differ by a negative sign. But after considering the differences in the Lagrangian, the results are then consistent. We list all the isospin factors in TABLE \ref{tab:2}.

	\renewcommand{\arraystretch}{1.5}
	\begin{table*}[htbp]
		\scriptsize
		\begin{center}
			\caption{\label{tab:2}  The isospin factors for different channels}
			
			\begin{tabular}{ c|ccccc}\toprule[1pt]
				\multicolumn{1}{c|}{Channels} & \multicolumn{1}{c}{$C_\pi$} &
				\multicolumn{1}{c}{$C_\eta$} &
				\multicolumn{1}{c}{$C_\rho$} &
				\multicolumn{1}{c}{$C_\omega$} &
				\multicolumn{1}{c}{$C_\sigma$} \\
				\midrule[1pt]
				$[D_1 D_1]_{I=1}\longleftrightarrow [D_1 D_1]_{I=1}$ & $\frac{1}{2}$ & $\frac{1}{6}$ & $\frac{1}{2}$ & $\frac{1}{2}$ & 1 \\				
				$[D_1 D_1]_{I=0}\longleftrightarrow [D_1 D_1]_{I=0}$ & $-\frac{3}{2}$ & $\frac{1}{6}$ & $-\frac{3}{2}$ & $\frac{1}{2}$ & 1 \\		
				$[D_1 \bar D_1]_{I=1}\longleftrightarrow [D_1 \bar D_1]_{I=1}$ & $-\frac{1}{2}$ & $\frac{1}{6}$ & $\frac{1}{2}$ & $-\frac{1}{2}$ & 1 \\				
				$[D_1 \bar D_1]_{I=0}\longleftrightarrow [D_1 \bar D_1]_{I=0}$ & $\frac{3}{2}$ & $\frac{1}{6}$ & $-\frac{3}{2}$ & $-\frac{1}{2}$ & 1  \\

				\bottomrule[1pt]
				
			\end{tabular}
		\end{center}
	\end{table*}
	\par 
	The coupling constant relate to the pseudoscalar meson exchange $g'=0.8$ is extracted from the QCD sum rule approach (QSR) \cite{ref44}. $f_{\pi}=132$ MeV is the pion decay constant \cite{ref45}. Based on previous research, we also get the coupling constants relate to the vector meson exchange. The results are $\beta=0.64$ and $\lambda=0.66$ GeV$^{-1}$ \cite{ref27}. The parameter $g_{V}$ can be fixed as $g_{V}=5.8$ by considering the electromagnetic couplings \cite{ref46}. The coupling constant of scalar exchange is $g_{\sigma}=g_{\pi}/2\sqrt{6}$, where $g_{\pi}=3.73$ \cite{ref47}. We should mention that although our data are extracted from the studies of $D_1$, these results also apply to $B_1$ for the heavy quark hardly participate in the interactions. All the results are shown in TABLE \ref{tab:3}.

	\renewcommand{\arraystretch}{1.5}	
	\begin{table*}[htbp]
		\scriptsize
		\begin{center}
			\caption{\label{tab:3}  The masses and the coupling constants of the heavy mesons and the exchanged light mesons. Our data are based on the Particle Data Group’s and the previous research's results \cite{ref48} }
			\begin{tabular}{l c|c} \toprule[1pt]
				\multicolumn{2}{c|}{Mass(MeV)} & {Coupling Constants} \\\midrule[1pt]
				Pseudoscalar & $m_\pi=134.98$ & $g'=0.8$   \\
				%\midrule[1pt]
				$$     &$m_\eta=547.86$ &$f_\pi=132$ MeV  \\
				
				Scalar     &$m_\sigma=600$ &$g_\pi=3.73$   \\
				
				Vector     &$m_\rho=775.26$ &$g_\sigma=g_\pi / 2\sqrt{6}$   \\
				
				$$     &$m_\omega=782.66$ &$g_V=5.8$ \\ 
				
				Heavy flavor     &$m_{D_1}=2422.1$ &$\beta=0.64$    \\
				
				$$     &$m_{B_1}=5726.1$ &$\lambda=0.66$ GeV$^{-1}$ \\

				\bottomrule[1pt]
				
			\end{tabular}
		\end{center}
	\end{table*}
	\par
	According to the Feynman rules, the external line of $D_1$ or $\bar D_1$ will turn into a polarization vector. In the center of mass frame, the polarization vector reads
	\begin{eqnarray}\label{13}
		\epsilon &=& (0,\boldsymbol{\epsilon}). 
	\end{eqnarray}
	Making a Lorentz boost to Eqs.(\ref{13}), we get the polarization vector in the laboratory frame
	\begin{eqnarray}
		\epsilon ^{lab}&=&\left(\frac{\boldsymbol{p}\cdot \boldsymbol{\epsilon }}{m}, \boldsymbol{\epsilon }+\frac{\boldsymbol{p}(\boldsymbol{p}\cdot \boldsymbol{\epsilon })}{m(p_0+m)}  \right).
	\end{eqnarray}

    \subsection{Effective potentials}\label{subsecC}
    Since the Lagrangian is constructed, we can get the Feynman rules of the vertices and calculate the scattering amplitudes. Now we try to find the relationship between the effective potentials and the scattering amplitudes.
    \begin{center}
    	\begin{tikzpicture}
    		\begin{feynman}		
    			\vertex (a) ;
    			\vertex [right=2.5 of a](b) ;
    			\vertex [above=1.8 of b](f4) {4} ;
    			\vertex [below=1.8 of b](f2) {2} ;
    			\vertex [above=1.8 of a](f3) {3};
    			\vertex [below=1.8 of a](f1) {1} ;
    			\diagram* {
    				(f2) -- [fermion, edge label=$P_2$] (b) -- [scalar, momentum=$q$] (a) -- [fermion, edge label=$P_3$] (f3),
    				(f1) -- [fermion, edge label=$P_1$] (a), 
    				(b)-- [fermion, edge label=$P_4$] (f4),
    			};
    		\end{feynman}   
    	\end{tikzpicture}	
    \end{center}
    \par
    Consider a general 2\--{}2 scattering process at tree level. According to the scattering theory, the $S$-matrix elements read
    \begin{eqnarray}
    	\left\langle  f \mid S \mid i \right\rangle &=&\delta_{f i}-i 2 \pi \delta(E_{f}-E_{i})\langle f \mid t\mid i\rangle
    \end{eqnarray}
    with $t$ satisfies Lippmann-Schwinger equation
    \begin{eqnarray}
    	t&=&V+V\frac{1}{E-H+i\epsilon } V.
    \end{eqnarray}
    Apply Born approximation to the first order $t=V$
    \begin{eqnarray}
       		\left\langle  f \mid S \mid i \right\rangle &=&\delta_{f i}-i 2 \pi \delta(E_{f}-E_{i})\langle f \mid V\mid i\rangle.
    \end{eqnarray}
    Expand $\left\langle  f \mid V \mid i \right\rangle$ by plane waves
    \begin{eqnarray}\label{18}
    	\left\langle  f \mid V \mid i \right\rangle&=&\frac{1}{(2 \pi)^{6}} \iint \mathrm d^{3} \boldsymbol{r}_{\boldsymbol{1}} \mathrm d^{3} \boldsymbol{r}_{\boldsymbol{2}} \mathrm d^{3} \boldsymbol{r}_{\boldsymbol{3}} \mathrm d^{3} \boldsymbol{r}_{\boldsymbol{4}} e^{-i\left(\boldsymbol{p}_{3} \cdot \boldsymbol{r}_{\boldsymbol{3}}+\boldsymbol{p}_{4} \cdot \boldsymbol{r}_{4}\right)}  V\left(\boldsymbol{r}_{\boldsymbol{1}}-\boldsymbol{r}_{\boldsymbol{2}}, \nabla_{r_1}, \nabla_{r_2}\right) e^{i\left(\boldsymbol{p}_{1} \cdot \boldsymbol{r}_{\boldsymbol{1}}+\boldsymbol{p}_{2} \cdot \boldsymbol{r}_{\boldsymbol{2}}\right)}.
   \end{eqnarray}
   According to the momentum conservation law, $\left\langle  f \mid V \mid i \right\rangle$ can be defined as 
   \begin{eqnarray}\label{19}	
    	\left\langle  f \mid V \mid i \right\rangle&\equiv &\frac{1}{(2\pi)^3} \delta^{3}\left(\boldsymbol{p}_{\boldsymbol{3}}+\boldsymbol{p}_{\boldsymbol{4}}-\boldsymbol{p}_{\boldsymbol{1}}-\boldsymbol{p}_{\boldsymbol{2}}\right) T_{f i},
    \end{eqnarray} 
    where $T_{fi}$ is the $T$-matrix element.
    \par
    We introduce relative coordinate, center-of-mass coordinate, relative momentum and center-of-mass momentum
    \begin{eqnarray}
    	\begin{array}{ll}
    		\boldsymbol{r}=\boldsymbol{x_1}-\boldsymbol{x_2},&\boldsymbol{r}^{\prime}=\boldsymbol{x_3}-\boldsymbol{x_4},\\
    		\boldsymbol{R}=\frac{1}{2}(\boldsymbol{x_1}+\boldsymbol{x_2}),&\boldsymbol{R}^{\prime}=\frac{1}{2}(\boldsymbol{x_3}+\boldsymbol{x_4})
    	\end{array}
   \end{eqnarray}
   and
   \begin{eqnarray}
   	    \begin{array}{ll}
   	       \boldsymbol{P}=\boldsymbol{p_1}+\boldsymbol{p_2},&\boldsymbol{P}^{\prime}=\boldsymbol{p_3}+\boldsymbol{p_4},\\
   	       \boldsymbol{p}=\frac{1}{2}(\boldsymbol{p_1}-\boldsymbol{p_2}),&\boldsymbol{p}^{\prime}=\frac{1}{2}(\boldsymbol{p_3}-\boldsymbol{p_4}).  	
   	    \end{array}
   \end{eqnarray}	
    Then we combine the Eqs.(\ref{18}) and (\ref{19}) to get
    \begin{eqnarray}
    	\delta(\boldsymbol{P}'-\boldsymbol{P})T_{fi}=\frac{1}{(2\pi)^3}\iint \mathrm{d}^3\boldsymbol{r}\mathrm{d}^3\boldsymbol{r}'\mathrm{d}^3\boldsymbol{R}\mathrm{d}^3\boldsymbol{R}' e^{-i(\boldsymbol{p}'\cdot \boldsymbol{r}'-\boldsymbol{p}\cdot \boldsymbol{r}+\boldsymbol{P}'\cdot \boldsymbol{R}'-\boldsymbol{P}\cdot \boldsymbol{R} )}V(\boldsymbol{r},\boldsymbol{r}',\nabla _r,\nabla _{r'}).  
    \end{eqnarray}
    In order to simplify the calculation and get some meaningful results, we introduce a new group of variables to describe our functions
    \begin{eqnarray}
    	\boldsymbol{q}&=& \boldsymbol{P_3}-\boldsymbol{P_1},\\
    	\boldsymbol{k}&=& \frac{1}{2}(\boldsymbol{P_3}+\boldsymbol{P_1}).
    \end{eqnarray}
    We also need to add a form factor into the integral to avoid the divergence in the high momentum region. In this work, we choose the mono pole form factor because it works very well in explaining deuteron
    \begin{eqnarray*}
    	 F(q)=\frac{\Lambda^2-m^2 }{\Lambda^2-q^2}.
    \end{eqnarray*}
    Then we get the relation between the $T$-matrix element and the effective potential $V$
    \begin{eqnarray}\label{25}
    	T_{fi}=\frac{1}{(2\pi)^3} \iint \mathrm{d}^3\boldsymbol{r}\mathrm{d}^3\boldsymbol{r}'F(q)^2 V(\boldsymbol{r},\boldsymbol{r}',\nabla _r,\nabla _{r'}) e^{i\boldsymbol{k} \cdot (\boldsymbol{r}'-\boldsymbol{r})+i\frac{1}{2}\boldsymbol{q} \cdot (\boldsymbol{r}'+\boldsymbol{r})}.
    \end{eqnarray}
    In Eqs.(\ref{25}), we notice that $T_{fi}$ is actually the effective potential in the momentum representation $V(\boldsymbol{q},\boldsymbol{k})$.
    \par 
    The relation between $V(\boldsymbol{q},\boldsymbol{k})$ and $\mathcal{M}$ is obtained from the scattering theory 
    \begin{eqnarray}
    	V(\boldsymbol{q},\boldsymbol{k} )=-\frac{\mathcal{M} }{\sqrt{2E_{1}}\sqrt{2E_{2}}\sqrt{2E_{3}}\sqrt{2E_{4}} }.
    \end{eqnarray}
    Finally, we get the effective potential in the coordinate representation
    \begin{eqnarray}
    	 V(\boldsymbol{r},\boldsymbol{r}',\nabla _r,\nabla _{r'})=\frac{1}{(2\pi)^3} \iint \mathrm{d}^3\boldsymbol{q}\mathrm{d}^3\boldsymbol{k}F(q)^2 V(\boldsymbol{q},\boldsymbol{k} ) e^{i\boldsymbol{k} \cdot (\boldsymbol{r}'-\boldsymbol{r})+i\frac{1}{2}\boldsymbol{q} \cdot (\boldsymbol{r}'+\boldsymbol{r})}.
    \end{eqnarray}
    \par 
    Here we list the effective potentials of $\sigma$, $\pi$, $\eta$, $\rho$ and $\omega$ exchange. The $1/m_{D_1}$ and $1/m^2_{D_1}$ terms are derived from the recoil corrections. There actually exists more higher order terms, but we discard them for they are severe depressed. If the building blocks of the system are $B_1$ mesons, we just need to replace all the $m_{D_1}$ by $m_{B_1}$. All the Fourier transformations are listed in the Appendix.  
    \par
    The effective potential of $\sigma$ exchange in the coordinate space is
    \begin{eqnarray}
    	V\left(r \right )=
    	&&-C_{\sigma}\frac{m_\sigma}{4 \pi} g_\sigma^2H_0(\boldsymbol{\epsilon}_1 \cdot \boldsymbol{\epsilon}_3^\dagger )(\boldsymbol{\epsilon}_2 \cdot \boldsymbol{\epsilon}_4^\dagger )  \nonumber
    	-C_{\sigma}\frac{m_\sigma}{4 \pi m^2_{D_1}} g_\sigma^2H_0 (\boldsymbol{\epsilon}_1 \cdot \boldsymbol{\epsilon}_3^\dagger )(\boldsymbol{\epsilon}_2 \cdot \boldsymbol{\epsilon}_4^\dagger )\nabla_r^2  \nonumber
    	-C_{\sigma}\frac{m_\sigma^3}{4 \pi m^2_{D_1}} g_\sigma^2H_2(\boldsymbol{\epsilon}_1 \cdot \boldsymbol{\epsilon}_3^\dagger )(\boldsymbol{\epsilon}_2 \cdot \boldsymbol{\epsilon}_4^\dagger )(\boldsymbol{r} \cdot \nabla_r )  \nonumber\\
    	&&-C_{\sigma}\frac{im_\sigma^3}{8 \pi m^2_{D_1}} g_\sigma^2 H_2(\boldsymbol{\epsilon}_2 \cdot \boldsymbol{\epsilon}_4^\dagger)[(\boldsymbol{\epsilon}_1 \times  \boldsymbol{\epsilon}_3^\dagger)\cdot  \boldsymbol{L}]  \nonumber
    	-C_{\sigma}\frac{im_\sigma^3}{8 \pi m^2_{D_1}} g_\sigma^2 H_2(\boldsymbol{\epsilon}_1 \cdot \boldsymbol{\epsilon}_3^\dagger)[(\boldsymbol{\epsilon}_2 \times  \boldsymbol{\epsilon}_4^\dagger)\cdot  \boldsymbol{L}]  \nonumber\\
    	&&+C_{\sigma}\frac{m_\sigma^3}{24\pi m^2_{D_1}}g_\sigma^2 H_3 S(\boldsymbol{\hat{r}},\boldsymbol{\epsilon}_1,\boldsymbol{\epsilon}_3^\dagger)(\boldsymbol{\epsilon}_2 \cdot \boldsymbol{\epsilon}_4^\dagger)  \nonumber
    	+C_{\sigma}\frac{m_\sigma^3}{24\pi m^2_{D_1}}g_\sigma^2 H_3 S(\boldsymbol{\hat{r}},\boldsymbol{\epsilon}_2,\boldsymbol{\epsilon}_4^\dagger)(\boldsymbol{\epsilon}_1 \cdot \boldsymbol{\epsilon}_3^\dagger)  \nonumber\\
    	&&+C_{\sigma}\frac{m_\sigma^3}{24\pi m^2_{D_1}}g_\sigma^2 H_1(\boldsymbol{\epsilon}_1 \cdot \boldsymbol{\epsilon}_3^\dagger )(\boldsymbol{\epsilon}_2 \cdot \boldsymbol{\epsilon}_4^\dagger).
    \end{eqnarray}
    \par
    The effective potential of $\pi$ exchange is
    \begin{eqnarray}
    	V\left(r \right )=
    	&&-C_\pi \frac{m_\pi^3}{12\pi f_{\pi}^{2}}g^{\prime 2} H_3 S(\boldsymbol{\hat{r}},\boldsymbol{\epsilon}_1\times \boldsymbol{\epsilon}_3^\dagger,\boldsymbol{\epsilon}_2\times \boldsymbol{\epsilon}_4^\dagger)
    	+C_\pi \frac{m_\pi^3}{12\pi f_{\pi}^{2}}g'^2 H_1 (\boldsymbol{\epsilon}_1 \times \boldsymbol{\epsilon}_3^\dagger )\cdot (\boldsymbol{\epsilon}_2 \times \boldsymbol{\epsilon}_4^\dagger ).
    \end{eqnarray}
    \par
    The effective potential of $\eta$ exchange is similar with that of $\pi$, we just need to replace $C_\pi$ by $C_\eta$ and $m_\pi$ by $m_\eta$.\\
    \par
    The effective potential of $\rho$ exchange is
    \begin{eqnarray}
    	V\left(r \right )=
    	&&C_\rho\frac{m_\rho}{4 \pi} \beta ^2g_V^2H_0(\boldsymbol{\epsilon}_1 \cdot \boldsymbol{\epsilon}_3^\dagger )(\boldsymbol{\epsilon}_2 \cdot \boldsymbol{\epsilon}_4^\dagger )\nonumber
    	+C_\rho\frac{m_\rho^3}{6 \pi}\lambda ^2 g_V^2H_1(\boldsymbol{\epsilon}_1 \times  \boldsymbol{\epsilon}_3^\dagger )\cdot (\boldsymbol{\epsilon}_2 \times  \boldsymbol{\epsilon}_4^\dagger )\nonumber
    	+C_\rho\frac{m_\rho^3}{12\pi}\lambda ^2 g_V^2H_3 S(\boldsymbol{\hat{r}},\boldsymbol{\epsilon}_1\times \boldsymbol{\epsilon}_3^\dagger,\boldsymbol{\epsilon}_2\times \boldsymbol{\epsilon}_4^\dagger)\nonumber\\
    	&&-C_\rho\frac{im_\rho^3}{2\pi m_{D_1}}\beta \lambda g_V^2H_2(\boldsymbol{\epsilon}_2 \cdot \boldsymbol{\epsilon}_4^\dagger)[(\boldsymbol{\epsilon}_1 \times \boldsymbol{\epsilon}_3^\dagger)\cdot  \boldsymbol{L} ] \nonumber    
    	-C_\rho\frac{im_\rho^3}{2\pi m_{D_1}}\beta \lambda g_V^2H_2(\boldsymbol{\epsilon}_1 \cdot \boldsymbol{\epsilon}_3^\dagger)[(\boldsymbol{\epsilon}_2 \times \boldsymbol{\epsilon}_4^\dagger)\cdot  \boldsymbol{L} ] \nonumber\\     
    	&&+C_\rho\frac{m_\rho^3}{12 \pi m_{D_1}}\beta \lambda  g_V^2H_3S(\boldsymbol{\hat{r}},\boldsymbol{\epsilon}_1,\boldsymbol{\epsilon}_3^\dagger)(\boldsymbol{\epsilon}_2 \cdot \boldsymbol{\epsilon}_4^\dagger)\nonumber
    	+C_\rho\frac{m_\rho^3}{12 \pi m_{D_1}}\beta \lambda  g_V^2H_3S(\boldsymbol{\hat{r}},\boldsymbol{\epsilon}_2,\boldsymbol{\epsilon}_4^\dagger)(\boldsymbol{\epsilon}_1 \cdot \boldsymbol{\epsilon}_3^\dagger)\nonumber\\
    	&&-C_\rho\frac{m_\rho^3}{6 \pi m_{D_1}}\beta \lambda  g_V^2H_1(\boldsymbol{\epsilon}_1 \cdot \boldsymbol{\epsilon}_3^\dagger )(\boldsymbol{\epsilon}_2 \cdot \boldsymbol{\epsilon}_4^\dagger )\nonumber
    	-C_\rho\frac{m_\rho^3}{24 \pi m^2_{D_1}}\beta^2  g_V^2H_3S(\boldsymbol{\hat{r}},\boldsymbol{\epsilon}_1,\boldsymbol{\epsilon}_3^\dagger)(\boldsymbol{\epsilon}_2 \cdot \boldsymbol{\epsilon}_4^\dagger)\nonumber\\
    	&&-C_\rho\frac{m_\rho^3}{24 \pi m^2_{D_1}}\beta^2  g_V^2H_3S(\boldsymbol{\hat{r}},\boldsymbol{\epsilon}_2,\boldsymbol{\epsilon}_4^\dagger)(\boldsymbol{\epsilon}_1 \cdot \boldsymbol{\epsilon}_3^\dagger)\nonumber
    	+C_\rho\frac{im_\rho^3}{8\pi m^2_{D_1}}\beta^2  g_V^2H_2(\boldsymbol{\epsilon}_2 \cdot \boldsymbol{\epsilon}_4^\dagger)[(\boldsymbol{\epsilon}_1 \times \boldsymbol{\epsilon}_3^\dagger)\cdot  \boldsymbol{L}] \nonumber\\    
    	&&+C_\rho\frac{im_\rho^3}{8\pi m^2_{D_1}}\beta^2  g_V^2H_2(\boldsymbol{\epsilon}_1 \cdot \boldsymbol{\epsilon}_3^\dagger)[(\boldsymbol{\epsilon}_2 \times \boldsymbol{\epsilon}_4^\dagger)\cdot  \boldsymbol{L}] \nonumber
    	-C_\rho\frac{m_\rho}{4 \pi m^2_{D_1}}\beta ^2 g_V^2H_0(\boldsymbol{\epsilon}_1 \cdot \boldsymbol{\epsilon}_3^\dagger )(\boldsymbol{\epsilon}_2 \cdot \boldsymbol{\epsilon}_4^\dagger )\nabla_r^2\nonumber\\
    	&&-C_\rho\frac{m_\rho^3}{4 \pi m^2_{D_1}}\beta ^2 g_V^2H_2(\boldsymbol{\epsilon}_1 \cdot \boldsymbol{\epsilon}_3^\dagger )(\boldsymbol{\epsilon}_2 \cdot \boldsymbol{\epsilon}_4^\dagger )(\boldsymbol{r} \cdot \nabla_r)
    	+C_\rho\frac{7m_\rho^3}{48\pi m^2_{D_1}}\beta^2 g_V^2H_1 (\boldsymbol{\epsilon}_1 \cdot \boldsymbol{\epsilon}_3^\dagger )(\boldsymbol{\epsilon}_2 \cdot \boldsymbol{\epsilon}_4^\dagger ).
    \end{eqnarray}
    \par
    Similarly, we can get the effective potential of $\omega$ by changing the isospin factor and the mass of light mesons.
    \par
    We list the definition of $S(\hat{\boldsymbol{r}},\boldsymbol{a},\boldsymbol{b})$ and the expressions of $H_{0}$, $H_{1}$ and $H_{3}$ in the Appendix.

    \subsection{Schr{\"o}dinger equation}\label{subsecd}
    We should solve the Schr{\"o}dinger equation to confirm if there exists a bound state in our systems
    \begin{eqnarray}
    	\left (H_0(\boldsymbol{r})+V(\boldsymbol{r})\right ) \mid \psi \rangle=E\mid \psi \rangle 
    \end{eqnarray}
    with
    \begin{eqnarray}
    	&& H_0(\boldsymbol{r})=-\frac{1}{2m} \nabla^{2},\\
    	&& \nabla^{2}=\frac{1}{r} \frac{\mathrm{d} ^{2}}{\mathrm{d } r^{2}} r-\frac{\boldsymbol{L}^{2}}{r^{2}}.
    \end{eqnarray}   
    \par
    Considering the $S$\--{}$D$ wave mixing effect, we unfold the Schr{\"o}dinger equation by Eqs (\ref{2})$-$(\ref{4})
    \begin{eqnarray}
    	\begin{pmatrix}
    		H_{0}^{SS}(\boldsymbol{r})+V^{S S}(\boldsymbol{r}) & V^{S D}(\boldsymbol{r}) \\
    		V^{D S}(\boldsymbol{r}) & H_{0}^{DD}(\boldsymbol{r})+V^{D D}(\boldsymbol{r})
    	\end{pmatrix}
    	\begin{pmatrix}
    		R_{S} \\
    		R_{D}
    	\end{pmatrix}
    	=E
    	\begin{pmatrix}
    		R_{S} \\
    		R_{D}
    	\end{pmatrix}
    \end{eqnarray} 
    with
    \begin{eqnarray}
    	\begin{array}{ll}
    		H_0^{S S}(\boldsymbol{r})=\left\langle{ }^{1} S_{0}|H_0(\boldsymbol{r})|{ }^{1} S_{0}\right\rangle, & H_0^{S D}(\boldsymbol{r})=\left\langle{ }^{1} S_{0}|H_0(\boldsymbol{r})|{ }^{5} D_{0}\right\rangle, \\
    		H_0^{D S}(\boldsymbol{r})=\left\langle{ }^{5} D_{0}|H_0(\boldsymbol{r})|{ }^{1} S_{0}\right\rangle, & H_0^{D D}(\boldsymbol{r})=\left\langle{ }^{5} D_{0}|H_0(\boldsymbol{r})|{ }^{5} D_{0}\right\rangle
    	\end{array}
    \end{eqnarray}
    and
    \begin{eqnarray}
    	\begin{array}{ll}
    		V^{S S}(\boldsymbol{r})=\left\langle{ }^{1} S_{0}|V(\boldsymbol{r})|{ }^{1} S_{0}\right\rangle, & V^{S D}(\boldsymbol{r})=\left\langle{ }^{1} S_{0}|V(\boldsymbol{r})|{ }^{5} D_{0}\right\rangle, \\
    		V^{D S}(\boldsymbol{r})=\left\langle{ }^{5} D_{0}|V(\boldsymbol{r})|{ }^{1} S_{0}\right\rangle, & V^{D D}(\boldsymbol{r})=\left\langle{ }^{5} D_{0}|V(\boldsymbol{r})|{ }^{5} D_{0}\right\rangle
    	\end{array}
    \end{eqnarray} 
    for $^1S_0\leftrightarrow {^5}D_0$ channel.
    \begin{eqnarray}
    	\begin{pmatrix}
    		H_{0}^{SS}(\boldsymbol{r})+V^{S S}(\boldsymbol{r}) & V^{S D}(\boldsymbol{r}) \\
    		V^{D S}(\boldsymbol{r}) & H_{0}^{DD}(\boldsymbol{r})+V^{D D}(\boldsymbol{r})
    	\end{pmatrix}
    	\begin{pmatrix}
    		R_{S}' \\
    		R_{D}'
    	\end{pmatrix}=E
    	\begin{pmatrix}
    		R_{S}' \\
    		R_{D}'
    	\end{pmatrix}    	
    \end{eqnarray}
    with 
    \begin{eqnarray}
    	\begin{array}{ll}
    		H_0^{S S}(\boldsymbol{r})=\left\langle{ }^{3} S_{1}| H_0(\boldsymbol{r})|{ }^{3} S_{1}\right\rangle, & H_0^{S D}(\boldsymbol{r})=\left\langle{ }^{3} S_{1}|H_0(\boldsymbol{r})|{ }^{3} D_{1}\right\rangle, \\
    		H_0^{D S}(\boldsymbol{r})=\left\langle{ }^{3} D_{1}|H_0(\boldsymbol{r})|{ }^{3} S_{1}\right\rangle, & H_0^{D D}(\boldsymbol{r})=\left\langle{ }^{3} D_{1}|H_0(\boldsymbol{r})|{ }^{3} D_{1}\right\rangle
    	\end{array}
    \end{eqnarray}
    and
    \begin{eqnarray}
    	\begin{array}{ll}
    		V^{S S}(\boldsymbol{r})=\left\langle{ }^{3} S_{1}|V(\boldsymbol{r})|{ }^{3} S_{1}\right\rangle, & V^{S D}(\boldsymbol{r})=\left\langle{ }^{3} S_{1}|V(\boldsymbol{r})|{ }^{3} D_{1}\right\rangle, \\
    		V^{D S}(\boldsymbol{r})=\left\langle{ }^{3} D_{1}|V(\boldsymbol{r})|{ }^{3} S_{1}\right\rangle, & V^{D D}(\boldsymbol{r})=\left\langle{ }^{3} D_{1}|V(\boldsymbol{r})|{ }^{3} D_{1}\right\rangle
    	\end{array}
    \end{eqnarray} 
    for $^3S_1\leftrightarrow {^3}D_1$ channel.
	\begin{eqnarray}
		\begin{pmatrix}
			H_{0}^{SS}(\boldsymbol{r})+V^{S S}(\boldsymbol{r}) & V^{S D_1}(\boldsymbol{r}) & V^{S D_2}(\boldsymbol{r})\\
			V^{D_1 S}(\boldsymbol{r}) & H_{0}^{D_1D_1}(\boldsymbol{r})+V^{D_1 D_1}(\boldsymbol{r}) &V^{D_1 D_2}(\boldsymbol{r})\\
			V^{D_2 S}(\boldsymbol{r}) & V^{D_2 D_1}(\boldsymbol{r}) & H_{0}^{D_2D_2}(\boldsymbol{r})+V^{D_2 D_2}(\boldsymbol{r})
		\end{pmatrix}
		\begin{pmatrix}
			R_{S}'' \\
			R_{D_1}''\\
			R_{D_2}''
		\end{pmatrix}=E
		\begin{pmatrix}
			R_{S}'' \\
			R_{D_1}''\\
			R_{D_2}''
		\end{pmatrix}
	\end{eqnarray}
	with
	\begin{eqnarray}
		\begin{array}{lll}
			H_0^{S S}(\boldsymbol{r})=\left\langle{ }^{5} S_{2}|H_0(\boldsymbol{r})|{ }^{5} S_{2}\right\rangle, & H_0^{S D_1}(\boldsymbol{r})=\left\langle{ }^{5} S_{2}|H_0(\boldsymbol{r})|{ }^{1} D_{2}  \right\rangle, & H_0^{S D_2}(\boldsymbol{r})=\left\langle{ }^{5} S_{2}|H_0(\boldsymbol{r})|{ }^{5} D_{2}  \right\rangle, \\
			H_0^{D_1 S}(\boldsymbol{r})=\left\langle{ }^{1} D_{2}|H_0(\boldsymbol{r})|{ }^{5} S_{2}\right\rangle, & H_0^{D_1 D_1}(\boldsymbol{r})=\left\langle{ }^{1} D_{2}|H_0(\boldsymbol{r})|{ }^{1} D_{2}\right\rangle, & H_0^{D_1 D_2}(\boldsymbol{r})=\left\langle{ }^{1} D_{2}|H_0(\boldsymbol{r})|{ }^{5} D_{2}\right\rangle, \\
			H_0^{D_2 S}(\boldsymbol{r})=\left\langle{ }^{5} D_{2}|H_0(\boldsymbol{r})|{ }^{5} S_{2}\right\rangle, & H_0^{D_2 D_1}(\boldsymbol{r})=\left\langle{ }^{5} D_{2}|H_0(\boldsymbol{r})|{ }^{1} D_{2}\right\rangle, & H_0^{D_2 D_2}(\boldsymbol{r})=\left\langle{ }^{5} D_{2}|H_0(\boldsymbol{r})|{ }^{5} D_{2}\right\rangle 
		\end{array}
	\end{eqnarray}
	and
	\begin{eqnarray}
		\begin{array}{lll}
			V^{S S}(\boldsymbol{r})=\left\langle{ }^{5} S_{2}|V(\boldsymbol{r})|{ }^{5} S_{2}\right\rangle, & V^{S D_1}(\boldsymbol{r})=\left\langle{ }^{5} S_{2}|V(\boldsymbol{r})|{ }^{1} D_{2}  \right\rangle, & V^{S D_2}(\boldsymbol{r})=\left\langle{ }^{5} S_{2}|V(\boldsymbol{r})|{ }^{5} D_{2}  \right\rangle, \\
			V^{D_1 S}(\boldsymbol{r})=\left\langle{ }^{1} D_{2}|V(\boldsymbol{r})|{ }^{5} S_{2}\right\rangle, & V^{D_1 D_1}(\boldsymbol{r})=\left\langle{ }^{1} D_{2}|V(\boldsymbol{r})|{ }^{1} D_{2}\right\rangle, & V^{D_1 D_2}(\boldsymbol{r})=\left\langle{ }^{1} D_{2}|V(\boldsymbol{r})|{ }^{5} D_{2}\right\rangle, \\
			V^{D_2 S}(\boldsymbol{r})=\left\langle{ }^{5} D_{2}|V(\boldsymbol{r})|{ }^{5} S_{2}\right\rangle, & V^{D_2 D_1}(\boldsymbol{r})=\left\langle{ }^{5} D_{2}|V(\boldsymbol{r})|{ }^{1} D_{2}\right\rangle, & V^{D_2 D_2}(\boldsymbol{r})=\left\langle{ }^{5} D_{2}|V(\boldsymbol{r})|{ }^{5} D_{2}\right\rangle 
		\end{array}
	\end{eqnarray}  			
    for $^5S_2\leftrightarrow {^1}D_2\leftrightarrow {^5}D_2$ channel.
	\par
	There are some angular-momentum related terms in the effective potentials. We label these operators by the mark $O_1$ to $O_7$ in the Appendix. $O_1$ and $O_2$ correspond to spin-spin interactions, $O_3$, $O_4$ and $O_5$ correspond to tensor force, while $O_6$ and $O_7$ correspond to spin-orbit force. We list the matrix elements of these operators in the Appendix.

	\section{Numerical results} \label{sec3}
	We diagonalize the total Hamiltonian matrix to get the
	eigenvalues and eigenvectors of our systems. If there exists a negative eigenvalue in the system, this system can be a bound state. Further, if the root-mean-square (RMS) radius of the system is in a proper range, it can be a molecule state. In our work, we take the cutoff value $\Lambda$ in the range of $0.65 \le \Lambda \le 4.90$ GeV.  
	\subsection{The $D_1 D_1$ system} \label{subsecA}
	In TABLE \ref{tab:4}-\ref{tab:6}, we present the numerical results of the $D_1 D_1$ with $I(J^P)=0(1^+)$, $I(J^P)=1(0^+)$ and $I(J^P)=1(2^+)$. In order to find out each meson's effect, we present the $S$ wave contributions from each meson. Besides, we plot the contributions from the $S$ and the $D$ wave, and the effect of $S$\--{}$D$ wave mixing together. We also present the wave functions if there exists a bound state solution.
	\par
\begin{figure}[htbp]
		\begin{minipage}{0.5\textwidth}
			\centering
			\includegraphics[scale=0.7]{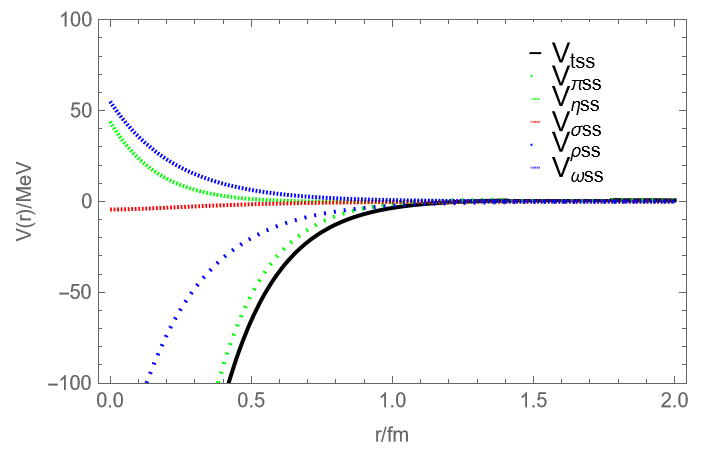}
			\caption{The $S$ wave potentials of the $D_1 D_1$ system with $I(J^P)=0(1^+)$ when the cutoff parameter is fixed at 1.05 GeV without recoil corrections}
			\label{fig:1}
		\end{minipage}
		\begin{minipage}{0.49\textwidth}
			\centering
			\includegraphics[scale=0.7]{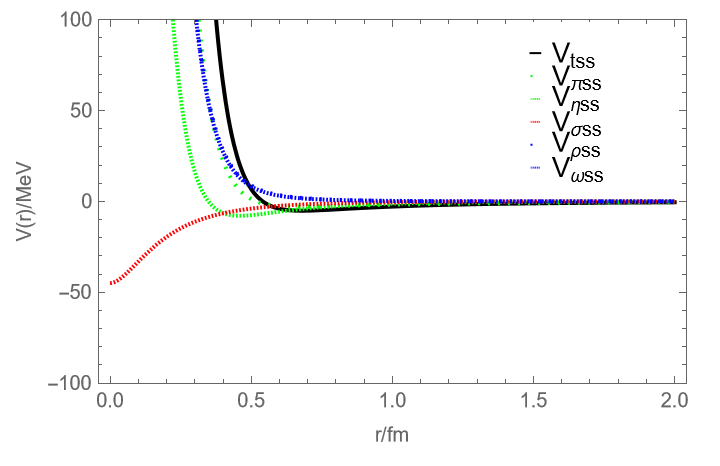}
				\caption{The $S$ wave potentials of the $D_1 D_1$ system with $I(J^P)=1(0^+)$ when the cutoff parameter is fixed at 3.03 GeV without recoil corrections}
				\label{fig:2}	
		\end{minipage}
    \end{figure}
\begin{figure}[htbp]	
    		\centering
    		\includegraphics[scale=0.7]{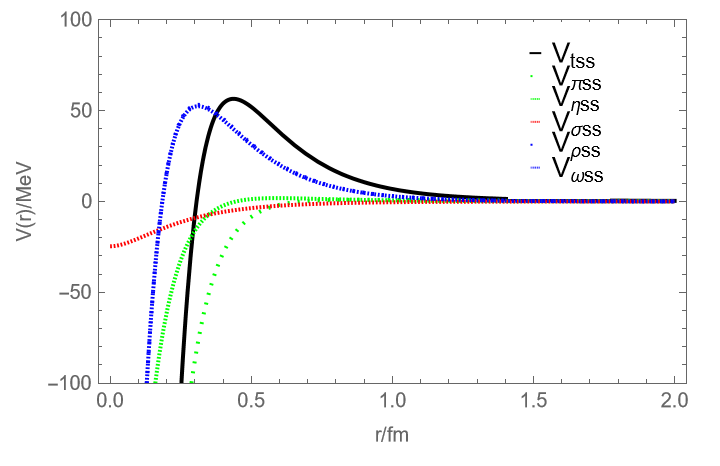}
    		\caption{The $S$ wave potentials of the $D_1 D_1$ system with $I(J^P)=1(2^+)$ when the cutoff parameter is fixed at 2.10 GeV without recoil corrections}
    		\label{fig:3}	
    \end{figure}
\begin{figure}[htbp]
		\begin{minipage}{0.49\textwidth}
			\centering
			\includegraphics[scale=0.7]{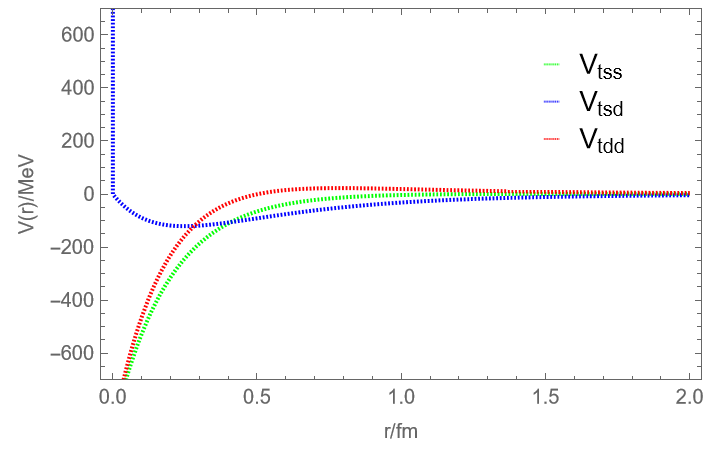}
			\caption{The $S$\--$D$ wave mixing effects of the $D_1 D_1$ system with $I(J^P)=0(1^+)$ when the cutoff parameter is fixed at 1.05 GeV without recoil corrections}
			\label{fig:4}
		\end{minipage}
		\begin{minipage}{0.49\textwidth}
			\centering
			\includegraphics[scale=0.7]{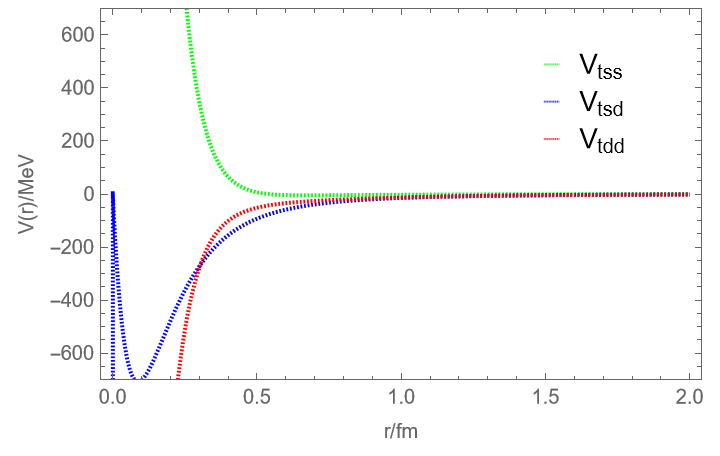}
			\caption{The $S$\--$D$ wave mixing effects of the $D_1 D_1$ system with $I(J^P)=1(0^+)$ when the cutoff parameter is fixed at 3.03 GeV without recoil corrections}
			\label{fig:5}	
		\end{minipage}
	\end{figure}
\begin{figure}[htbp]
			\centering
			\includegraphics[scale=0.7]{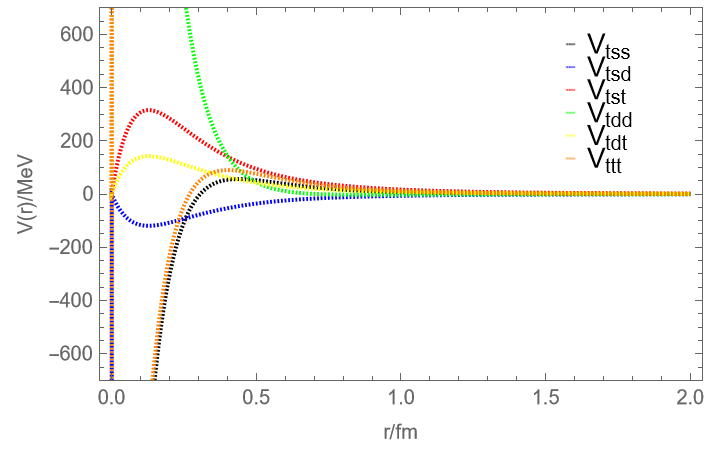}
			\caption{The $S$\--$D$ wave mixing effects of the $D_1 D_1$ system with $I(J^P)=1(2^+)$ when the cutoff parameter is fixed at 2.10 GeV without recoil corrections}
			\label{fig:6}
	\end{figure}
\begin{figure}[htbp]
		\begin{minipage}{0.49\textwidth}
			\centering
			\includegraphics[scale=0.7]{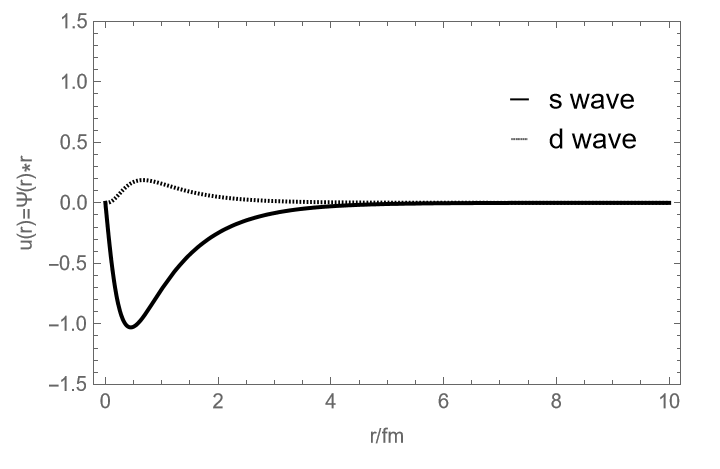}
			\caption{The wave functions of the $D_1 D_1$ system with $I(J^P)=0(1^+)$ when the cutoff parameter is fixed at 1.05 GeV without recoil corrections}
			\label{fig:7}
		\end{minipage}
		\begin{minipage}{0.49\textwidth}
			\centering
			\includegraphics[scale=0.7]{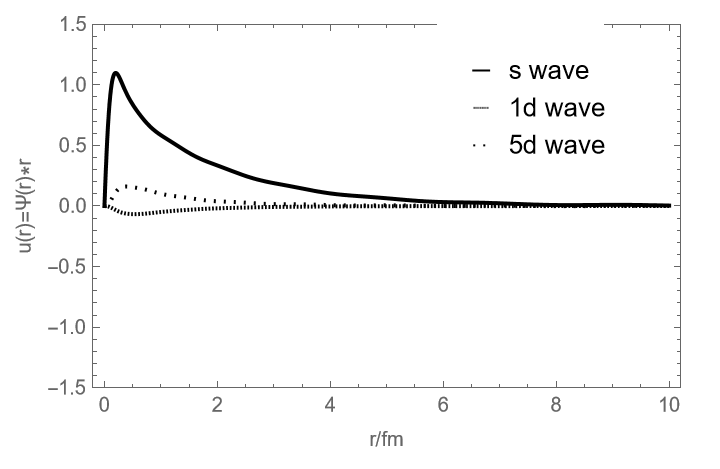}
			\caption{The wave functions of the $D_1 D_1$ system with $I(J^P)=1(2^+)$ when the cutoff parameter is fixed at 2.10 GeV without recoil corrections}
			\label{fig:8}	
		\end{minipage}
	\end{figure}
\begin{figure}[htbp]
		\begin{minipage}{0.5\textwidth}
			\centering
			\includegraphics[scale=0.7]{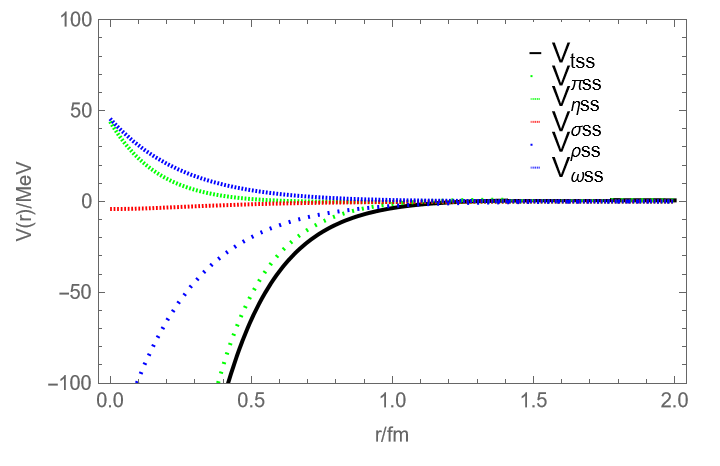}
			\caption{The $S$ wave potentials of the $D_1 D_1$ system with $I(J^P)=0(1^+)$ when the cutoff parameter is fixed at 1.05 GeV with recoil corrections}
			\label{fig:9}
		\end{minipage}
		\begin{minipage}{0.49\textwidth}
			\centering
			\includegraphics[scale=0.7]{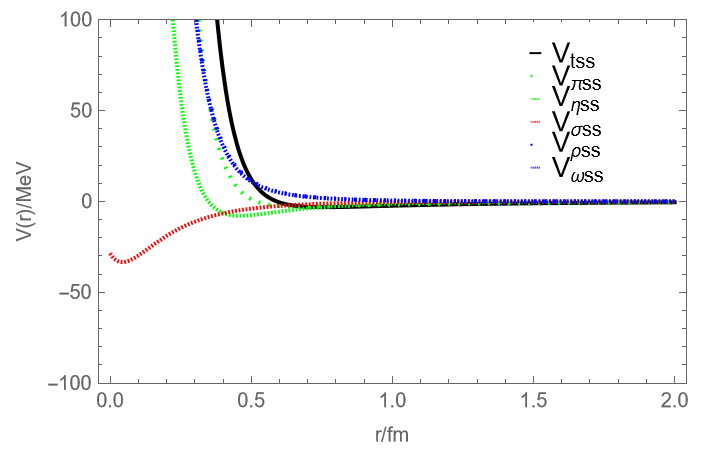}
			\caption{The $S$ wave potentials of the $D_1 D_1$ system with $I(J^P)=1(0^+)$ when the cutoff parameter is fixed at 3.03 GeV with recoil corrections}
			\label{fig:10}	
		\end{minipage}
	\end{figure}
\begin{figure}[htbp]	
		\centering
		\includegraphics[scale=0.7]{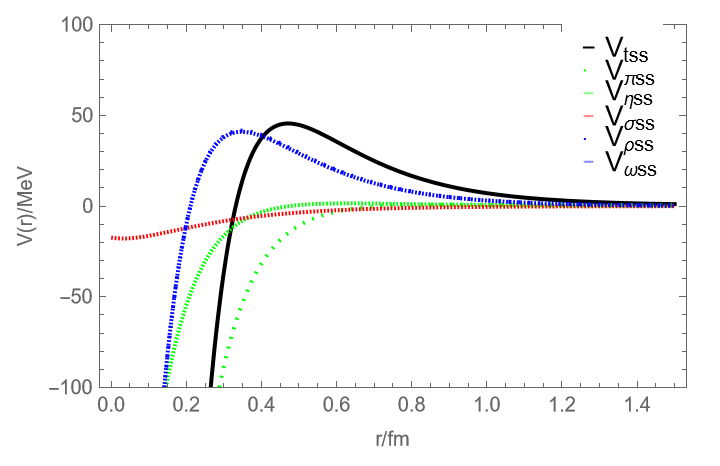}
		\caption{The $S$ wave potentials of the $D_1 D_1$ system with $I(J^P)=1(2^+)$ when the cutoff parameter is fixed at 1.92 GeV with recoil corrections}
		\label{fig:11}	
	\end{figure}
\begin{figure}[htbp]
		\begin{minipage}{0.49\textwidth}
			\centering
			\includegraphics[scale=0.7]{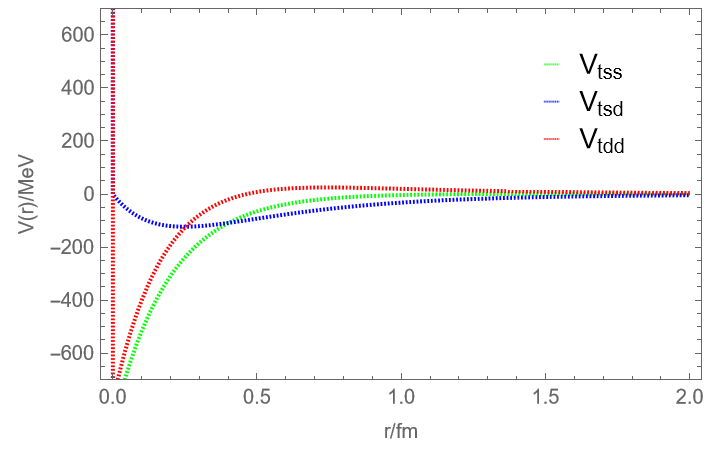}
			\caption{The $S$\--$D$ wave mixing effects of the $D_1 D_1$ system with $I(J^P)=0(1^+)$ when the cutoff parameter is fixed at 1.05 GeV with recoil corrections}
			\label{fig:12}
		\end{minipage}
		\begin{minipage}{0.49\textwidth}
			\centering
			\includegraphics[scale=0.7]{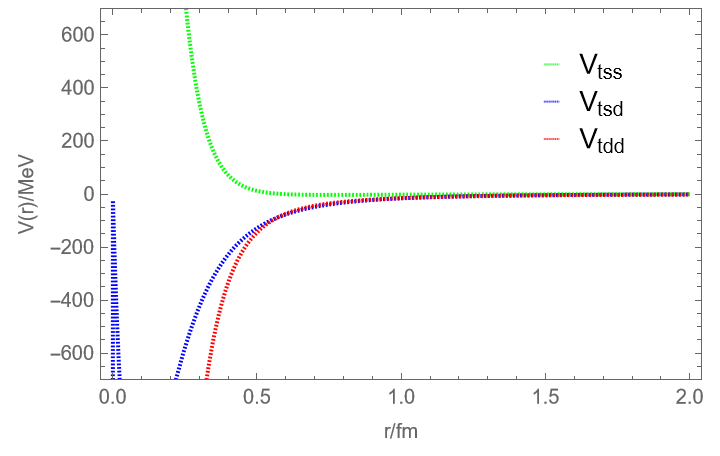}
			\caption{The $S$\--$D$ wave mixing effects of the $D_1 D_1$ system with $I(J^P)=1(0^+)$ when the cutoff parameter is fixed at 3.03 GeV with recoil corrections}
			\label{fig:13}	
		\end{minipage}
	\end{figure}
\begin{figure}[htbp]
		\centering
		\includegraphics[scale=0.7]{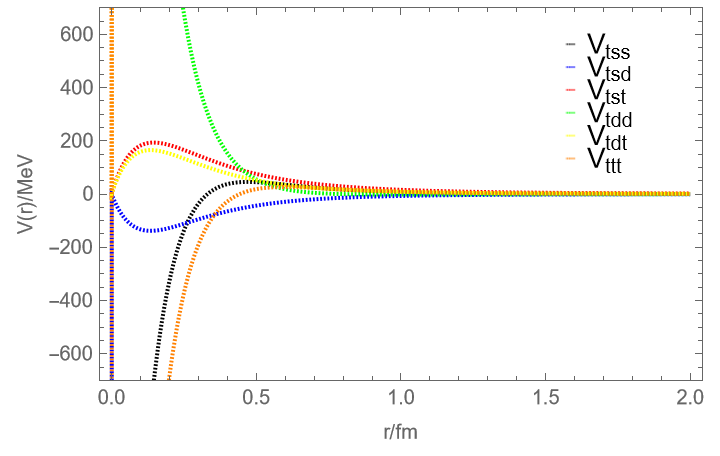}
		\caption{The $S$\--$D$ wave mixing effects the of $D_1 D_1$ system with $I(J^P)=1(2^+)$ when the cutoff parameter is fixed at 1.92 GeV with recoil corrections}
		\label{fig:14}
	\end{figure}
\begin{figure}[htbp]
		\begin{minipage}{0.49\textwidth}
			\centering
			\includegraphics[scale=0.7]{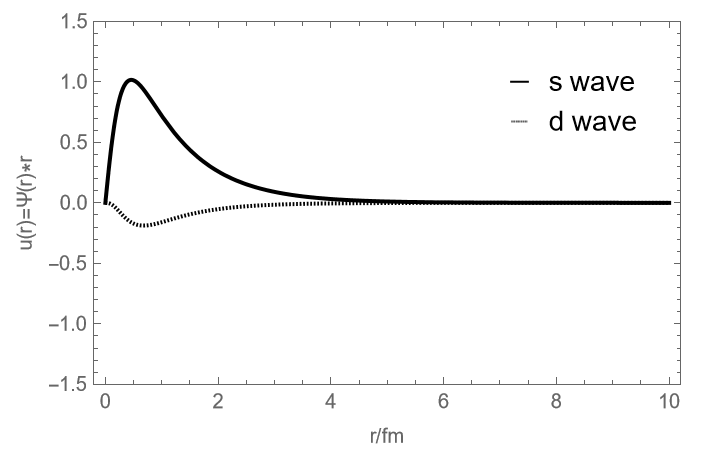}
			\caption{The wave functions of the $D_1 D_1$ system with $I(J^P)=0(1^+)$ when the cutoff parameter is fixed at 1.05 GeV with recoil corrections}
			\label{fig:15}
		\end{minipage}
		\begin{minipage}{0.49\textwidth}
			\centering
			\includegraphics[scale=0.7]{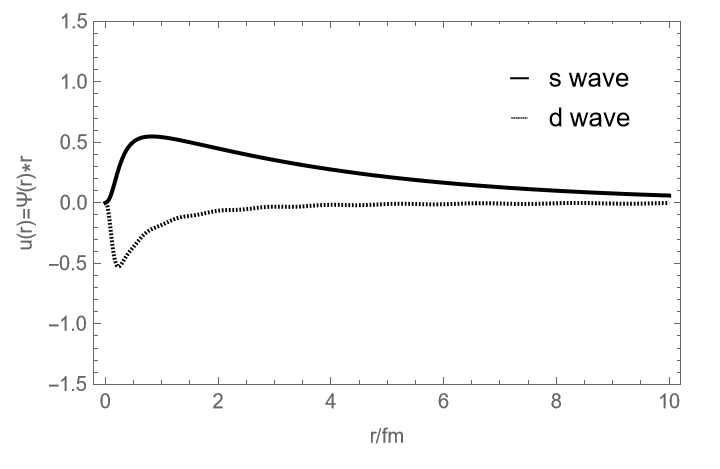}
			\caption{The wave functions of the $D_1 D_1$ system with $I(J^P)=1(0^+)$ when the cutoff parameter is fixed at 3.03 GeV with recoil corrections}
			\label{fig:16}	
		\end{minipage}
	\end{figure}
\begin{figure}[htbp]
		\centering
		\includegraphics[scale=0.7]{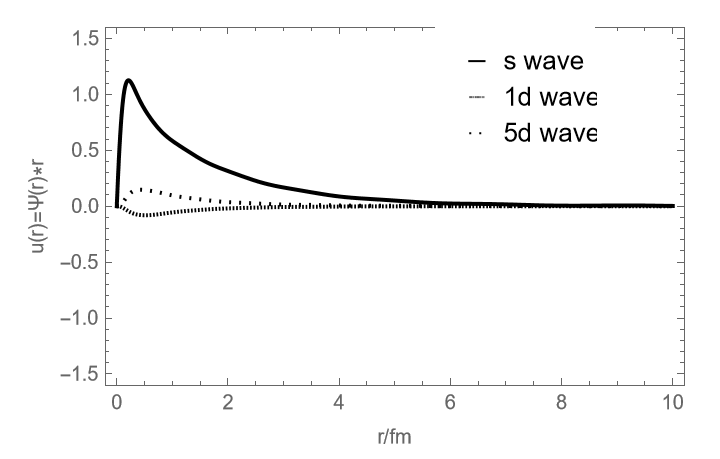}
		\caption{The wave functions of the $D_1 D_1$ system with $I(J^P)=1(2^+)$ when the cutoff parameter is fixed at 1.92 GeV with recoil corrections}
		\label{fig:17}
	\end{figure}

	For the cases without the recoil corrections, the $S$ wave contributions to the potential are shown in FIG. \ref{fig:1}-\ref{fig:3}. The contributions from the $S$ and the $D$ wave, and the effect of $S$\--{}$D$ wave mixing are shown in FIG. \ref{fig:4}-\ref{fig:6}. The wave functions are shown in FIG. \ref{fig:7}-\ref{fig:8}.
	For the cases with the recoil corrections, the $S$ wave contributions are shown in FIG. \ref{fig:9}-\ref{fig:11}. The contributions from the $S$ and the $D$ wave, and the effect of $S$\--{}$D$ wave mixing are shown in FIG. \ref{fig:12}-\ref{fig:14}. The wave functions are shown in FIG. \ref{fig:15}-\ref{fig:17}.
	\par
	For the $I(J^P)=0(1^+)$ channel in TABLE \ref{tab:4}, without regard to the recoil corrections, the bound state appears when taking the cutoff value $\Lambda=0.85$ GeV with the binding energy $0.34$ MeV and the RMS radius $4.68$ fm. The $\pi$ and $\rho$ meson exchange provide $S$ wave with a remarkable attractive potential. When the cutoff value is taken as $\Lambda=1.05$ GeV, the binding energy increases to $18.9$ MeV and the RMS radius decreases to $0.93$ fm. After considering the recoil corrections, the binding energy decreases while the RMS radius increases. When the cutoff value is taken as $\Lambda=1.05$ GeV, the binding energy is $17.8$ MeV and the RMS radius is $0.95$ fm.
		\begin{table*}[htbp]
		\scriptsize
		\begin{center}
			\caption{\label{tab:4} The numerical results of  $D_1 D_1 $ $I(J^P)=0(1^+)$ system }
			\begin{tabular}{l cccc|ccccc}\toprule[1pt]
				\multicolumn{5}{c|}{Without recoil corrections} & \multicolumn{5}{c}{With recoil corrections}\\
				\midrule[1pt]
				$\Lambda$(GeV) & B.E.(MeV) & RMS(fm)  & $^3S_1 $($\%$)  & $^3D_1$($\%$) & $\Lambda$(GeV) & B.E.(MeV) & RMS(fm)  & $^3S_1 $($\%$)  & $^3D_1$($\%$) \\  
				0.85 & 0.34 & 4.68 & 98.36 & 1.64 & 0.85 & 0.33 & 4.72 & 98.37 & 1.63 \\
				0.90 & 2.03 & 2.30 & 97.39 & 2.61 & 0.90 & 1.96 & 2.33 & 97.40 & 2.60 \\
				0.95 & 5.46 & 1.52 & 96.97 & 3.03 & 0.95 & 5.23 & 1.55 & 96.97 & 3.03 \\
				1.00 & 11.0 & 1.15 & 96.81 & 3.19 & 1.00 & 9.67 & 1.21 & 96.78 & 3.22 \\
				1.05 & 18.9 & 0.93 & 96.79 & 3.21 & 1.05 & 17.8 & 0.95 & 96.74 & 3.26 \\
				\bottomrule[1pt]				
			\end{tabular}
		\end{center}
	\end{table*} 
	\par
	For the $I(J^P)=1(0^+)$ channel in TABLE \ref{tab:5}, we fail to find a bound state in the heavy quark limit. The  $S$ wave attractive potential is quite shallow but the $D$ wave contribute a strong attractive potential. Besides, the $S$\--{}$D$ wave mixing effect is pronounced, which increases the proportion of the $D$ wave. After considering the recoil corrections, the bound state appears when taking the cutoff value $\Lambda=3.03$ GeV, with the binding energy $1.03$ MeV and the RMS radius $3.19$ fm. When the cutoff value is taken as $\Lambda=3.07$ GeV, the binding energy increases to $10.3$ MeV and the RMS radius decreases to $1.21$ fm.
		\begin{table*}[htbp]
		\scriptsize
		\begin{center}
			\caption{\label{tab:5} The numerical results of  $D_1 D_1 $ $I(J^P)=1(0^+)$ system }
			\begin{tabular}{lccccc}\toprule[1pt]
			    &\multicolumn{5}{c}{With recoil corrections}\\
				\midrule[1pt]
				&$\Lambda$(GeV) & B.E.(MeV) & RMS(fm)  & $^1S_0 $($\%$)  & $^5D_0$($\%$)  \\  
			& 3.03 & 1.03 & 3.19 & 85.78 & 14.22 \\
			& 3.04 & 2.33 & 2.24 & 79.48 & 20.52 \\
			& 3.05 & 4.27 & 1.73 & 72.96 & 27.04 \\
			& 3.06 & 6.91 & 1.41 & 67.07 & 32.93 \\
			& 3.07 & 10.3 & 1.21 & 61.06 & 38.94 \\

				\bottomrule[1pt]
				
			\end{tabular}
		\end{center}
	\end{table*} 
	\par 
	For the $I(J^P)=1(2^+)$ channel in TABLE \ref{tab:6}, without regard to the recoil corrections, the bound state appears when taking the cutoff value $\Lambda=2.06$ GeV, with the binding energy $0.56$ MeV and the RMS radius $3.66$ fm. All the five mesons' exchanges provide $S$ wave with an attractive potential. When the cutoff value is taken as $\Lambda=2.10$ GeV, the binding energy increases to $5.09$ MeV and the RMS radius decreases to $1.28$ fm. After considering the recoil corrections, the binding energy increases while the RMS radius decreases. The bound state appears when the cutoff value is $\Lambda=1.88$ GeV, with the binding energy $0.89$ MeV and the RMS radius $2.95$ fm.	
	\begin{table*}[htbp]
		\scriptsize
		\begin{center}
			\caption{\label{tab:6} The numerical results of  $D_1 D_1 $ $I(J^P)=1(2^+)$ system }
			\begin{tabular}{l ccccc|ccccccc}\toprule[1pt]
				\multicolumn{6}{c|}{Without recoil corrections} & \multicolumn{6}{c}{With recoil corrections}\\
				\midrule[1pt]
				$\Lambda$(GeV) & B.E.(MeV) & RMS(fm)  & $^5S_2 $($\%$)  & $^1D_2$($\%$) & $^5D_2$($\%$) & $\Lambda$(GeV) & B.E.(MeV) & RMS(fm)  & $^5S_2 $($\%$)  & $^1D_2$($\%$) & $^5D_2$($\%$)\\ 
				2.06 & 0.56 & 3.66 & 98.30 & 0.28 & 1.42 & 1.88 & 0.89 & 2.95 & 98.21 & 0.41 &1.38\\ 
				2.07 & 1.32 & 2.45 & 97.86 & 0.35 & 1.79 & 1.89 & 1.83 & 2.07 & 97.90 & 0.48 &1.62\\
				2.08 & 2.34 & 1.84 & 97.60 & 0.39 & 2.01 & 1.90 & 3.04 & 1.61 & 97.72 & 0.52 &1.76\\
				2.09 & 3.60 & 1.50 & 97.45 & 0.41 & 2.14 & 1.91 & 4.50 & 1.34 & 97.62 & 0.55 &1.83\\
				2.10 & 5.09 & 1.28 & 97.36 & 0.42 & 2.22 & 1.92 & 6.21 & 1.15 & 97.57 & 0.56 &1.87\\
				\bottomrule[1pt]	
			\end{tabular}
		\end{center}
	\end{table*}
	\par 
	From the numerical results, we find that the $I(J^P)=0(1^+)$ channel is most likely to form a loosely bound state. For the isospin singlets, the recoil corrections provide a slight negative contribution in forming bound states. While for the isospin triplets, the recoil corrections provide a remarkable positive contribution. Particularly, the recoil corrections make it possible for the $I(J^P)=1(0^+)$ channel to form a loosely bound state. Besides, the $S$ wave is the main component of the wave function in all channels.\\
    

	\clearpage

		\subsection{The $D_1 \bar D_1$ system} \label{subsecB}
	In TABLE \ref{tab:7}-\ref{tab:11}, we present the numerical results of the $D_1 \bar D_1$ with $I(J^P)=0(0^+)$, $I(J^P)=0(1^+)$, $I(J^P)=0(2^+)$, $I(J^P)=1(0^+)$ and $I(J^P)=1(1^+)$. We don't present the results of the $I(J^P)=1(2^+)$ channel because we fail to find a bound state solution whether considering the recoil corrections or not.
	\par
\begin{figure}[htbp]
    	\begin{minipage}{0.49\textwidth}
    		\centering
    		\includegraphics[scale=0.7]{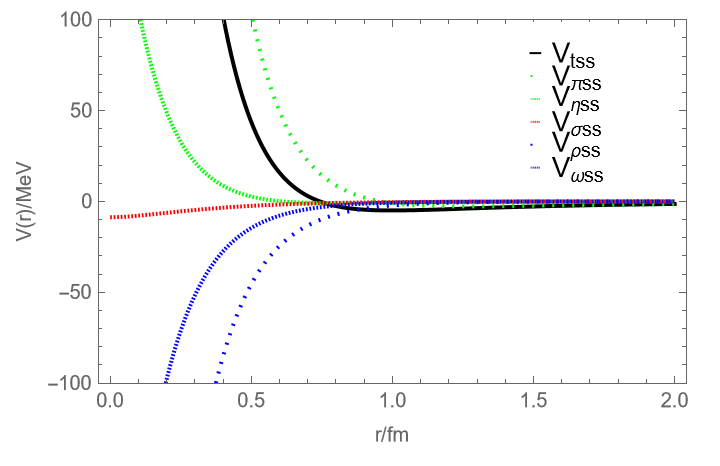}
    		\caption{The $S$ wave potentials of the $D_1 \bar D_1$ system with $I(J^P)=0(0^+)$ when the cutoff parameter is fixed at 1.30 GeV without recoil corrections }
    		\label{fig:18}
    	\end{minipage}
    	\begin{minipage}{0.49\textwidth}
    		\centering
    		\includegraphics[scale=0.7]{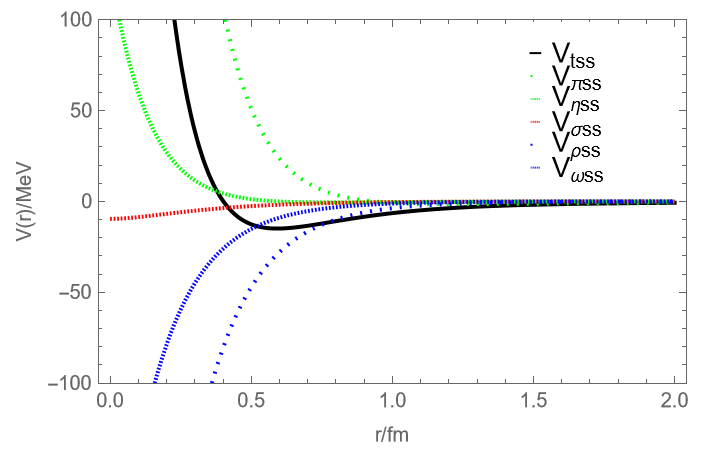}
    		\caption{The $S$ wave potentials of the $D_1 \bar D_1$ system with $I(J^P)=0(1^+)$ when the cutoff parameter is fixed at 1.35 GeV without recoil corrections}
    		\label{fig:19}	
    	\end{minipage}
        \end{figure}
\begin{figure}[htbp]
    		\centering
    		\includegraphics[scale=0.7]{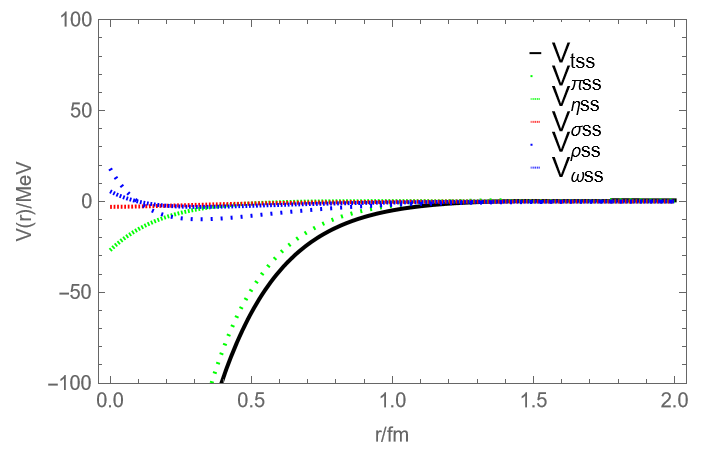}
    		\caption{The $S$ wave potentials of the $D_1 \bar D_1$ system with $I(J^P)=0(2^+)$ when the cutoff parameter is fixed at 0.95 GeV without recoil corrections}
    		\label{fig:20}
        \end{figure}
\begin{figure}[htbp]
    	\begin{minipage}{0.49\textwidth}
    		\centering
    		\includegraphics[scale=0.7]{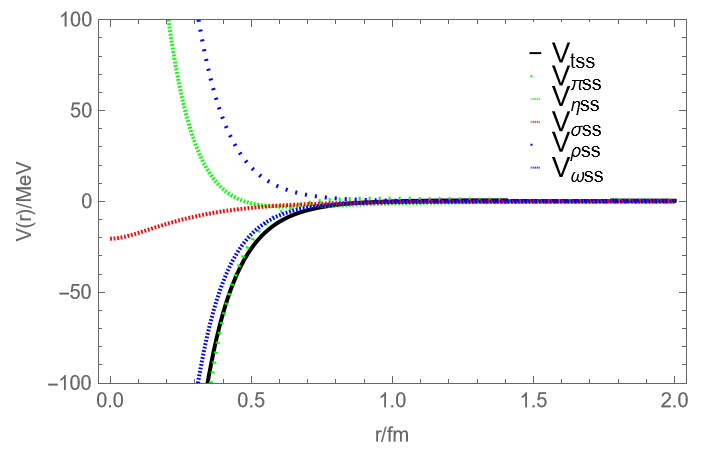}
    		\caption{The $S$ wave potentials of the $D_1 \bar D_1$ system with $I(J^P)=1(0^+)$ when the cutoff parameter is fixed at 1.90 GeV without recoil corrections}
    		\label{fig:21}	
    	\end{minipage}
    	\begin{minipage}{0.49\textwidth}
    		\centering
    		\includegraphics[scale=0.7]{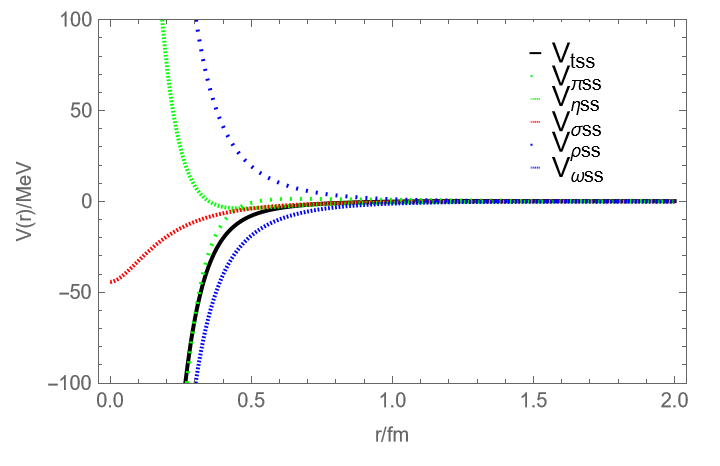}
    		\caption{The $S$ wave potentials of the $D_1 \bar D_1$ system with $I(J^P)=1(1^+)$ when the cutoff parameter is fixed at 3.00 GeV without recoil corrections}
    		\label{fig:22}
    	\end{minipage} 	
       \end{figure}
\begin{figure}[htbp]
    	\begin{minipage}{0.49\textwidth}
    		\centering
    		\includegraphics[scale=0.7]{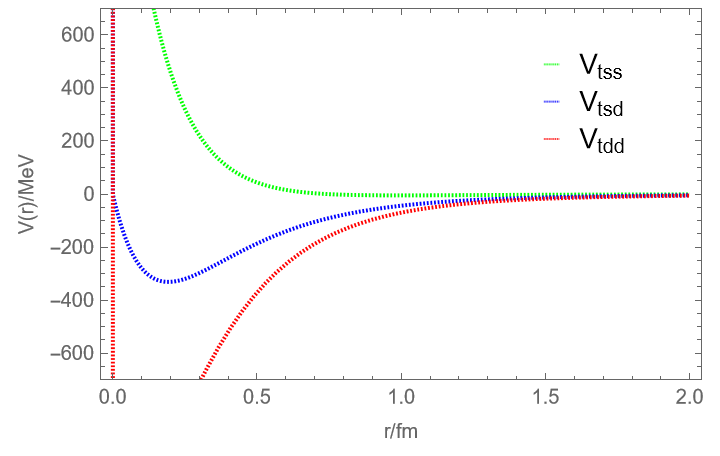}
    		\caption{The $S$\--$D$ wave mixing effects of the $D_1 \bar D_1$ system with $I(J^P)=0(0^+)$ when the cutoff parameter is fixed at 1.30 GeV without recoil corrections}
    		\label{fig:23}
    	\end{minipage}
    	\begin{minipage}{0.49\textwidth}
    		\centering
    		\includegraphics[scale=0.7]{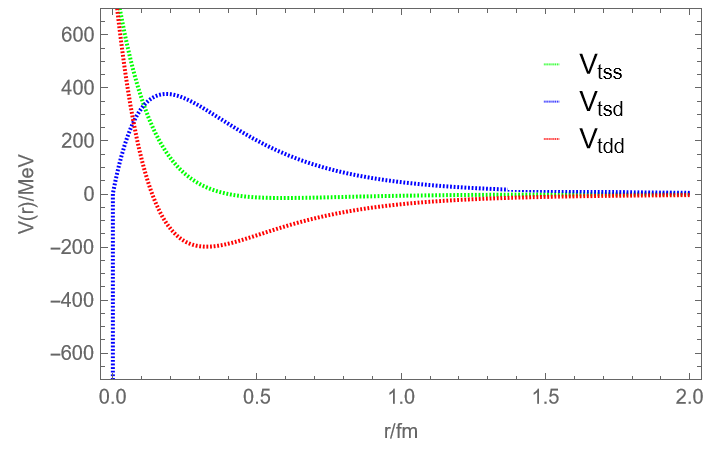}
    		\caption{The $S$\--$D$ wave mixing effects of the $D_1 \bar D_1$ system with $I(J^P)=0(1^+)$ when the cutoff parameter is fixed at 1.35 GeV without recoil corrections}
    		\label{fig:24}	
    	\end{minipage}
        \end{figure}
\begin{figure}[htbp]
    		\centering
    		\includegraphics[scale=0.7]{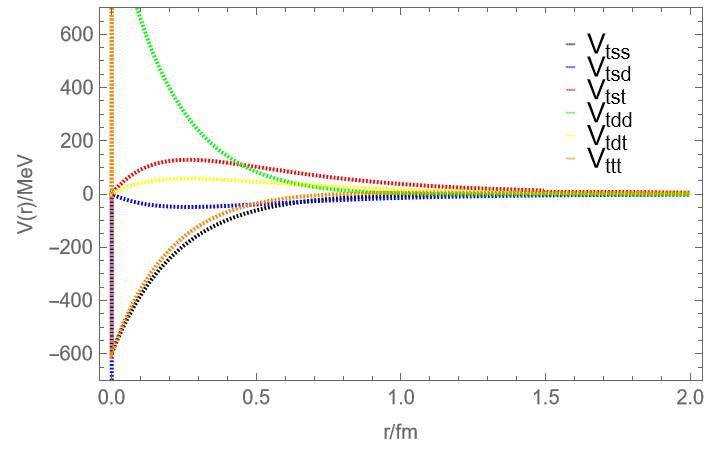}
    		\caption{The $S$\--$D$ wave mixing effects of the $D_1 \bar D_1$ system with $I(J^P)=0(2^+)$ when the cutoff parameter is fixed at 0.95 GeV without recoil corrections}
    		\label{fig:25}
        \end{figure}
\begin{figure}[htbp]
    	\begin{minipage}{0.49\textwidth}
    		\centering
    		\includegraphics[scale=0.7]{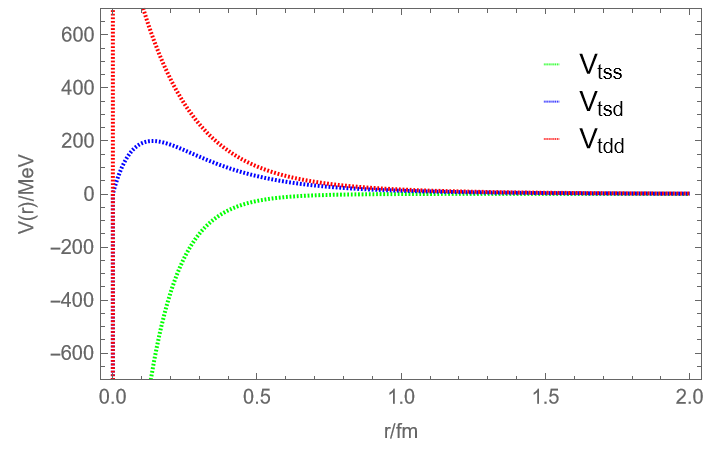}
    		\caption{The $S$\--$D$ wave mixing effects of the $D_1 \bar D_1$ system with $I(J^P)=1(0^+)$ when the cutoff parameter is fixed at 1.90 GeV without recoil corrections}
    		\label{fig:26}	
    	\end{minipage}
    	\begin{minipage}{0.49\textwidth}
    		\centering
    		\includegraphics[scale=0.7]{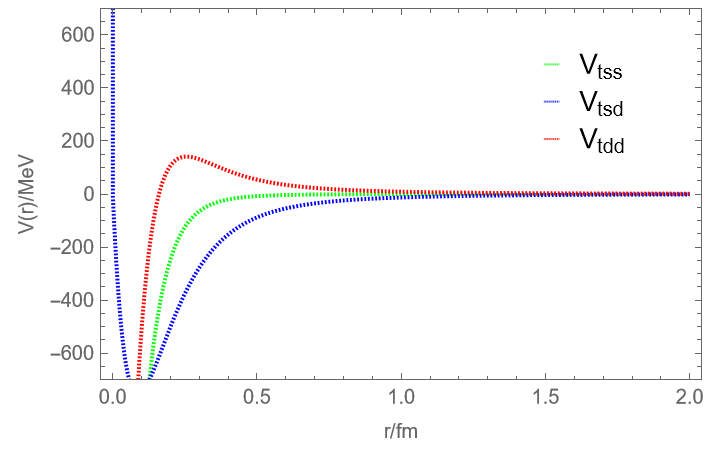}
    		\caption{The $S$\--$D$ wave mixing effects of the $D_1 \bar D_1$ system with $I(J^P)=1(1^+)$ when the cutoff parameter is fixed at 3.00 GeV without recoil corrections}
    		\label{fig:27}
    	\end{minipage} 
        \end{figure}
\begin{figure}[htbp]
    	\begin{minipage}{0.49\textwidth}
    		\centering
    		\includegraphics[scale=0.7]{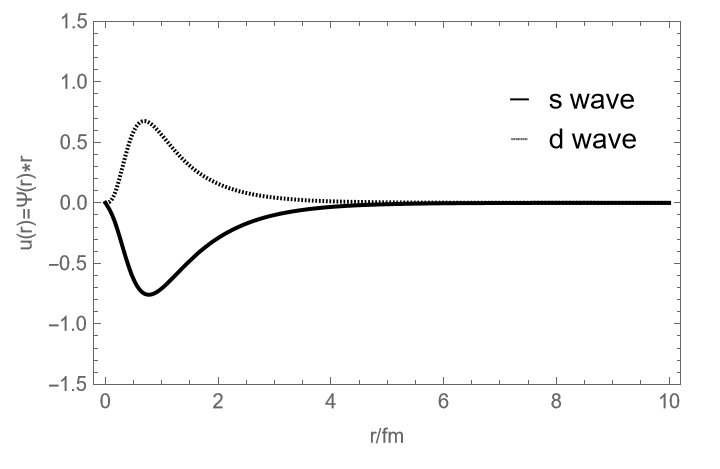}
    		\caption{The wave functions of the $D_1 \bar D_1$ system with $I(J^P)=0(0^+)$ when the cutoff parameter is fixed at 1.30 GeV without recoil corrections}
    		\label{fig:28}
    	\end{minipage}
    	\begin{minipage}{0.49\textwidth}
    		\centering
    		\includegraphics[scale=0.7]{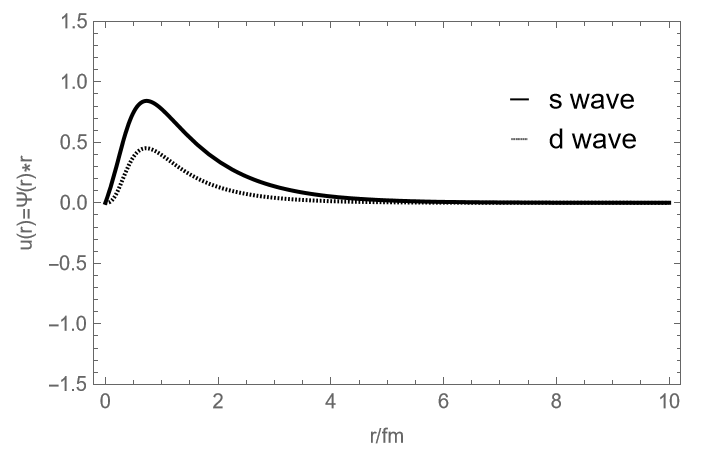}
    		\caption{The wave functions of the $D_1 \bar D_1$ system with $I(J^P)=0(1^+)$ when the cutoff parameter is fixed at 1.35 GeV without recoil corrections}
    		\label{fig:29}	
    	\end{minipage}
        \end{figure}
\begin{figure}[htbp]
    		\centering
    		\includegraphics[scale=0.7]{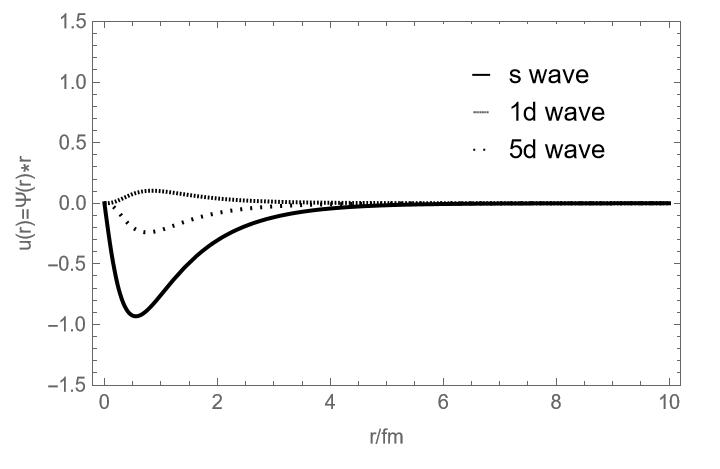}
    		\caption{The wave functions of the $D_1 \bar D_1$ system with $I(J^P)=0(2^+)$ when the cutoff parameter is fixed at 0.95 GeV without recoil corrections}
    		\label{fig:30}
    	\end{figure}
\begin{figure}[htbp]
    	\begin{minipage}{0.49\textwidth}
    		\centering
    		\includegraphics[scale=0.7]{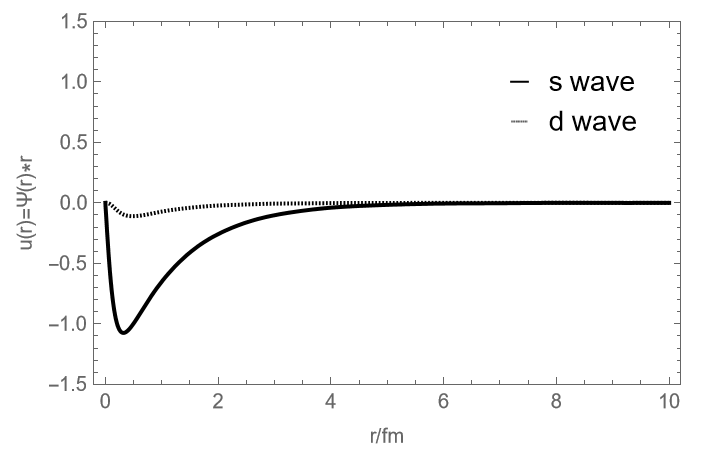}
    		\caption{The wave functions of the $D_1 \bar D_1$ system with $I(J^P)=1(0^+)$ when the cutoff parameter is fixed at 1.90 GeV without recoil corrections}
    		\label{fig:31}	
    	\end{minipage}
    	\begin{minipage}{0.49\textwidth}
    		\centering
    		\includegraphics[scale=0.7]{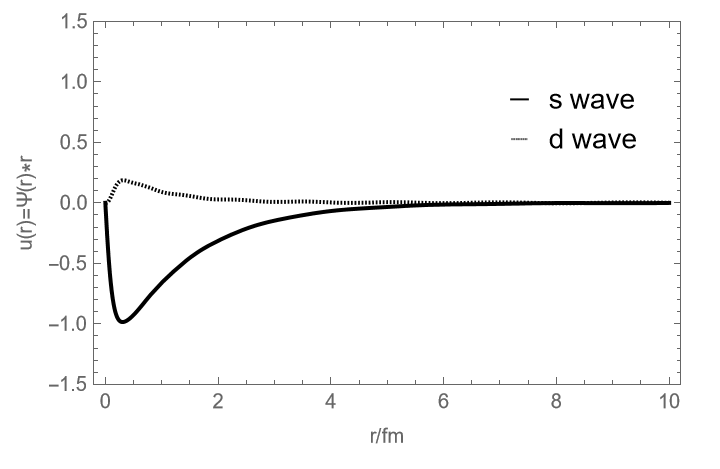}
    		\caption{The wave functions of the $D_1 \bar D_1$ system with $I(J^P)=1(1^+)$ when the cutoff parameter is fixed at 3.00 GeV without recoil corrections}
    		\label{fig:32}
    	\end{minipage} 
        \end{figure}
\begin{figure}[htbp]
        \begin{minipage}{0.49\textwidth}
        	\centering
        	\includegraphics[scale=0.7]{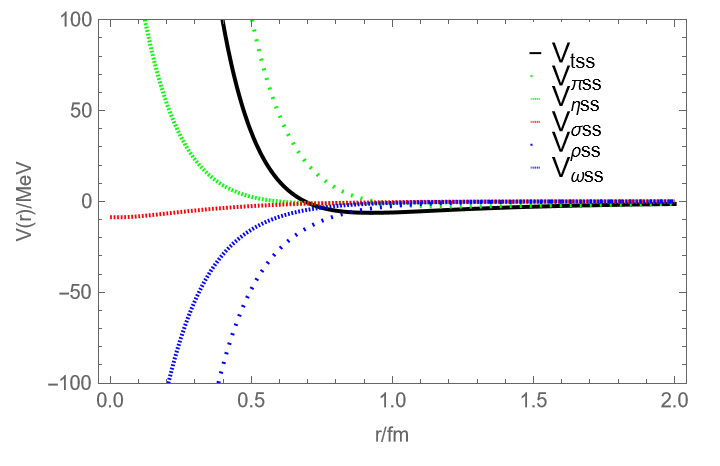}
        	\caption{The $S$ wave potentials of the $D_1 \bar D_1$ system with $I(J^P)=0(0^+)$ when the cutoff parameter is fixed at 1.35 GeV with recoil corrections}
        	\label{fig:33}
        	\end{minipage}
        \begin{minipage}{0.49\textwidth}
        	\centering
        	\includegraphics[scale=0.7]{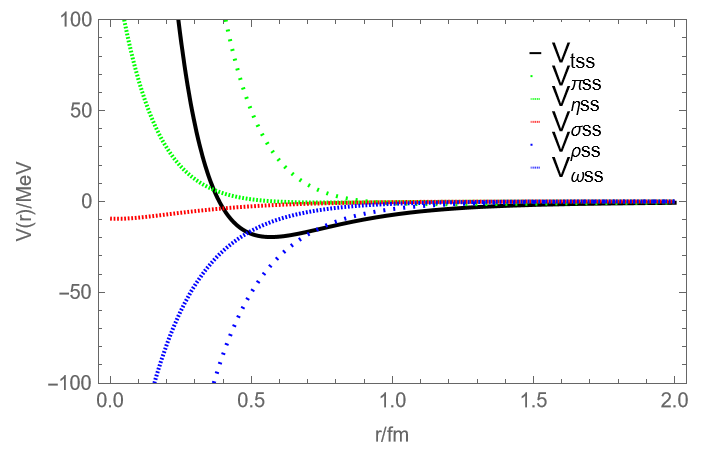}
        	\caption{The $S$ wave potentials of the $D_1 \bar D_1$ system with $I(J^P)=0(1^+)$ when the cutoff parameter is fixed at 1.40 GeV with recoil corrections}
        	\label{fig:34}	
        	\end{minipage}
        \end{figure}
\begin{figure}[htbp]
        	\centering
        	\includegraphics[scale=0.7]{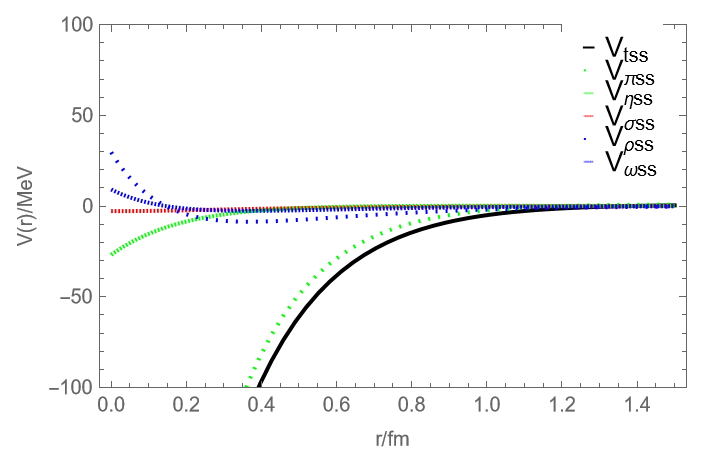}
        	\caption{The $S$ wave potentials of the $D_1 \bar D_1$ system with $I(J^P)=0(2^+)$ when the cutoff parameter is fixed at 0.95 GeV with recoil corrections}
        	\label{fig:35}
        \end{figure}
\begin{figure}[htbp]
        	\begin{minipage}{0.49\textwidth}
        		\centering
        		\includegraphics[scale=0.7]{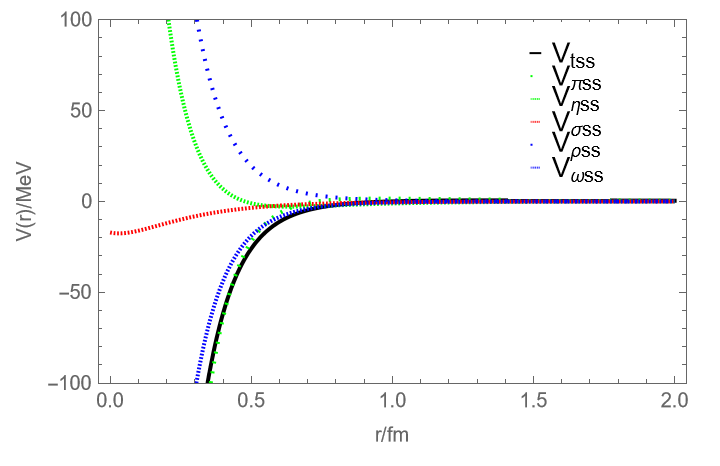}
        		\caption{The $S$ wave potentials of the $D_1 \bar D_1$ system with $I(J^P)=1(0^+)$ when the cutoff parameter is fixed at 1.90 GeV with recoil corrections}
        		\label{fig:36}	
        	\end{minipage}
        	\begin{minipage}{0.49\textwidth}
        		\centering
        		\includegraphics[scale=0.7]{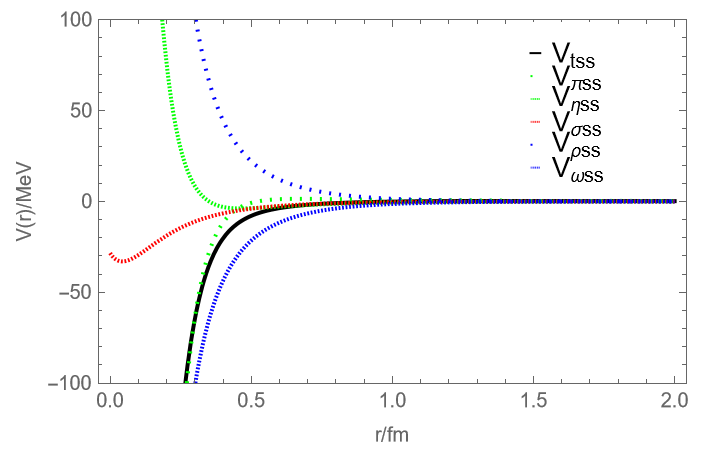}
        		\caption{The $S$ wave potentials of the $D_1 \bar D_1$ system with $I(J^P)=1(1^+)$ when the cutoff parameter is fixed at 3.00 GeV with recoil corrections}
        		\label{fig:37}
        	\end{minipage} 	
        \end{figure}
\begin{figure}[htbp]
        	\begin{minipage}{0.49\textwidth}
        		\centering
        		\includegraphics[scale=0.7]{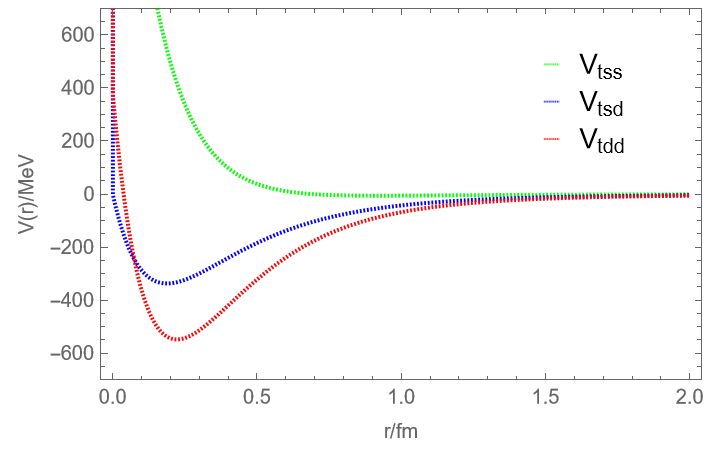}
        		\caption{The $S$\--$D$ wave mixing effects of the $D_1 \bar D_1$ system with $I(J^P)=0(0^+)$ when the cutoff parameter is fixed at 1.35 GeV with recoil corrections}
        		\label{fig:38}
        	\end{minipage}
        	\begin{minipage}{0.49\textwidth}
        		\centering
        		\includegraphics[scale=0.7]{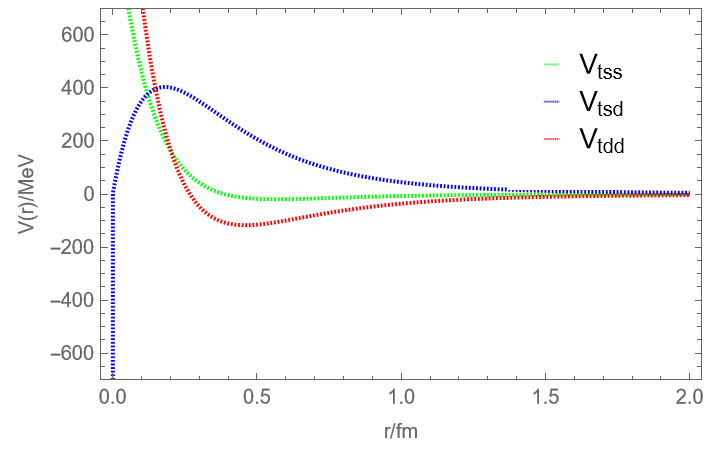}
        		\caption{The $S$\--$D$ wave mixing effects of the $D_1 \bar D_1$ system with $I(J^P)=0(1^+)$ when the cutoff parameter is fixed at 1.40 GeV with recoil corrections}
        		\label{fig:39}	
        	\end{minipage}
        \end{figure}
\begin{figure}[htbp]
        	\centering
        	\includegraphics[scale=0.7]{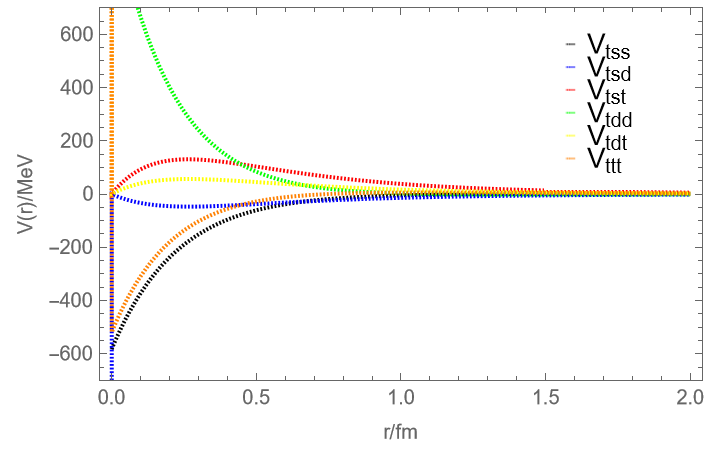}
        	\caption{The $S$\--$D$ wave mixing effects of the $D_1 \bar D_1$ system with $I(J^P)=0(2^+)$ when the cutoff parameter is fixed at 0.95 GeV with recoil corrections}
        	\label{fig:40}
        \end{figure}
\begin{figure}[htbp]
        	\begin{minipage}{0.49\textwidth}
        		\centering
        		\includegraphics[scale=0.7]{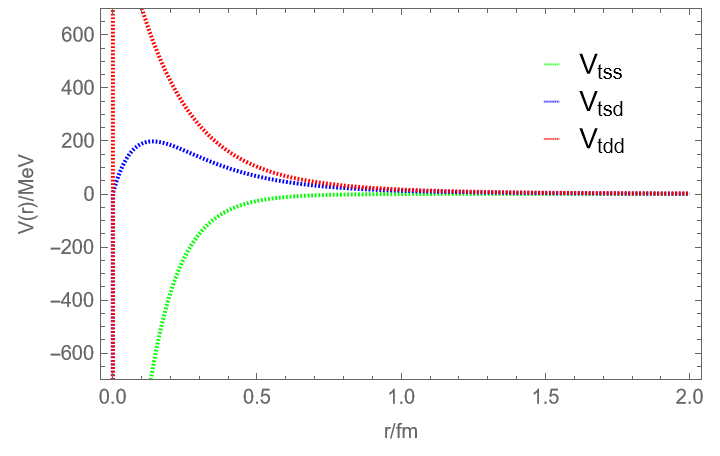}
        		\caption{The $S$\--$D$ wave mixing effects of the $D_1 \bar D_1$ system with $I(J^P)=1(0^+)$ when the cutoff parameter is fixed at 1.90 GeV with recoil corrections}
        		\label{fig:41}	
        	\end{minipage}
        	\begin{minipage}{0.49\textwidth}
        		\centering
        		\includegraphics[scale=0.7]{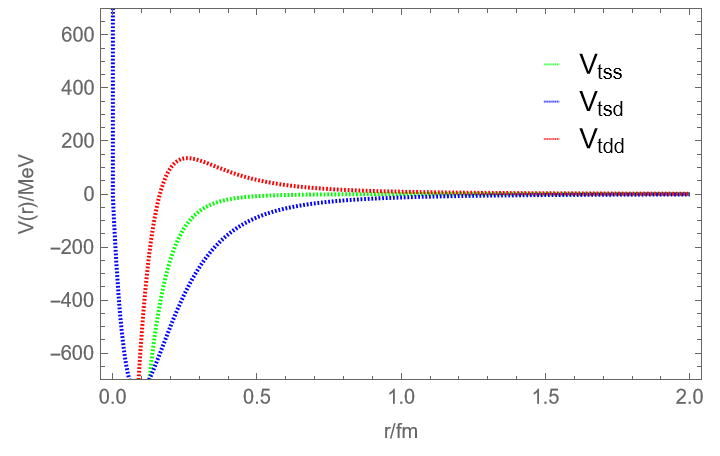}
        		\caption{The $S$\--$D$ wave mixing effects of the $D_1 \bar D_1$ system with $I(J^P)=1(1^+)$ when the cutoff parameter is fixed at 3.00 GeV with recoil corrections}
        		\label{fig:42}
        	\end{minipage} 
        \end{figure}
\begin{figure}[htbp]
        	\begin{minipage}{0.49\textwidth}
        		\centering
        		\includegraphics[scale=0.7]{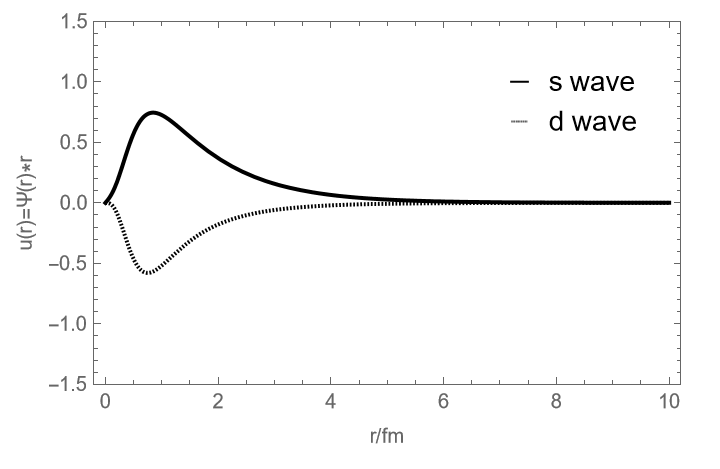}
        		\caption{The wave functions of the $D_1 \bar D_1$ system with $I(J^P)=0(0^+)$ when the cutoff parameter is fixed at 1.35 GeV with recoil corrections}
        		\label{fig:43}
        	\end{minipage}
        	\begin{minipage}{0.49\textwidth}
        		\centering
        		\includegraphics[scale=0.7]{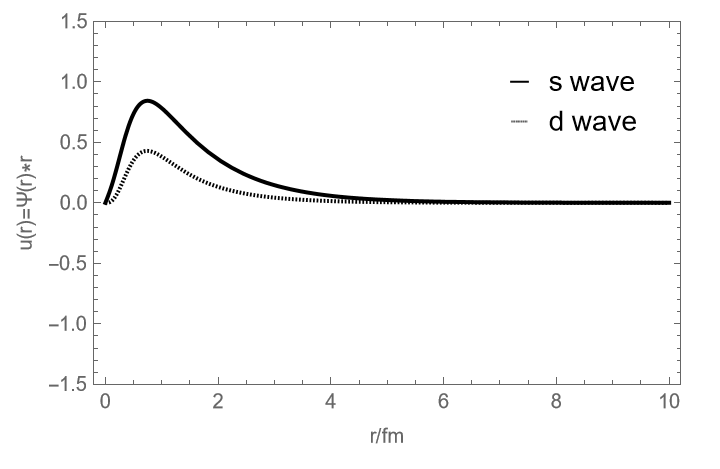}
        		\caption{The wave functions of the $D_1 \bar D_1$ system with $I(J^P)=0(1^+)$ when the cutoff parameter is fixed at 1.40 GeV with recoil corrections}
        		\label{fig:44}	
        	\end{minipage}
        \end{figure}
\begin{figure}[htbp]
        	\centering
        	\includegraphics[scale=0.7]{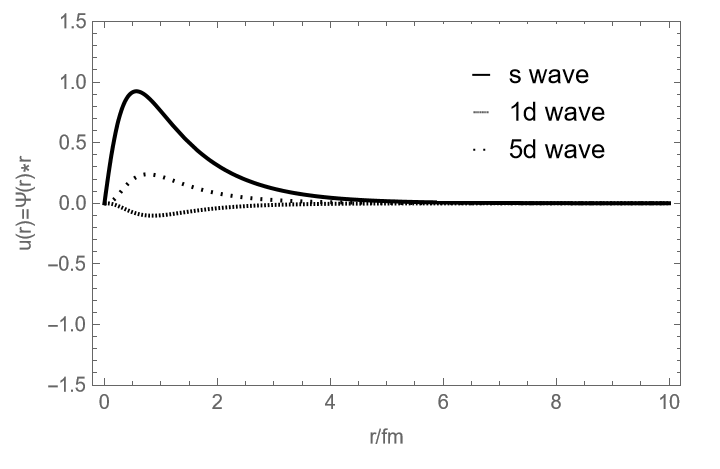}
        	\caption{The wave functions of the $D_1 \bar D_1$ system with $I(J^P)=0(2^+)$ when the cutoff parameter is fixed at 0.95 GeV with recoil corrections}
        	\label{fig:45}
        \end{figure}
\begin{figure}[htbp]
        	\begin{minipage}{0.49\textwidth}
        		\centering
        		\includegraphics[scale=0.7]{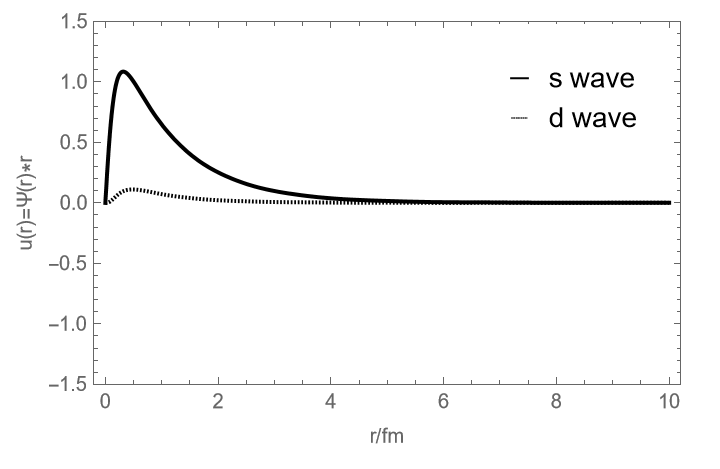}
        		\caption{The wave functions of the $D_1 \bar D_1$ system with $I(J^P)=1(0^+)$ when the cutoff parameter is fixed at 1.90 GeV with recoil corrections}
        		\label{fig:46}	
        	\end{minipage}
        	\begin{minipage}{0.49\textwidth}
        		\centering
        		\includegraphics[scale=0.7]{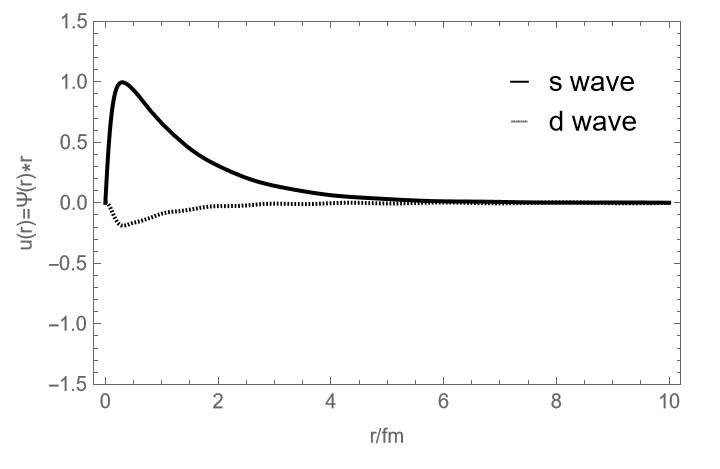}
        		\caption{The wave functions of the $D_1 \bar D_1$ system with $I(J^P)=1(1^+)$ when the cutoff parameter is fixed at 3.00 GeV with recoil corrections}
        		\label{fig:47}
        	\end{minipage} 
        \end{figure}

	For the cases without the recoil corrections, the $S$ wave contributions from each meson are shown in FIG. \ref{fig:18}-\ref{fig:22}. The contributions from the $S$ and the $D$ wave, and the effect of $S$\--{}$D$ wave mixing are shown in FIG. \ref{fig:23}-\ref{fig:27}. The wave functions are plotted in FIG. \ref{fig:28}-\ref{fig:32}.
	For the cases with the recoil corrections, the $S$ wave contributions are shown in FIG. \ref{fig:33}-\ref{fig:37}. The contributions from the $S$ and the $D$ wave, and the effect of $S$\--{}$D$ wave mixing are shown in FIG. \ref{fig:38}-\ref{fig:42}. The wave functions are plotted in FIG. \ref{fig:43}-\ref{fig:47}.
	\par 
	For the $I(J^P)=0(0^+)$ channel in TABLE \ref{tab:7}, regardless of the recoil corrections, the bound state appears when taking the cutoff value $\Lambda=1.10$ GeV, with the binding energy 0.16 MeV and the RMS radius 6.12 fm. Although the $S$ wave provides little attractive potential, the $D$ wave potential helps to form the bound state. Moreover, the $S$\--{}$D$ wave mixing effect rises the proportion of the $D$ wave. When the cutoff value rises to $\Lambda=1.30$ GeV, the binding energy increases to 19.3 MeV and the RMS radius decreases to 1.21 fm. After considering the recoil corrections, the bound state becomes looser. The bound state appears when the cutoff value is taken as $\Lambda=1.15$ GeV, with the binding energy 0.39 MeV and the RMS radius 5.04 fm.
	\begin{table*}[htbp]
		\scriptsize
		\begin{center}
			\caption{\label{tab:7} The numerical results of  $D_1 \bar D_1 $ $I(J^P)=0(0^+)$ system }
			
			\begin{tabular}{l cccc|cccccc}\toprule[1pt]
				\multicolumn{5}{c|}{Without recoil corrections} & \multicolumn{5}{c}{With recoil corrections}\\
				\midrule[1pt]
				$\Lambda$(GeV) & B.E.(MeV) & RMS(fm)  & $^1S_0 $($\%$)  & $^5D_0$($\%$) & $\Lambda$(GeV) & B.E.(MeV) & RMS(fm)  & $^1S_0 $($\%$)  & $^5D_0$($\%$) \\  
				1.10 & 0.16 & 6.12 & 94.72 & 5.28  & 1.15 & 0.39 & 5.04 & 92.84 & 7.16 \\
				1.15 & 1.44 & 3.16 & 85.90 & 14.10 & 1.20 & 1.60 & 3.02 & 86.02 & 13.98 \\
				1.20 & 4.63 & 2.01 & 76.54 & 23.46 & 1.25 & 3.94 & 2.14 & 79.45 & 20.55 \\
				1.25 & 10.4 & 1.50 & 68.47 & 31.53 & 1.30 & 7.59 & 1.68 & 73.74 & 26.26 \\
				1.30 & 19.3 & 1.21 & 61.93 & 38.07 & 1.35 & 12.7 & 1.40 & 68.97 & 31.03 \\

				\bottomrule[1pt]
				
			\end{tabular}
		\end{center}
	\end{table*}
	\par 
	For the $I(J^P)=0(1^+)$ channel in TABLE \ref{tab:8}, the results are similar to those of the $I(J^P)=0(0^+)$ channel. We can find a loosely bound state whether we consider the recoil corrections or not. The recoil corrections make the bound state looser. Regardless of the recoil corrections, the bound state appears when taking the cutoff value $\Lambda=1.15$ GeV, with the binding energy 0.39 MeV and the RMS radius 4.89 fm. When the cutoff value rises to $\Lambda=1.35$ GeV, the binding energy increases to 14.6 MeV and the RMS radius decreases to 1.24 fm. After considering the recoil corrections, the bound state appears when the cutoff value is taken as $\Lambda=1.20$ GeV, with the binding energy 0.95 MeV and the RMS radius 3.56 fm.
	\begin{table*}[htbp]
		\scriptsize
		\begin{center}
			\caption{\label{tab:8} The numerical results of  $D_1 \bar D_1 $ $I(J^P)=0(1^+)$ system }
			
			\begin{tabular}{l cccc|cccccc}\toprule[1pt]
				\multicolumn{5}{c|}{Without recoil corrections} & \multicolumn{5}{c}{With recoil corrections}\\
				\midrule[1pt]
				$\Lambda$(GeV) & B.E.(MeV) & RMS(fm)  & $^3S_1 $($\%$)  & $^3D_1$($\%$) & $\Lambda$(GeV) & B.E.(MeV) & RMS(fm)  & $^3S_1 $($\%$)  & $^3D_1$($\%$) \\  
				1.15 & 0.39 & 4.89 & 94.95 & 5.05  & 1.20 & 0.95 & 3.56 & 92.95 & 7.05 \\
				1.20 & 1.77 & 2.78 & 90.61 & 9.39  & 1.25 & 2.60 & 2.39 & 89.64 & 10.36 \\
				1.25 & 4.43 & 1.94 & 86.94 & 13.06 & 1.30 & 5.22 & 1.82 & 86.93 & 13.07 \\
				1.30 & 8.63 & 1.51 & 83.92 & 16.08 & 1.35 & 8.91 & 1.49 & 84.71 & 15.29 \\
				1.35 & 14.6 & 1.24 & 81.40 & 18.60 & 1.40 & 13.7 & 1.27 & 82.86 & 17.14 \\

				\bottomrule[1pt]
				
			\end{tabular}
		\end{center}
	\end{table*}
	\par 
	For the $I(J^P)=0(2^+)$ channel in TABLE \ref{tab:9}, the $S$ wave is the major component of the ground state. The $\pi$ exchange provides a remarkable attractive potential for the $S$ wave. Regardless of the recoil corrections, the bound state appears at the cutoff value $\Lambda=0.75$ GeV, with the binding energy 0.07 MeV and the RMS radius 6.28 fm. When the cutoff value increases to $\Lambda=0.95$ GeV, the binding energy increases to 15.6 MeV and the RMS radius decreases to 1.08 fm. The recoil corrections have little effect on this channel. The bound state still appears when the cutoff value is taken as $\Lambda=0.75$ GeV with the binding energy 0.07 MeV.
	\clearpage
	\begin{table*}[htbp]
		\scriptsize
		\begin{center}
			\caption{\label{tab:9} The numerical results of  $D_1 \bar D_1 $ $I(J^P)=0(2^+)$ system }
			
			\begin{tabular}{l ccccc|ccccccc}\toprule[1pt]
				\multicolumn{6}{c|}{Without recoil corrections} & \multicolumn{6}{c}{With recoil corrections}\\
				\midrule[1pt]
				$\Lambda$(GeV) & B.E.(MeV) & RMS(fm)  & $^5S_2 $($\%$)  & $^1D_2$($\%$) & $^5D_2$($\%$) & $\Lambda$(GeV) & B.E.(MeV) & RMS(fm)  & $^5S_2 $($\%$)  & $^1D_2$($\%$) & $^5D_2$($\%$)\\  
				0.75 & 0.07 & 6.28 & 98.02 & 0.34 & 1.64 & 0.75 & 0.07 & 6.29 & 98.02 & 0.34 &1.64\\
				0.80 & 0.91 & 3.35 & 96.10 & 0.68 & 3.22 & 0.80 & 0.91 & 3.35 & 96.10 & 0.68 &3.22\\
				0.85 & 3.36 & 1.94 & 94.60 & 0.93 & 4.47 & 0.85 & 3.31 & 1.96 & 94.60 & 0.93 &4.47\\
				0.90 & 8.10 & 1.37 & 93.71 & 1.06 & 5.23 & 0.90 & 7.87 & 1.39 & 93.71 & 1.05 &5.24\\
				0.95 & 15.6 & 1.08 & 93.12 & 1.13 & 5.75 & 0.95 & 14.9 & 1.10 & 93.11 & 1.11 &5.78\\

				\bottomrule[1pt]
				
			\end{tabular}
		\end{center}
	\end{table*}
	\par 
	For the $I(J^P)=1(0^+)$ channel in TABLE \ref{tab:10}, regardless of the recoil corrections, the bound state appears when the cutoff value is taken as $\Lambda=1.50$ GeV, with the binding energy 0.07 MeV and the RMS radius 6.02 fm. When the cutoff value rises to $\Lambda=1.90$ GeV, the binding energy increases to 13.6 MeV and the RMS radius decreases to 0.92 fm. The $\pi$ and $\omega$ meson exchanges are the main contributions of the attractive potential in the $S$ wave. The $\rho$ exchange potential and the $\omega$ exchange potential almost cancel out, so the $S$ wave potential is similar to the $\pi$ exchange potential. After considering the recoil corrections, the binding energy has a slight increase. When the cutoff value is taken as $\Lambda=1.50$ GeV, the binding energy increases to 0.10 MeV while the RMS radius decreases to 5.78 fm.
		\begin{table*}[htbp]
		\scriptsize
		\begin{center}
			\caption{\label{tab:10} The numerical results of  $D_1 \bar D_1 $ $I(J^P)=1(0^+)$ system }
			
			\begin{tabular}{l cccc|cccccc}\toprule[1pt]
				\multicolumn{5}{c|}{Without recoil corrections} & \multicolumn{5}{c}{With recoil corrections}\\
				\midrule[1pt]
				$\Lambda$(GeV) & B.E.(MeV) & RMS(fm)  & $^1S_0 $($\%$)  & $^5D_0$($\%$) & $\Lambda$(GeV) & B.E.(MeV) & RMS(fm)  & $^1S_0 $($\%$)  & $^5D_0$($\%$) \\  
				1.50 & 0.07 & 6.02 & 99.71 & 0.29 & 1.50 & 0.10 & 5.78 & 99.70 & 0.30 \\
				1.60 & 1.22 & 2.69 & 99.39 & 0.61 & 1.60 & 1.33 & 2.58 & 99.39 & 0.61 \\
				1.70 & 3.71 & 1.62 & 99.21 & 0.79 & 1.70 & 3.94 & 1.58 & 99.21 & 0.79 \\
				1.80 & 7.75 & 1.17 & 99.10 & 0.90 & 1.80 & 8.14 & 1.15 & 99.10 & 0.90 \\
				1.90 & 13.6 & 0.92 & 99.03 & 0.97 & 1.90 & 14.2 & 0.90 & 99.04 & 0.96 \\

				\bottomrule[1pt]
				
			\end{tabular}
		\end{center}
	\end{table*}
	\par
	For the $I(J^P)=1(1^+)$ channel in TABLE \ref{tab:11}, the results are similar to those of the $I(J^P)=1(0^+)$ channel. We can find a loosely bound state whether we consider the recoil corrections or not. the recoil corrections provide an insignificant positive contribution in forming the bound state. The $S$ wave provides a remarkable attractive potential. It is similar to the $\pi$ exchange potential because the $\rho$ exchange potential and the $\omega$ exchange potential almost cancel out. Without regard to the recoil corrections, the bound state appears at $\Lambda=2.60$ GeV, with the binding energy 0.23 MeV and the RMS radius 5.07 fm. When the cutoff value increases to $\Lambda=3.00$ GeV, the binding energy increases to 9.08 MeV and the RMS radius decreases to 1.09 fm. After considering the recoil corrections, the bound state still appears when the cutoff value is taken as $\Lambda=2.60$ GeV. The binding energy increases to 0.29 MeV and the RMS radius decreases to 4.78 fm.
		\begin{table*}[htbp]
		\scriptsize
		\begin{center}
			\caption{\label{tab:11} The numerical results of  $D_1 \bar D_1 $ $I(J^P)=1(1^+)$ system }
			
			\begin{tabular}{l cccc|cccccc}\toprule[1pt]
				\multicolumn{5}{c|}{Without recoil corrections} & \multicolumn{5}{c}{With recoil corrections}\\
				\midrule[1pt]
				$\Lambda$(GeV) & B.E.(MeV) & RMS(fm)  & $^3S_1 $($\%$)  & $^3D_1$($\%$) & $\Lambda$(GeV) & B.E.(MeV) & RMS(fm)  & $^3S_1 $($\%$)  & $^3D_1$($\%$) \\  
				2.60 & 0.23 & 5.07 & 99.28 & 0.72 & 2.60 & 0.29 & 4.78 & 99.24 & 0.76 \\
				2.70 & 1.16 & 2.77 & 98.73 & 1.27 & 2.70 & 1.29 & 2.64 & 98.69 & 1.31 \\
				2.80 & 2.87 & 1.82 & 98.29 & 1.71 & 2.80 & 3.10 & 1.76 & 98.26 & 1.74 \\
				2.90 & 5.47 & 1.36 & 97.94 & 2.06 & 2.90 & 5.81 & 1.33 & 97.92 & 2.08 \\
				3.00 & 9.08 & 1.09 & 97.64 & 2.36 & 3.00 & 9.56 & 1.07 & 97.63 & 2.37 \\

				\bottomrule[1pt]
				
			\end{tabular}
		\end{center}
	\end{table*}
	\par

	From the numerical results, we find that the $I(J^P)=0(2^+)$ channel is most likely to form a loosely bound state.  For the isospin singlets, except the $I(J^P)=0(2^+)$ channel, the recoil corrections provide a relatively obvious negative effect in forming the bound state. For the isospin triplets, the recoil corrections provide a slight positive effect in forming the bound state.  Considering the states with the same isospin, the $J=0$ and the $J=1$ channels exhibit similar dynamical behavior. Besides, the $S$ wave is the main component of the wave function of all channels.  	
        
        \clearpage

    \subsection{The $B_1 B_1$ system} \label{subsecC}
\begin{figure}[htbp]
    	\begin{minipage}{0.49\textwidth}
    		\centering
    		\includegraphics[scale=0.62]{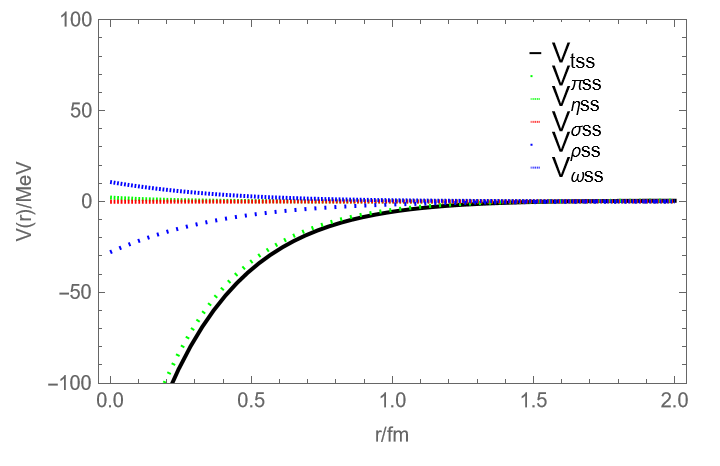}
    		\caption{The $S$ wave potentials of the $B_1 B_1$ system with $I(J^P)=0(1^+)$ when the cutoff parameter is fixed at 0.67 GeV}
    		\label{fig:48}
    	\end{minipage}
    	\begin{minipage}{0.49\textwidth}
    		\centering
    		\includegraphics[scale=0.62]{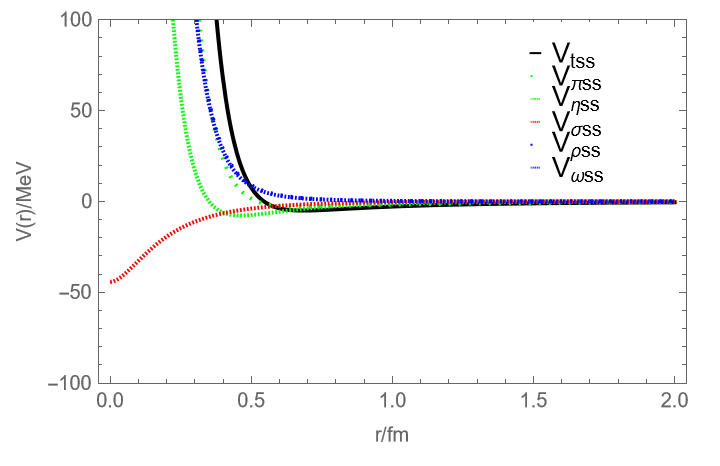}
    		\caption{The $S$ wave potentials of the $B_1 B_1$ system with $I(J^P)=1(0^+)$ when the cutoff parameter is fixed at 3.00 GeV}
    		\label{fig:49}	
    	\end{minipage}
        \end{figure}

\begin{figure}[htbp]
    		\centering
    		\includegraphics[scale=0.62]{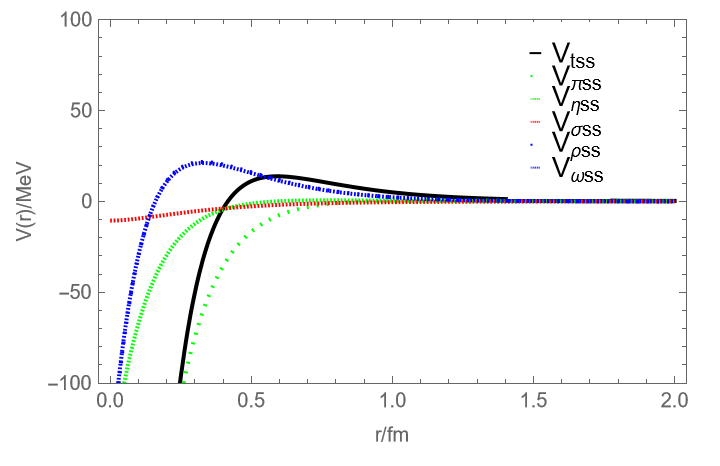}
    		\caption{The $S$ wave potentials of the $B_1 B_1$ system with $I(J^P)=1(2^+)$ when the cutoff parameter is fixed at 1.40 GeV}
    		\label{fig:50}
    \end{figure}

\begin{figure}[htbp]
    	\begin{minipage}{0.49\textwidth}
    		\centering
    		\includegraphics[scale=0.62]{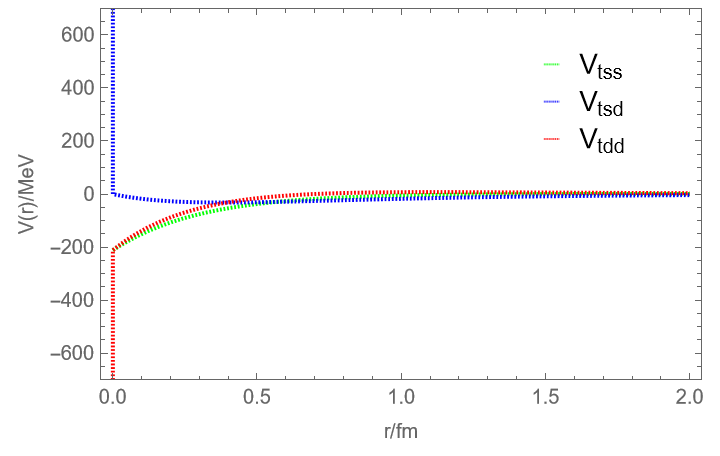}
    		\caption{The $S$\--$D$ wave mixing effects of the $B_1 B_1$ system with $I(J^P)=0(1^+)$ when the cutoff parameter is fixed at 0.67 GeV}
    		\label{fig:51}
    	\end{minipage}
    	\begin{minipage}{0.49\textwidth}
    		\centering
    		\includegraphics[scale=0.62]{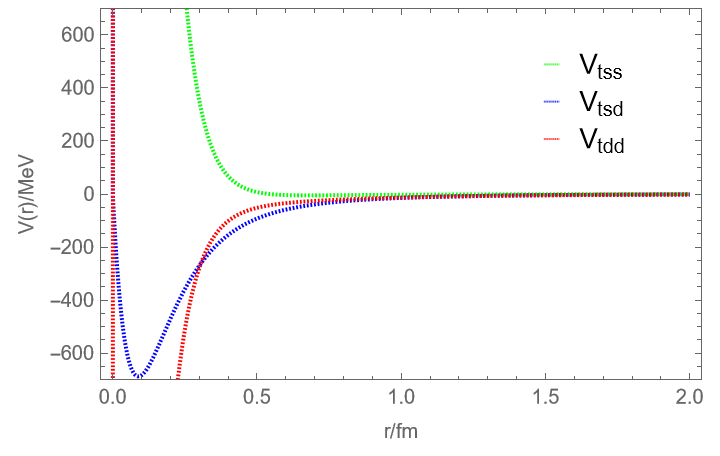}
    		\caption{The $S$\--$D$ wave mixing effects of the $B_1 B_1$ system with $I(J^P)=1(0^+)$ when the cutoff parameter is fixed at 3.00 GeV}
    		\label{fig:52}	
    	\end{minipage}
    	\end{figure}

\begin{figure}[htbp] 
    		\centering
    		\includegraphics[scale=0.67]{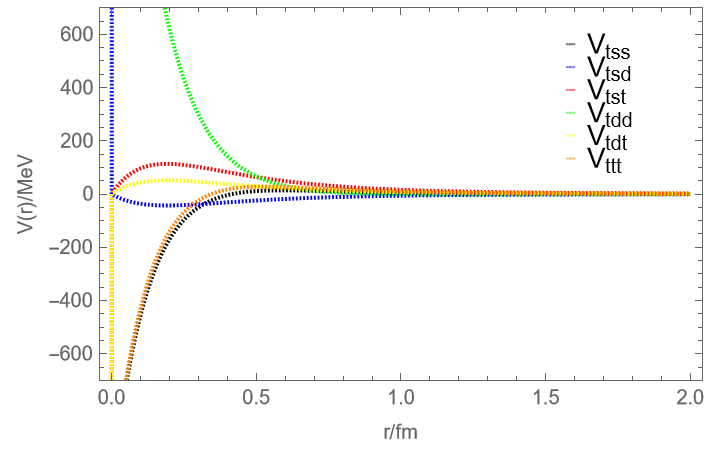}
    		\caption{The $S$\--$D$ wave mixing effects of the $B_1 B_1$ system with $I(J^P)=1(2^+)$ when the cutoff parameter is fixed at 1.40 GeV}
    		\label{fig:53}
    \end{figure}

\begin{figure}[htbp]
    	\begin{minipage}{0.49\textwidth}
    		\centering
    		\includegraphics[scale=0.67]{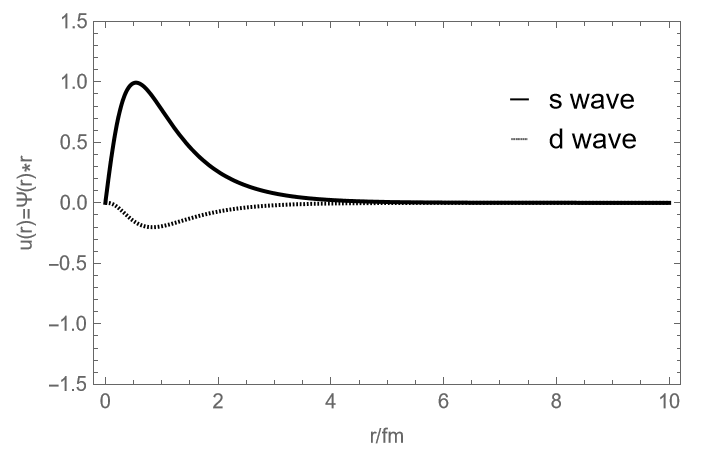}
    		\caption{The wave functions of the $B_1 B_1$ system with $I(J^P)=0(1^+)$ when the cutoff parameter is fixed at 0.67 GeV}
    		\label{fig:54}
    	\end{minipage}
    	\begin{minipage}{0.49\textwidth}
    		\centering
    		\includegraphics[scale=0.67]{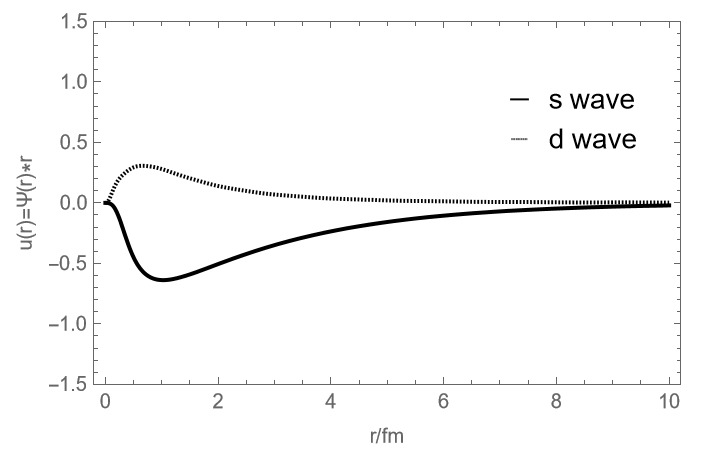}
    		\caption{The wave functions of the $B_1 B_1$ system with $I(J^P)=1(0^+)$ when the cutoff parameter is fixed at 3.00 GeV}
    		\label{fig:55}	
    	\end{minipage}
    	\end{figure}

\begin{figure}[htbp] 
    		\centering
    		\includegraphics[scale=0.67]{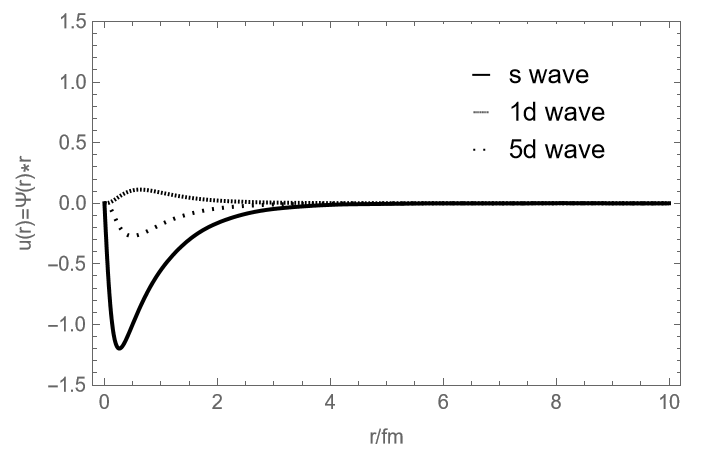}
    		\caption{The wave functions of the $B_1 B_1$ system with $I(J^P)=1(2^+)$ when the cutoff parameter is fixed at 1.40 GeV}
    		\label{fig:56}
    \end{figure}

    In TABLE \ref{tab:13}-\ref{tab:15}, we present the numerical results of $B_1 B_1$ with $I(J^P)=0(1^+)$, $I(J^P)=1(0^+)$ and $I(J^P)=1(2^+)$. The $S$ wave contributions are shown in FIG. \ref{fig:48}-\ref{fig:50}. The contributions from the $S$ and the $D$ wave, and the effect of $S$\--{}$D$ wave mixing are given in FIG. \ref{fig:51}-\ref{fig:53}. The wave functions are plotted in FIG. \ref{fig:54}-\ref{fig:56}. We draw the graphs without the recoil corrections, because we can find a bound state solution whether taking the recoil corrections into account or not.
    \par 
    For the $I(J^P)=0(1^+)$ channel in TABLE \ref{tab:13}, we manage to find a bound state at $\Lambda=0.67$ GeV, with the binding energy 9.79 MeV and RMS radius 0.97 fm. When $\Lambda$ increases to 0.75 GeV, the binding energy is 14.6 MeV and the RMS radius is 0.83 fm. The recoil corrections have little effect on the formation of the bound states.    
    \begin{table*}[htbp]
    	\scriptsize
    	\begin{center}
    		\caption{\label{tab:13} The numerical results of  $B_1 B_1 $ $I(J^P)=0(1^+)$ system }
    		\begin{tabular}{l cccc|cccccc}\toprule[1pt]
    			\multicolumn{5}{c|}{Without recoil corrections} & \multicolumn{5}{c}{With recoil corrections}\\
    			\midrule[1pt]
    			$\Lambda$(GeV) & B.E.(MeV) & RMS(fm)  & $^3S_1 $($\%$)  & $^3D_1$($\%$) & $\Lambda$(GeV) & B.E.(MeV) & RMS(fm)  & $^3S_1 $($\%$)  & $^3D_1$($\%$) \\  
    			0.67 & 9.79 & 0.97 & 95.86 & 4.14 & 0.67 & 9.71 & 0.97 & 95.85 & 4.15 \\
    			0.69 & 10.4 & 0.94 & 95.58 & 4.42 & 0.69 & 10.3 & 0.95 & 95.57 & 4.43 \\
    			0.71 & 11.3 & 0.91 & 95.36 & 4.64 & 0.71 & 11.3 & 0.91 & 95.36 & 4.64 \\
    			0.73 & 12.7 & 0.87 & 95.21 & 4.79 & 0.73 & 12.7 & 0.87 & 95.21 & 4.79 \\
    			0.75 & 14.6 & 0.83 & 95.11 & 4.89 & 0.75 & 14.6 & 0.83 & 95.11 & 4.89 \\
    			\bottomrule[1pt]				
    		\end{tabular}
    	\end{center}
    \end{table*}
    \par 
    Unlike the $D_1D_1$ system, the $I(J^P)=1(0^+)$ channel of the $B_1B_1$ system can form a loosely bound state without the recoil corrections. When the cutoff value $\Lambda=2.60$ GeV, the binding energy is 0.06 MeV and the RMS radius is 6.43 fm. After considering the recoil corrections, we can find a bound state solution at a smaller cutoff value.
    \begin{table*}[htbp]
    	\scriptsize
    	\begin{center}
    		\caption{\label{tab:14} The numerical results of  $B_1 B_1 $ $I(J^P)=1(0^+)$ system }
    		\begin{tabular}{l cccc|cccccc}\toprule[1pt]
    			\multicolumn{5}{c|}{Without recoil corrections} & \multicolumn{5}{c}{With recoil corrections}\\
    			\midrule[1pt]
    			$\Lambda$(GeV) & B.E.(MeV) & RMS(fm)  & $^3S_1 $($\%$)  & $^3D_1$($\%$) & $\Lambda$(GeV) & B.E.(MeV) & RMS(fm)  & $^3S_1 $($\%$)  & $^3D_1$($\%$) \\  
    			2.60 & 0.06 & 6.43 & 96.70 & 3.30 & 2.05 & 0.29 & 4.13 & 91.19 & 8.81 \\
    			2.80 & 0.34 & 3.83 & 92.92 & 7.08 & 2.10 & 0.71 & 2.87 & 85.94 & 14.06 \\
    			3.00 & 1.11 & 2.39 & 87.77 & 12.23 & 2.15 & 1.57 & 2.10 & 79.15 & 20.85 \\
    			3.20 & 3.26 & 1.56 & 79.93 & 20.07 & 2.20 & 3.25 & 1.59 & 70.49 & 29.51 \\
    			3.40 & 11.7 & 0.96 & 62.67 & 37.33 & 2.25 & 6.51 & 1.23 & 59.85 & 40.15 \\
    			\bottomrule[1pt]				
    		\end{tabular}
    	\end{center}
    \end{table*} 
    \par 
    For the $I(J^P)=1(2^+)$ channel in TABLE \ref{tab:15}, the bound state solution appears where $\Lambda=1.00$ GeV, with the binding energy 0.14 MeV and RMS radius 4.74 fm. The binding energy is very sensitive to the cutoff value. When $\Lambda$ increases to 1.40 GeV, the binding energy surges to 10.6 MeV. The recoil corrections provide a positive effect in forming the bound state. 
    
    \begin{table*}[htbp]
    	\scriptsize
    	\begin{center}
    		\caption{\label{tab:15} The numerical results of  $B_1 B_1 $ $I(J^P)=1(2^+)$ system }
    		\begin{tabular}{l ccccc|ccccccc}\toprule[1pt]
    			\multicolumn{6}{c|}{Without recoil corrections} & \multicolumn{6}{c}{With recoil corrections}\\
    			\midrule[1pt]
    			$\Lambda$(GeV) & B.E.(MeV) & RMS(fm)  & $^5S_2 $($\%$)  & $^1D_2$($\%$) & $^5D_2$($\%$) & $\Lambda$(GeV) & B.E.(MeV) & RMS(fm)  & $^5S_2 $($\%$)  & $^1D_2$($\%$) & $^5D_2$($\%$)\\  
    			1.00 & 0.14 & 4.74 & 96.94 & 0.55 & 2.51 & 1.00 & 0.18 & 4.37 & 96.71 & 0.60 & 2.69 \\
    			1.10 & 0.40 & 3.18 & 95.38 & 0.85 & 3.77 & 1.10 & 0.57 & 2.70 & 94.96 & 0.96 & 4.08 \\
    			1.20 & 1.20 & 1.93 & 93.72 & 1.17 & 5.11 & 1.20 & 1.85 & 1.59 & 93.42 & 1.26 & 5.32 \\
    			1.30 & 3.73 & 1.14 & 92.90 & 1.27 & 5.83 & 1.30 & 5.79 & 0.94 & 93.11 & 1.27 & 5.62 \\
    			1.40 & 10.6 & 0.71 & 93.47 & 1.06 & 5.47 & 1.40 & 15.9 & 0.60 & 94.07 & 1.00 & 4.93 \\
    			\bottomrule[1pt]	
    		\end{tabular}
    	\end{center}
    \end{table*}
    From the numerical results, we find out that all the channels can form a bound state for $B_1B_1$ system. For the $I(J^P)=1(0^+)$ channel, the recoil corrections markedly contribute to the formation of the bound state. For the $I(J^P)=1(2^+)$ channel, the recoil corrections slightly contribute to the formation of the bound state. For the $I(J^P)=0(1^+)$ channel, the recoil corrections can be ignored. In addition, for all the channels, the $S$ wave is the main component of the wave function. 

    \clearpage

     \subsection{The $B_1 \bar B_1$ system} \label{subsecD}
\begin{figure}[htbp]
    	\begin{minipage}{0.49\textwidth}
    		\centering
    		\includegraphics[scale=0.7]{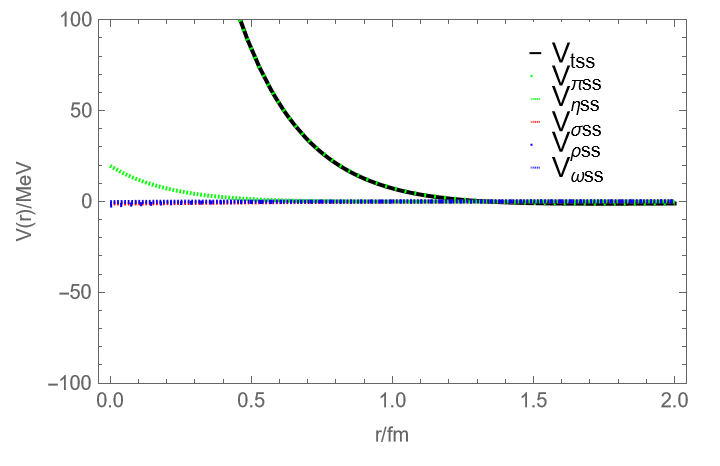}
    		\caption{The $S$ wave potentials of the $B_1 \bar B_1$ system with $I(J^P)=0(0^+)$ when the cutoff parameter is fixed at 0.80 GeV}
    		\label{fig:57}
    	\end{minipage}
    	\begin{minipage}{0.49\textwidth}
    		\centering
    		\includegraphics[scale=0.7]{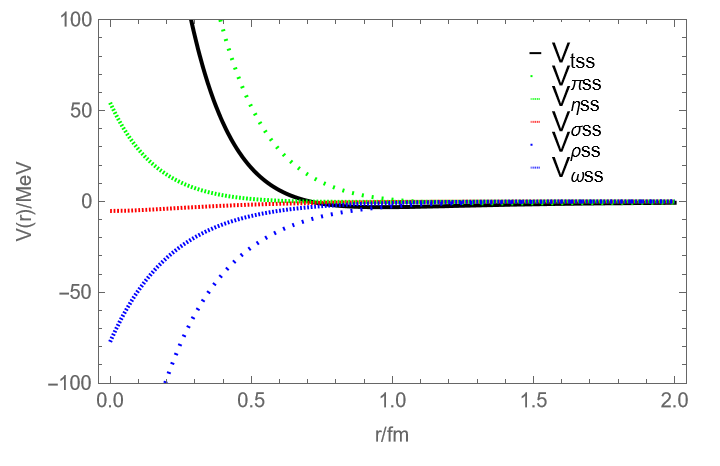}
    		\caption{The $S$ wave potentials of the $B_1 \bar B_1$ system with $I(J^P)=0(1^+)$ when the cutoff parameter is fixed at 1.10 GeV}
    		\label{fig:58}	
    	\end{minipage}
    \end{figure}

\begin{figure}[htbp]
    	\begin{minipage}{0.49\textwidth}
    		\centering
    		\includegraphics[scale=0.7]{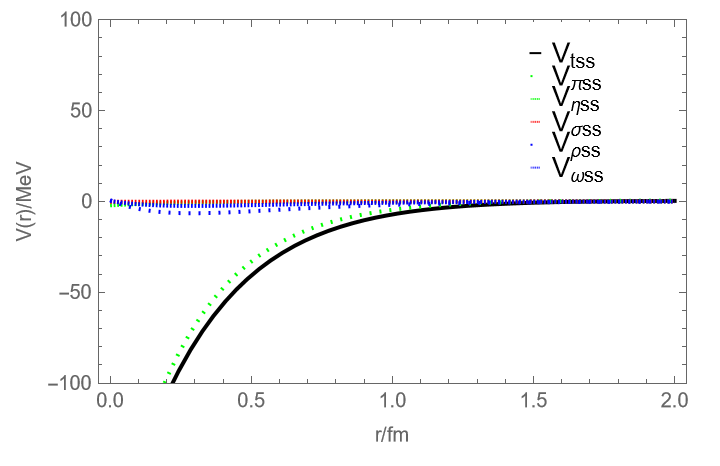}
    		\caption{The $S$ wave potentials of the $B_1 \bar B_1$ system with $I(J^P)=0(2^+)$ when the cutoff parameter is fixed at 0.67 GeV}
    		\label{fig:59}
    	\end{minipage}
    	\begin{minipage}{0.49\textwidth}
    		\centering
    		\includegraphics[scale=0.7]{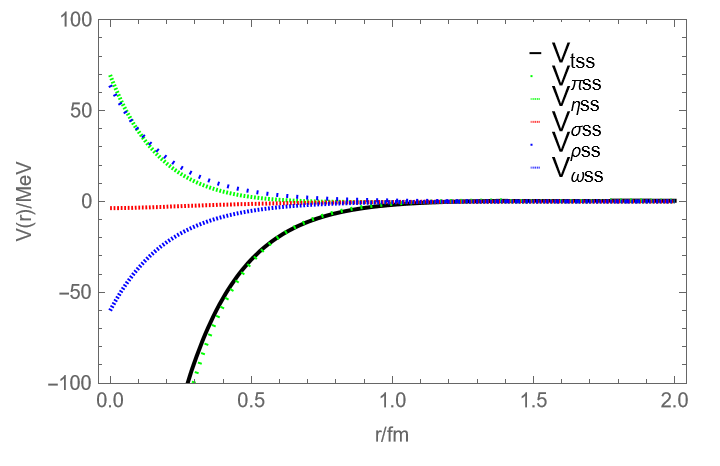}
    		\caption{The $S$ wave potentials of the $B_1 \bar B_1$ system with $I(J^P)=1(0^+)$ when the cutoff parameter is fixed at 1.00 GeV}
    		\label{fig:60}	
    	\end{minipage}
    \end{figure}

\begin{figure}[htbp]
    	\begin{minipage}{0.49\textwidth}
    		\centering
    		\includegraphics[scale=0.7]{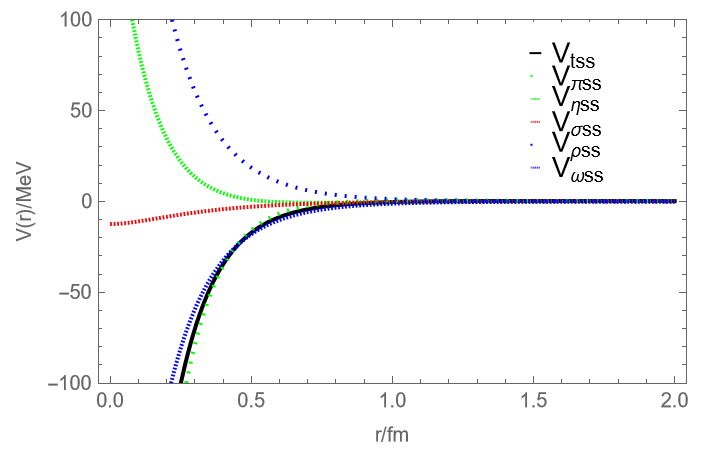}
    		\caption{The $S$ wave potentials of the $B_1 \bar B_1$ system with $I(J^P)=1(1^+)$ when the cutoff parameter is fixed at 1.50 GeV}
    		\label{fig:61}
    	\end{minipage}
       \begin{minipage}{0.49\textwidth}
    		\centering
    		\includegraphics[scale=0.7]{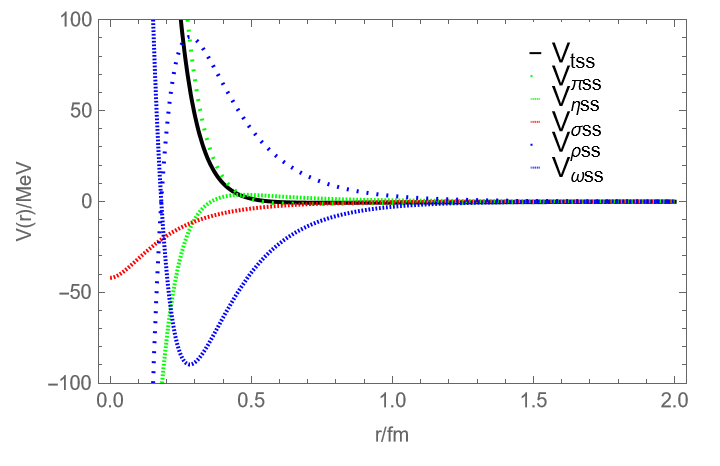}
    		\caption{The $S$ wave potentials of the $B_1 \bar B_1$ system with $I(J^P)=1(2^+)$ when the cutoff parameter is fixed at 2.90 GeV}
    		\label{fig:62}
    	\end{minipage}
    \end{figure}

\begin{figure}[htbp]
    	\begin{minipage}{0.49\textwidth}
    		\centering
    		\includegraphics[scale=0.7]{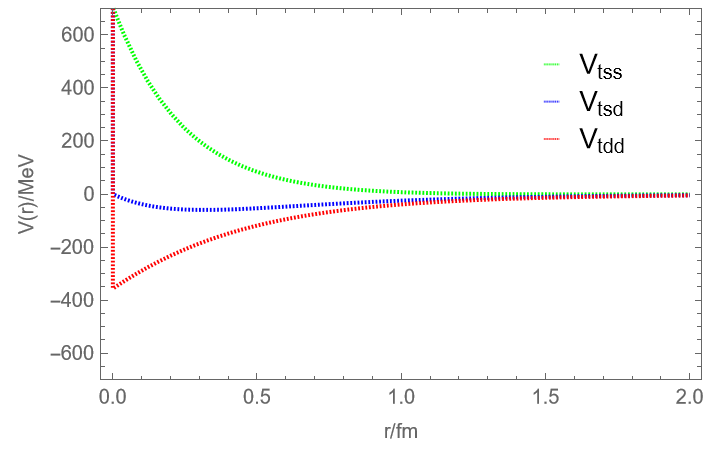}
    		\caption{The $S$\--$D$ wave mixing effects of the $B_1 \bar B_1$ system with $I(J^P)=0(0^+)$ when the cutoff parameter is fixed at 0.80 GeV}
    		\label{fig:63}
    	\end{minipage}
    	\begin{minipage}{0.49\textwidth}
    		\centering
    		\includegraphics[scale=0.7]{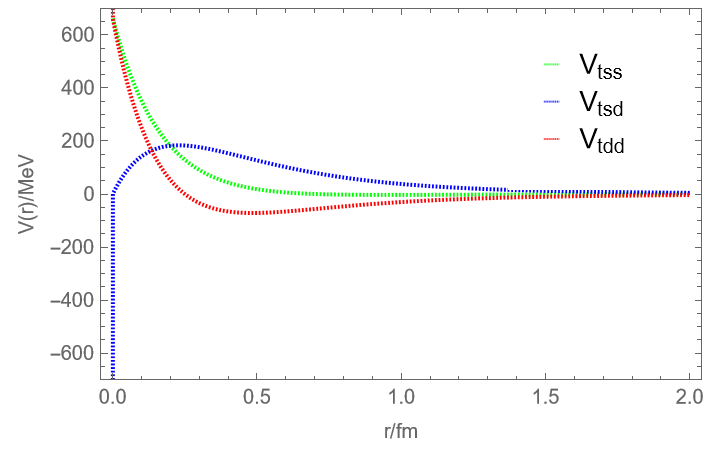}
    		\caption{The $S$\--$D$ wave mixing effects of the $B_1 \bar B_1$ system with $I(J^P)=0(1^+)$ when the cutoff parameter is fixed at 1.10 GeV}
    		\label{fig:64}	
    	\end{minipage}
    \end{figure}

\begin{figure}[htbp]
    	\begin{minipage}{0.49\textwidth}
    		\centering
    		\includegraphics[scale=0.7]{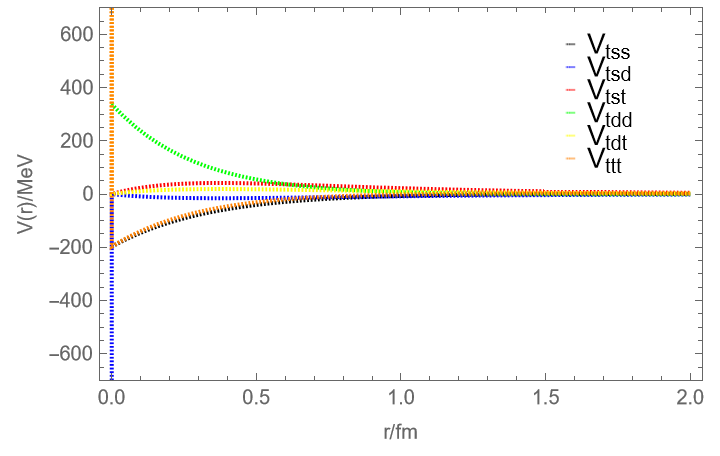}
    		\caption{The $S$\--$D$ wave mixing effects of the $B_1 \bar B_1$ system with $I(J^P)=0(2^+)$ when the cutoff parameter is fixed at 0.67 GeV}
    		\label{fig:65}
    	\end{minipage}
    	\begin{minipage}{0.49\textwidth}
    		\centering
    		\includegraphics[scale=0.7]{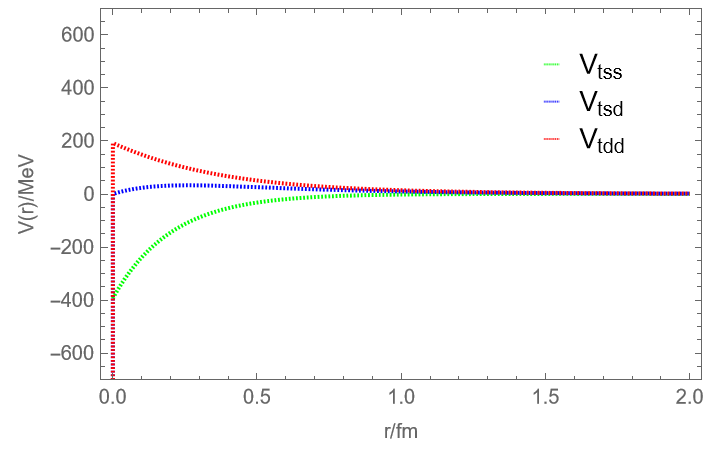}
    		\caption{The $S$\--$D$ wave mixing effects of the $B_1 \bar B_1$ system with $I(J^P)=1(0^+)$ when the cutoff parameter is fixed at 1.00 GeV}
    		\label{fig:66}	
    	\end{minipage}
    \end{figure}

\begin{figure}[htbp]
    	\begin{minipage}{0.49\textwidth}
    		\centering
    		\includegraphics[scale=0.7]{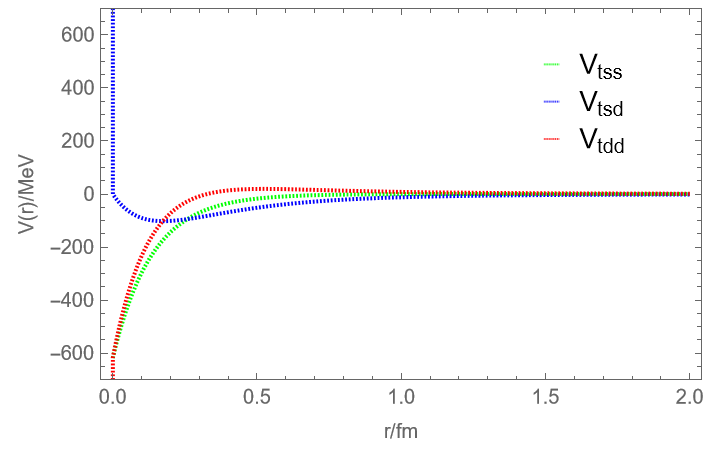}
    		\caption{The $S$\--$D$ wave mixing effects of the $B_1 \bar B_1$ system with $I(J^P)=1(1^+)$ when the cutoff parameter is fixed at 1.50 GeV}
    		\label{fig:67}
    	\end{minipage}
    	\begin{minipage}{0.49\textwidth}
    		\centering
    		\includegraphics[scale=0.7]{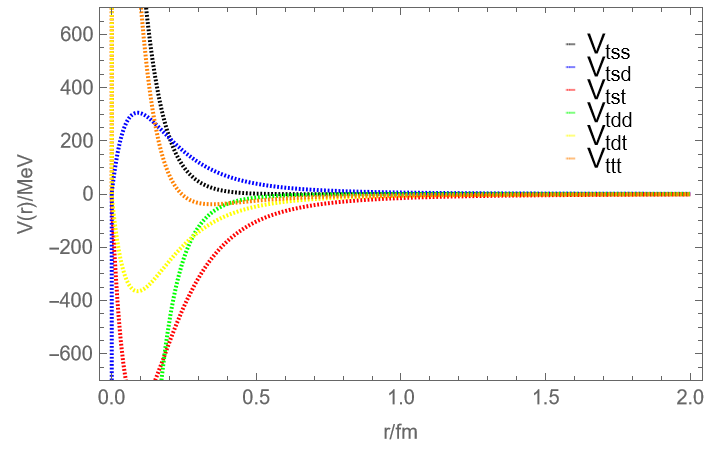}
    		\caption{The $S$\--$D$ wave mixing effects of the $B_1 \bar B_1$ system with $I(J^P)=1(2^+)$ when the cutoff parameter is fixed at 2.90 GeV}
    		\label{fig:68}
    	\end{minipage}
    \end{figure}

\begin{figure}[htbp]
    	\begin{minipage}{0.49\textwidth}
    		\centering
    		\includegraphics[scale=0.7]{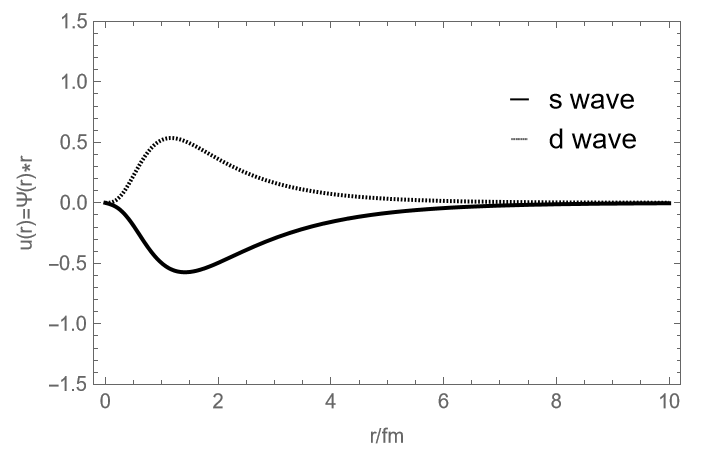}
    		\caption{The wave functions of the $B_1 \bar B_1$ system with $I(J^P)=0(0^+)$ when the cutoff parameter is fixed at 0.80 GeV}
    		\label{fig:69}
    	\end{minipage}
    	\begin{minipage}{0.49\textwidth}
    		\centering
    		\includegraphics[scale=0.7]{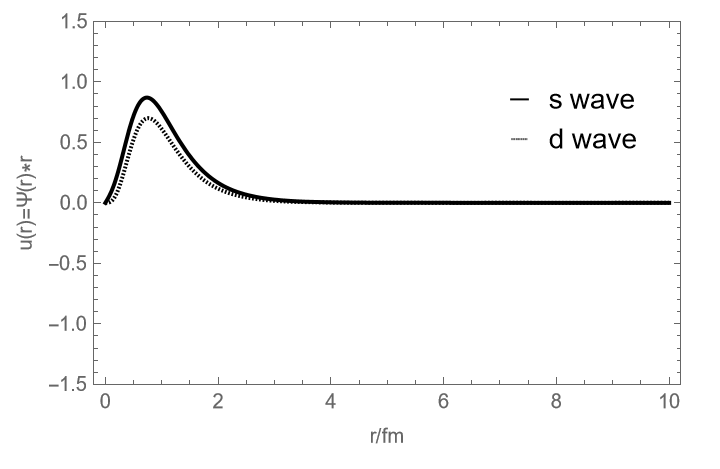}
    		\caption{The wave functions of the $B_1 \bar B_1$ system with $I(J^P)=0(1^+)$ when the cutoff parameter is fixed at 1.10 GeV}
    		\label{fig:70}	
    	\end{minipage}
    \end{figure}

\begin{figure}[htbp]
    	\begin{minipage}{0.49\textwidth}
    		\centering
    		\includegraphics[scale=0.7]{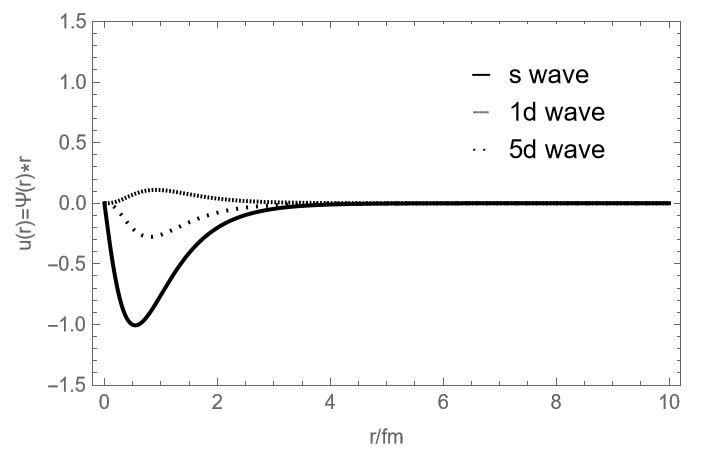}
    		\caption{The wave functions of the $B_1 \bar B_1$ system with $I(J^P)=0(2^+)$ when the cutoff parameter is fixed at 0.67 GeV}
    		\label{fig:71}
    	\end{minipage}
    	\begin{minipage}{0.49\textwidth}
    		\centering
    		\includegraphics[scale=0.7]{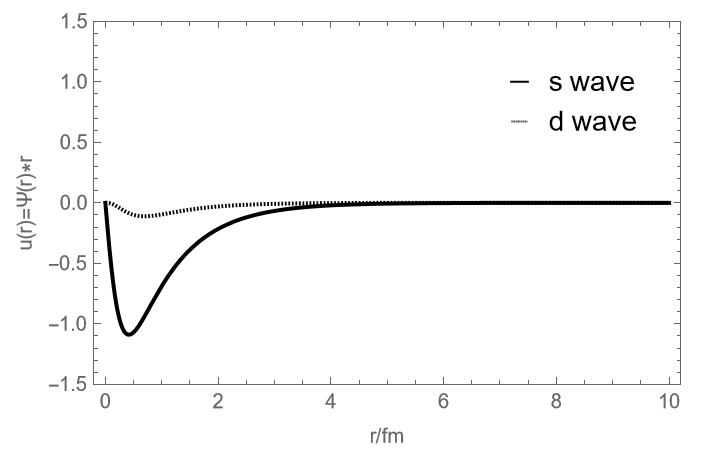}
    		\caption{The wave functions of the $B_1 \bar B_1$ system with $I(J^P)=1(0^+)$ when the cutoff parameter is fixed at 1.00 GeV}
    		\label{fig:72}	
    	\end{minipage}
    \end{figure}

\begin{figure}[htbp]
    	\begin{minipage}{0.49\textwidth}
    		\centering
    		\includegraphics[scale=0.7]{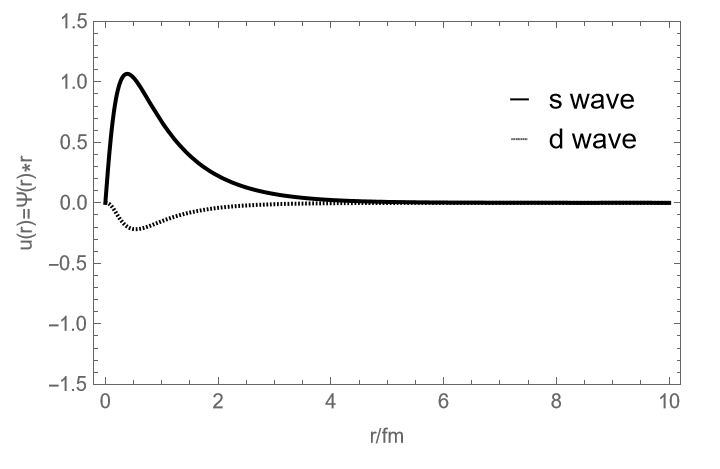}
    		\caption{The wave functions of the $B_1 \bar B_1$ system with $I(J^P)=1(1^+)$ when the cutoff parameter is fixed at 1.50 GeV}
    		\label{fig:73}
    	\end{minipage}
    	\begin{minipage}{0.49\textwidth}
    		\centering
    		\includegraphics[scale=0.7]{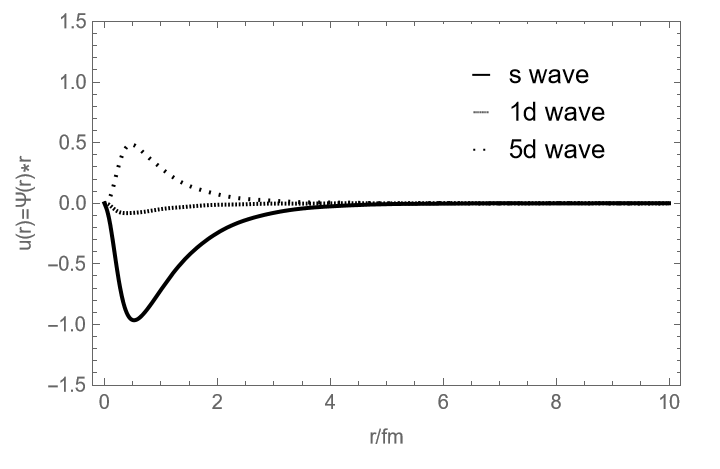}
    		\caption{The wave functions of the $B_1 \bar B_1$ system with $I(J^P)=1(2^+)$ when the cutoff parameter is fixed at 2.90 GeV}
    		\label{fig:74}
    	\end{minipage}
    \end{figure}

     In TABLE \ref{tab:16}-\ref{tab:21}, we present the numerical results of $B_1 \bar  B_1$ with $I(J^P)=0(0^+)$, $I(J^P)=0(1^+)$, $I(J^P)=0(2^+)$, $I(J^P)=1(0^+)$, $I(J^P)=1(1^+)$ and $I(J^P)=1(2^+)$. The $S$ wave contributions of each meson are shown in FIG. \ref{fig:57}-\ref{fig:62}. The contributions from the $S$ and the $D$ wave, and the effect of $S$\--{}$D$ wave mixing are given in FIG. \ref{fig:63}-\ref{fig:68}. The wave functions are plotted in FIG. \ref{fig:69}-\ref{fig:74}.
     \par 
     For the $I(J^P)=0(0^+)$ channel, we find a bound state solution when $\Lambda=0.68$ GeV, with the binding energy 0.48 MeV and RMS radius 3.93 fm. When $\Lambda=0.80$ GeV, the binding energy increases to 2.91 MeV.
    \begin{table*}[htbp]
    	\scriptsize
    	\begin{center}
    		\caption{\label{tab:16} The numerical results of  $B_1 \bar B_1 $ $I(J^P)=0(0^+)$ system }
    		\begin{tabular}{l cccc|cccccc}\toprule[1pt]
    			\multicolumn{5}{c|}{Without recoil corrections} & \multicolumn{5}{c}{With recoil corrections}\\
    			\midrule[1pt]
    			$\Lambda$(GeV) & B.E.(MeV) & RMS(fm)  & $^3S_1 $($\%$)  & $^3D_1$($\%$) & $\Lambda$(GeV) & B.E.(MeV) & RMS(fm)  & $^3S_1 $($\%$)  & $^3D_1$($\%$) \\  
    			0.68 & 0.48 & 3.93 & 80.68 & 19.32 & 0.68 & 0.44 & 4.04 & 81.49 & 18.51 \\
    			0.71 & 0.68 & 3.48 & 77.09 & 22.91 & 0.71 & 0.66 & 3.53 & 77.48 & 22.52 \\
    			0.74 & 1.08 & 2.96 & 71.90 & 28.10 & 0.74 & 1.08 & 2.97 & 72.01 & 27.99 \\
    			0.77 & 1.79 & 2.50 & 65.87 & 34.13 & 0.77 & 1.79 & 2.50 & 65.88 & 34.12 \\
    			0.80 & 2.91 & 2.12 & 59.70 & 40.30 & 0.80 & 2.91 & 2.12 & 59.73 & 40.27 \\
    			\bottomrule[1pt]				
    		\end{tabular}
    	\end{center}
    \end{table*} 
    \par 
    For the $I(J^P)=0(1^+)$ channel, the binding energy is particularly sensitive to the cutoff value. The bound state appears when $\Lambda=0.70$ GeV, with binding energy 0.05 MeV. When $\Lambda$ increases to 1.10 GeV, the binding energy surges to 23.9 MeV.
    
    \begin{table*}[htbp]
    	\scriptsize
    	\begin{center}
    		\caption{\label{tab:17} The numerical results of  $B_1 \bar B_1 $ $I(J^P)=0(1^+)$ system }
    		\begin{tabular}{l cccc|cccccc}\toprule[1pt]
    			\multicolumn{5}{c|}{Without recoil corrections} & \multicolumn{5}{c}{With recoil corrections}\\
    			\midrule[1pt]
    			$\Lambda$(GeV) & B.E.(MeV) & RMS(fm)  & $^3S_1 $($\%$)  & $^3D_1$($\%$) & $\Lambda$(GeV) & B.E.(MeV) & RMS(fm)  & $^3S_1 $($\%$)  & $^3D_1$($\%$) \\  
    			0.70 & 0.05 & 6.67 & 93.97 & 6.03 & 0.70 & 0.04 & 6.71 & 94.06 & 5.94 \\
    			0.80 & 0.50 & 3.69 & 85.25 & 14.75 & 0.80 & 0.50 & 3.70 & 85.26 & 14.74 \\
    			0.90 & 2.92 & 1.98 & 74.17 & 25.83 & 0.90 & 2.85 & 2.00 & 74.40 & 25.60 \\
    			1.00 & 9.76 & 1.32 & 66.97 & 33.03 & 1.00 & 9.24 & 1.34 & 67.55 & 32.45 \\
    			1.10 & 23.9 & 0.97 & 62.61 & 37.39 & 1.10 & 21.9 & 1.00 & 63.59 & 36.41 \\
    			\bottomrule[1pt]				
    		\end{tabular}
    	\end{center}
    \end{table*} 
    \par 
    For the $I(J^P)=0(2^+)$ channel, even when the cutoff value is 0.67 GeV, the binding energy reaches to 15.6 MeV and the RMS radius is 0.89 fm.
    
    \begin{table*}[htbp]
    	\scriptsize
    	\begin{center}
    		\caption{\label{tab:18} The numerical results of  $B_1 \bar B_1 $ $I(J^P)=0(2^+)$ system }
    		\begin{tabular}{l ccccc|ccccccc}\toprule[1pt]
    			\multicolumn{6}{c|}{Without recoil corrections} & \multicolumn{6}{c}{With recoil corrections}\\
    			\midrule[1pt]
    			$\Lambda$(GeV) & B.E.(MeV) & RMS(fm)  & $^5S_2 $($\%$)  & $^1D_2$($\%$) & $^5D_2$($\%$) & $\Lambda$(GeV) & B.E.(MeV) & RMS(fm)  & $^5S_2 $($\%$)  & $^1D_2$($\%$) & $^5D_2$($\%$)\\  
    			0.67 & 15.6 & 0.89 & 91.70 & 1.22 & 7.08 & 0.67 & 15.4 & 0.89 & 91.67 & 1.21 & 7.12 \\
    			0.69 & 16.0 & 0.87 & 91.27 & 1.29 & 7.44 & 0.69 & 15.9 & 0.88 & 91.25 & 1.29 & 7.46 \\
    			0.71 & 17.1 & 0.84 & 90.95 & 1.33 & 7.72 & 0.71 & 17.1 & 0.85 & 90.94 & 1.33 & 7.73 \\
    			0.73 & 18.9 & 0.81 & 90.72 & 1.36 & 7.92 & 0.73 & 18.9 & 0.81 & 90.71 & 1.36 & 7.93 \\
    			0.75 & 21.4 & 0.77 & 90.57 & 1.36 & 8.07 & 0.75 & 21.4 & 0.77 & 90.57 & 1.36 & 8.07 \\
    			\bottomrule[1pt]	
    		\end{tabular}
    	\end{center}
    \end{table*}
    \clearpage
    \par 
    For the $I(J^P)=1(0^+)$ channel, when $\Lambda=0.72$ GeV, the binding energy is 0.27 MeV and the RMS radius is 3.67 fm. A bound state with the binding energy 9.20 MeV and the RMS radius 0.85 fm is found at $\Lambda=1.00$ GeV.
    
    \begin{table*}[htbp]
    	\scriptsize
    	\begin{center}
    		\caption{\label{tab:19} The numerical results of  $B_1 \bar B_1 $ $I(J^P)=1(0^+)$ system }
    		\begin{tabular}{l cccc|cccccc}\toprule[1pt]
    			\multicolumn{5}{c|}{Without recoil corrections} & \multicolumn{5}{c}{With recoil corrections}\\
    			\midrule[1pt]
    			$\Lambda$(GeV) & B.E.(MeV) & RMS(fm)  & $^3S_1 $($\%$)  & $^3D_1$($\%$) & $\Lambda$(GeV) & B.E.(MeV) & RMS(fm)  & $^3S_1 $($\%$)  & $^3D_1$($\%$) \\  
    			0.72 & 0.27 & 3.67 & 99.25 & 0.75 & 0.72 & 0.27 & 3.67 & 99.25 & 0.75  \\
    			0.79 & 1.25 & 1.91 & 98.99 & 1.01 & 0.79 & 1.25 & 1.91 & 98.99 & 1.01 \\
    			0.86 & 2.99 & 1.33 & 98.90 & 1.10 & 0.86 & 3.00 & 1.33 & 98.90 & 1.10 \\
    			0.93 & 5.61 & 1.03 & 98.87 & 1.13 & 0.93 & 5.62 & 1.03 & 98.87 & 1.13 \\
    			1.00 & 9.20 & 0.85 & 98.87 & 1.13 & 1.00 & 9.22 & 0.85 & 98.87 & 1.13 \\
    			\bottomrule[1pt]				
    		\end{tabular}
    	\end{center}
    \end{table*}
    
    \par 
    For the $I(J^P)=1(1^+)$ channel, the bound state solution with the binding energy 0.19 MeV, and the RMS radius 4.21 fm appears at $\Lambda=1.10$ GeV.

    \begin{table*}[htbp]
    	\scriptsize
    	\begin{center}
    		\caption{\label{tab:20} The numerical results of  $B_1 \bar B_1 $ $I(J^P)=1(1^+)$ system }
    		\begin{tabular}{l cccc|cccccc}\toprule[1pt]
    			\multicolumn{5}{c|}{Without recoil corrections} & \multicolumn{5}{c}{With recoil corrections}\\
    			\midrule[1pt]
    			$\Lambda$(GeV) & B.E.(MeV) & RMS(fm)  & $^3S_1 $($\%$)  & $^3D_1$($\%$) & $\Lambda$(GeV) & B.E.(MeV) & RMS(fm)  & $^3S_1 $($\%$)  & $^3D_1$($\%$) \\  
    			1.10 & 0.19 & 4.21 & 98.41 & 1.59 & 1.10 & 0.20 & 4.19 & 98.41 & 1.59 \\
    			1.20 & 1.04 & 2.08 & 97.43 & 2.57 & 1.20 & 1.05 & 2.07 & 97.43 & 2.57 \\
    			1.30 & 2.63 & 1.40 & 96.87 & 3.13 & 1.30 & 2.65 & 1.40 & 96.87 & 3.13 \\
    			1.40 & 5.08 & 1.07 & 96.50 & 3.50 & 1.40 & 5.11 & 1.06 & 96.50 & 3.50 \\
    			1.50 & 8.54 & 0.86 & 96.24 & 3.76 & 1.50 & 8.59 & 0.86 & 96.24 & 3.76 \\
    			\bottomrule[1pt]				
    		\end{tabular}
    	\end{center}
    \end{table*} 
    \par 
    Unlike the $D_1\bar D_1$ system, the $I(J^P)=1(2^+)$ channel of the $B_1\bar B_1$ system can form a bound state. When $\Lambda$ runs between 2.10 GeV and 2.90 GeV, the binding energy runs between 0.17 MeV and 9.09 MeV.
    
    \begin{table*}[htbp]
    	\scriptsize
    	\begin{center}
    		\caption{\label{tab:21} The numerical results of  $B_1 \bar B_1 $ $I(J^P)=1(2^+)$ system }
    		\begin{tabular}{l ccccc|ccccccc}\toprule[1pt]
    			\multicolumn{6}{c|}{Without recoil corrections} & \multicolumn{6}{c}{With recoil corrections}\\
    			\midrule[1pt]
    			$\Lambda$(GeV) & B.E.(MeV) & RMS(fm)  & $^5S_2 $($\%$)  & $^1D_2$($\%$) & $^5D_2$($\%$) & $\Lambda$(GeV) & B.E.(MeV) & RMS(fm)  & $^5S_2 $($\%$)  & $^1D_2$($\%$) & $^5D_2$($\%$)\\  
    			2.10 & 0.17 & 4.73 & 95.76 & 0.21 & 4.03 & 2.10 & 0.18 & 4.68 & 95.69 & 0.21 & 4.10  \\
    			2.30 & 0.88 & 2.50 & 91.87 & 0.35 & 7.78 & 2.30 & 0.90 & 2.48 & 91.80 & 0.35 & 7.85 \\
    			2.50 & 2.39 & 1.66 & 88.53 & 0.43 & 11.04 & 2.50 & 2.44 & 1.65 & 88.45 & 0.43 & 11.12 \\
    			2.70 & 5.02 & 1.23 & 85.69 & 0.48 & 13.83 & 2.70 & 5.09 & 1.23 & 85.60 & 0.48 & 13.92 \\
    			2.90 & 9.09 & 0.98 & 83.24 & 0.51 & 16.25 & 2.90 & 9.22 & 0.97 & 83.15 & 0.51 & 16.34 \\
    			\bottomrule[1pt]	
    		\end{tabular}
    	\end{center}
    \end{table*}
    \par 
    From the numerical results, we find that all the channels can form a bound state and the recoil corrections are not important. 
   % \clearpage

    

    \clearpage

    \section{Summary and Discussion} \label{sec4}
    We have discussed the $D_1 D_1$, $D_1 \bar D_1$, $B_1 B_1$ and $B_1 \bar B_1$ systems in the framework of the OBE model and isospin SU(2) symmetry, which makes it possible for the exchange of $\pi$, $\eta$, $\sigma$, $\rho$ and $\omega$. We have also considered the $S$\--{}$D$ wave mixing effect and the recoil corrections. In the effective potentials, there are spin-spin interaction terms, tensor force terms and spin-orbit force terms. Particularly, the spin-orbit force terms appear after considering the recoil corrections at $O(\frac{1}{{m_{D_1}}})$ and $O(\frac{1}{{m_{B_1}}^2})$. 
    \par 
    We have thoroughly studied the $D_1 D_1$ and the $B_1 B_1$ systems with three sets of quantum numbers: $I(J^P)=0(1^+)$, $I(J^P)=1(0^+)$ and $I(J^P)=1(2^+)$, the $D_1 \bar D_1$ and the $B_1 \bar B_1$ systems with six sets of quantum numbers: $I(J^P)=0(0^+)$, $I(J^P)=0(1^+)$, $I(J^P)=0(2^+)$, $I(J^P)=1(0^+)$, $I(J^P)=1(1^+)$ and $I(J^P)=1(2^+)$.         
    \par 
    After solving the coupled channel Schr{\"o}dinger equation, we get the numerical results. The results of the $D_{1}$ systems without the recoil corrections are as follows. For the $D_1 D_1$ system, states with $I(J^P)=0(1^+)$ and $I(J^P)=1(2^+)$ can be molecular states. For the $D_1 \bar D_1$ system, states with  $I(J^P)=0(0^+)$, $I(J^P)=0(1^+)$, $I(J^P)=0(2^+)$, $I(J^P)=1(0^+)$ and $I(J^P)=1(1^+)$ can be molecular states. Among them, the $I(J^P)=0(1^+)$ channel of the $D_1 D_1$ system and the $I(J^P)=0(2^+)$ channel of the $D_1 \bar D_1$ system are most likely to be a molecular state.
    \par 
    However, after considering the recoil corrections, we obtain a result of considerable importance that the $I(J^P)=1(0^+)$ channel of the $D_1 D_1$ system can be a molecular state. It is noted that this channel can't be a molecular state without the recoil corrections.
    \par 
    For the $B_1 B_1$ and the $B_1 \bar B_1$ systems, all the channels can be a molecular state without considering the recoil corrections.
    \par
\begin{figure}[htbp]
    	\begin{minipage}{0.49\textwidth}
    		\centering
    		\includegraphics[scale=0.65]{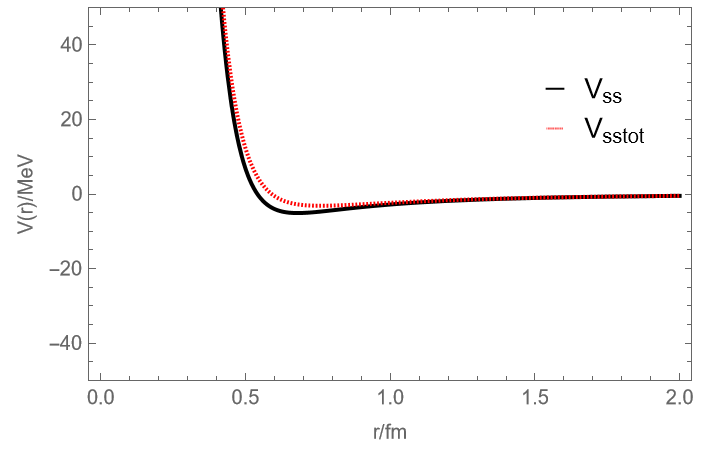}
    		\caption{The $S$ wave potential of the $D_1 D_1$ system with $I(J^P)=1(0^+)$ when the cutoff parameter is fixed at 3.03 GeV with or without recoil corrections }
    		\label{fig:75}
    	\end{minipage}
    	\begin{minipage}{0.49\textwidth}
    		\centering
    		\includegraphics[scale=0.65]{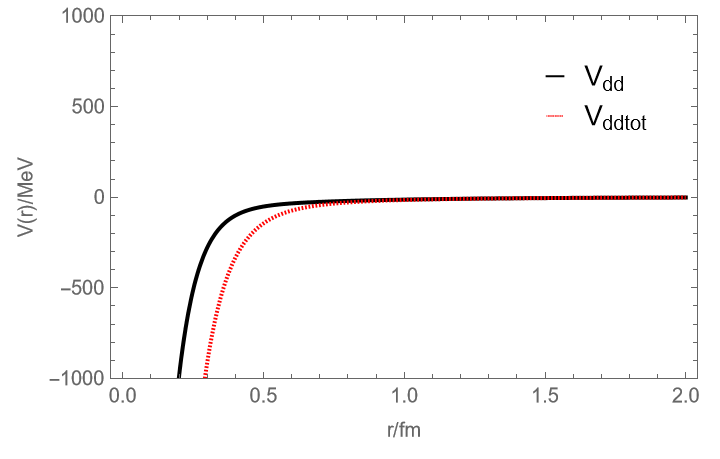}
    		\caption{The $D$ wave potential of the $D_1 D_1$ system with $I(J^P)=1(0^+)$ when the cutoff parameter is fixed at 3.03 GeV with or without recoil corrections }
    		\label{fig:76}
    	\end{minipage}
    \end{figure}
\begin{figure}[htbp]
    	\centering
    	\includegraphics[scale=0.65]{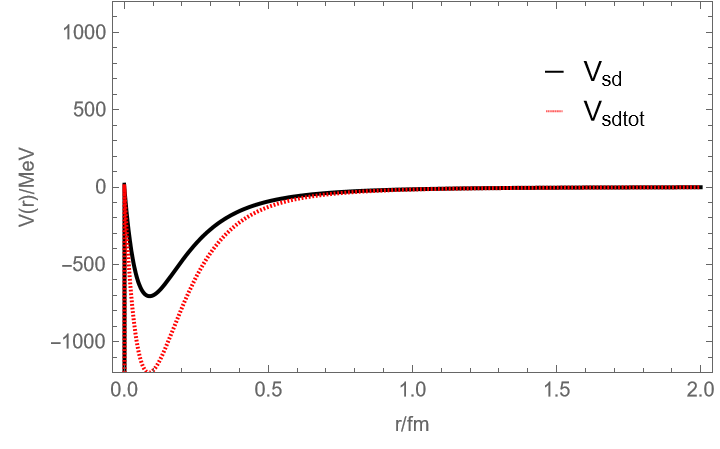}
    	\caption{The $S$\--$D$ wave mixing effects of the $D_1 D_1$ system with $I(J^P)=1(0^+)$ when the cutoff parameter is fixed at 3.03 GeV with or without recoil corrections }
    	\label{fig:77}
    \end{figure}
\begin{figure}[htbp]
    	\begin{minipage}{0.49\textwidth}
    		\centering
    		\includegraphics[scale=0.65]{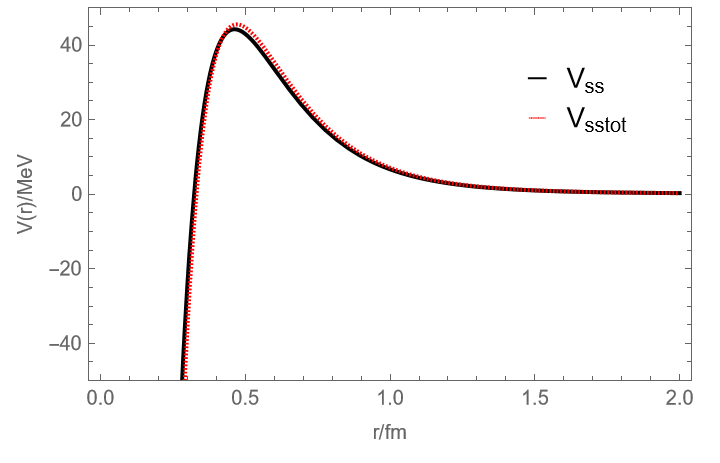}
    		\caption{The $S$ wave potential of the $D_1 D_1$ system with $I(J^P)=1(2^+)$ when the cutoff parameter is fixed at 1.92 GeV with or without recoil corrections }
    		\label{fig:78}
    	\end{minipage}
    	\begin{minipage}{0.49\textwidth}
    		\centering
    		\includegraphics[scale=0.65]{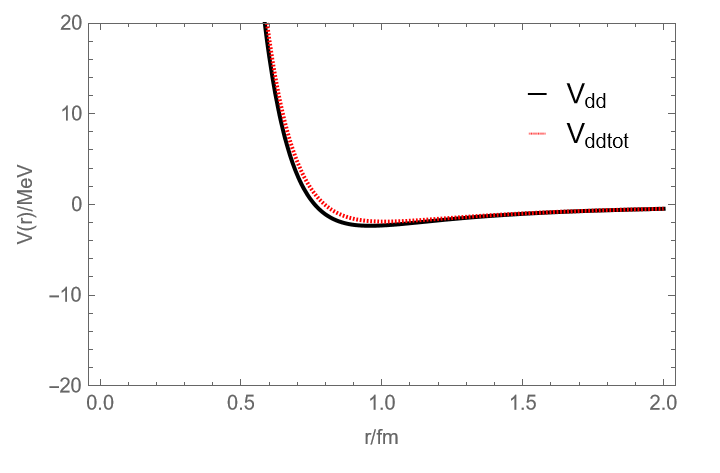}
    		\caption{The 1$D$ wave potential of the $D_1 D_1$ system with $I(J^P)=1(2^+)$ when the cutoff parameter is fixed at 1.92 GeV with or without recoil corrections }
    		\label{fig:79}	
    	\end{minipage}		
    \end{figure}
\begin{figure}[htbp]
    	\centering
    	\includegraphics[scale=0.65]{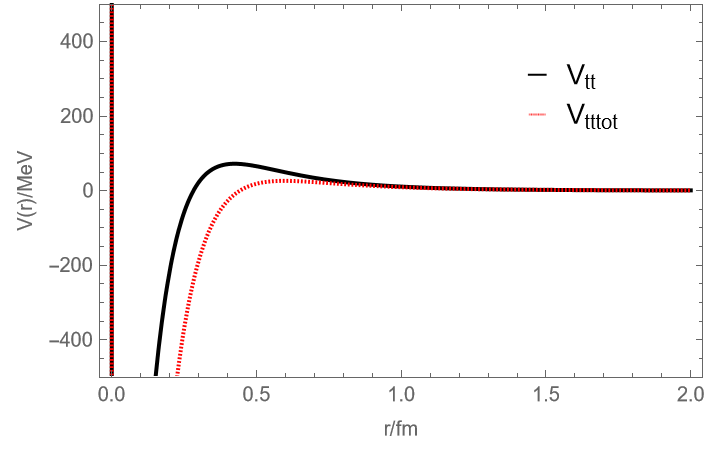}
    	\caption{The 5$D$ wave potential of the $D_1 D_1$ system with $I(J^P)=1(2^+)$ when the cutoff parameter is fixed at 1.92 GeV with or without recoil corrections }
    	\label{fig:80}
    \end{figure}
\begin{figure}[htbp]
    	\begin{minipage}{0.49\textwidth}
    		\centering
    		\includegraphics[scale=0.65]{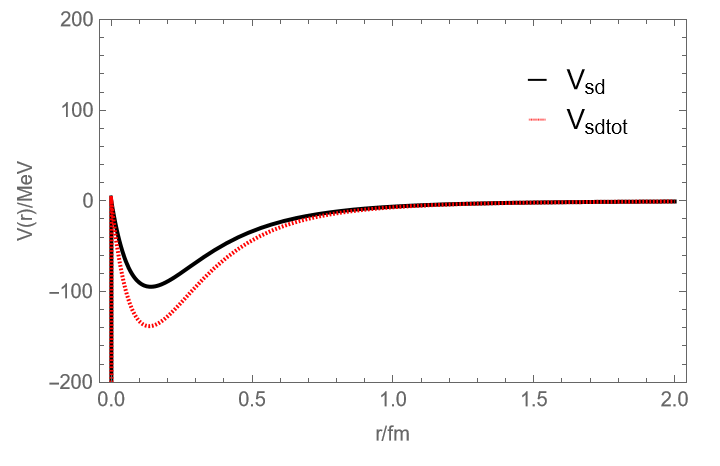}
    		\caption{The $S$\--1$D$ wave mixing effect of the $D_1 D_1$ system with $I(J^P)=1(2^+)$ when the cutoff parameter is fixed at 1.92 GeV with or without recoil corrections }
    		\label{fig:81}
    	\end{minipage}
    	\begin{minipage}{0.49\textwidth}
    		\centering
    		\includegraphics[scale=0.65]{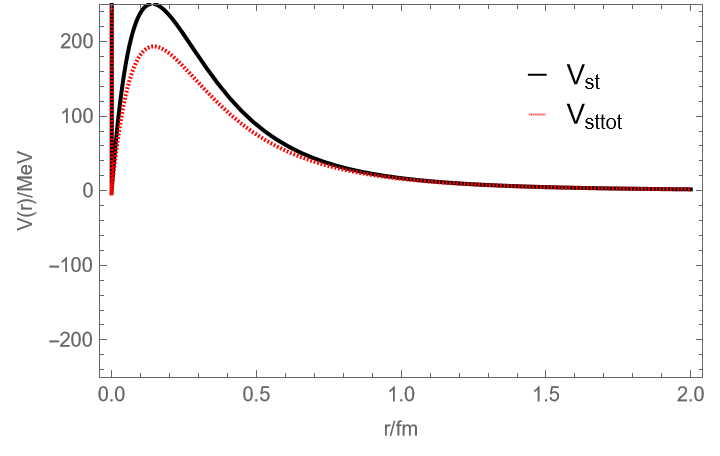}
    		\caption{The $S$\--5$D$ wave mixing effect of the $D_1 D_1$ system with $I(J^P)=1(2^+)$ when the cutoff parameter is fixed at 1.92 GeV with or without recoil corrections }
    		\label{fig:82}	
    	\end{minipage}		
    \end{figure}
\begin{figure}[htbp]
    	\centering
    	\includegraphics[scale=0.65]{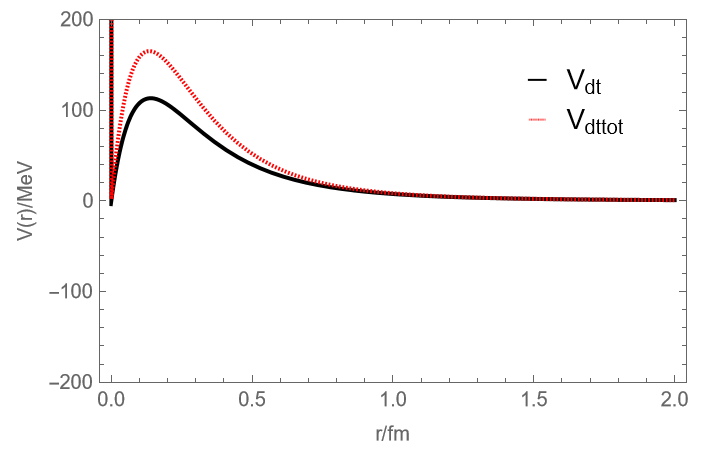}
    	\caption{The 1$D$\--5$D$ wave mixing effect of the $D_1 D_1$ system with $I(J^P)=1(2^+)$ when the cutoff parameter is fixed at 1.92 GeV with or without recoil corrections }
    	\label{fig:83}
    \end{figure}
\begin{figure}[htbp]
    	\begin{minipage}{0.49\textwidth}
    		\centering
    		\includegraphics[scale=0.65]{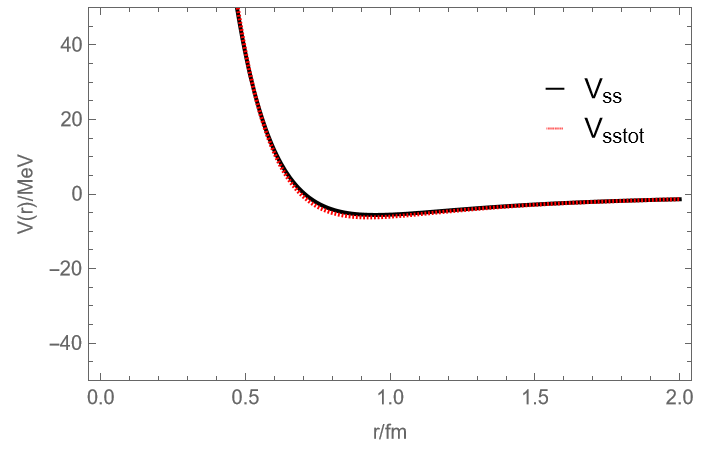}
    		\caption{The $S$ wave potential of the $D_1 \bar D_1$ system with $I(J^P)=0(0^+)$ when the cutoff parameter is fixed at 1.35 GeV with or without recoil corrections}
    		\label{fig:84}
    	\end{minipage}
    	\begin{minipage}{0.49\textwidth}
    		\centering
    		\includegraphics[scale=0.65]{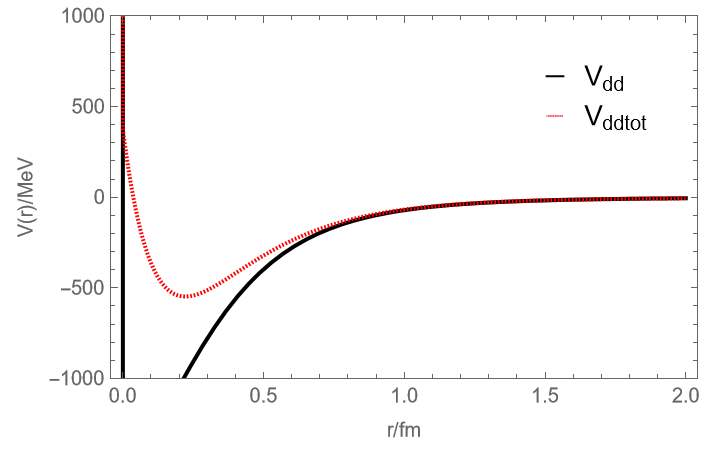}
    		\caption{The $D$ wave potential of the $D_1 \bar D_1$ system with $I(J^P)=0(0^+)$ when the cutoff parameter is fixed at 1.35 GeV with or without recoil corrections}
    		\label{fig:85}	
    	\end{minipage}
    \end{figure}
\begin{figure}[htbp]
    	\centering
    	\includegraphics[scale=0.65]{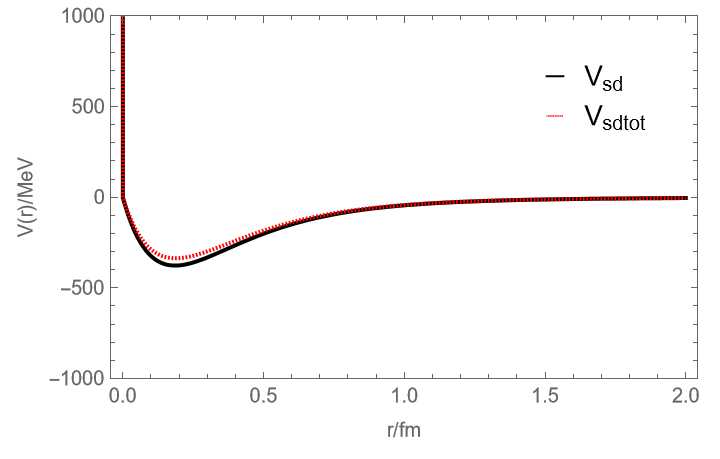}
    	\caption{The $S$\--$D$ wave mixing effects of the $D_1 \bar D_1$ system with $I(J^P)=0(0^+)$ when the cutoff parameter is fixed at 1.35 GeV with or without recoil corrections}
    	\label{fig:86}
    \end{figure}
\begin{figure}[htbp]
    	\begin{minipage}{0.49\textwidth}
    		\centering
    		\includegraphics[scale=0.65]{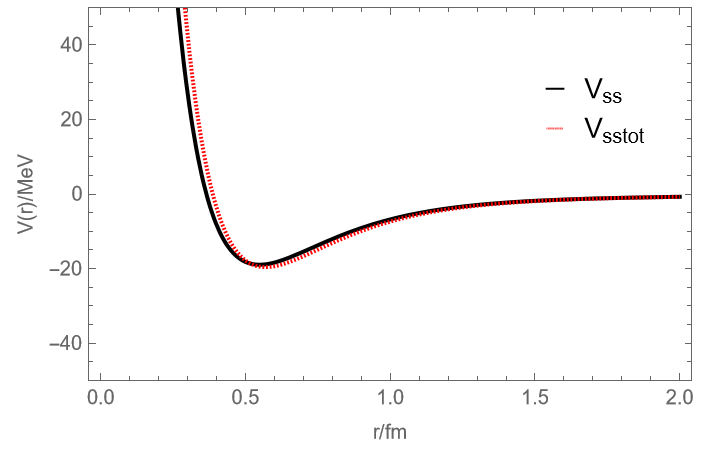}
    		\caption{The $S$ wave potential of the $D_1 \bar D_1$ system with $I(J^P)=0(1^+)$ when the cutoff parameter is fixed at 1.40 GeV with or without recoil corrections}
    		\label{fig:87}
    	\end{minipage}
    	\begin{minipage}{0.49\textwidth}
    		\centering
    		\includegraphics[scale=0.65]{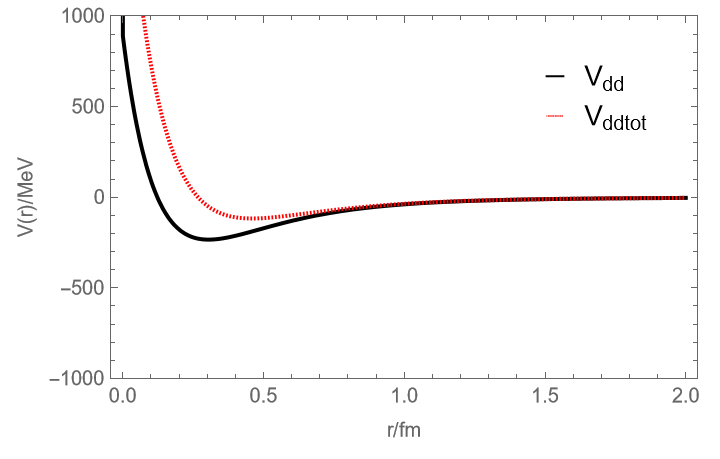}
    		\caption{The $D$ wave potential of the $D_1 \bar D_1$ system with $I(J^P)=0(1^+)$ when the cutoff parameter is fixed at 1.40 GeV with or without recoil corrections}
    		\label{fig:88}	
    	\end{minipage}
    \end{figure}
\begin{figure}[htbp]
    	\centering
    	\includegraphics[scale=0.65]{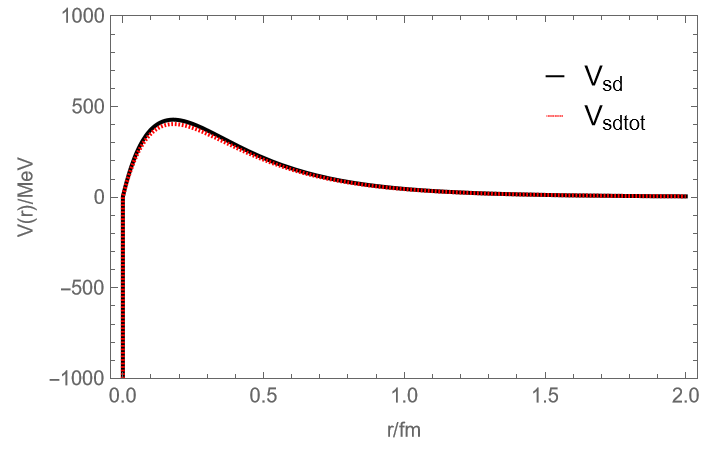}
    	\caption{The $S$\--$D$ wave mixing effect of the $D_1 \bar D_1$ system with $I(J^P)=0(1^+)$ when the cutoff parameter is fixed at 1.40 GeV with or without recoil corrections}
    	\label{fig:89}
    \end{figure}

    We notice that the recoil corrections have a great effect on the binding energy of several systems. Taking the $D_{1}$ systems as examples, the $I(J^{P})$=$1(0^{+})$ and $1(2^{+})$ $D_{1}D_{1}$ states, and the $I(J^{P})$=$0(0^{+})$ and $0(1^{+})$ $D_{1}\bar{D}_{1}$ states are such cases. Consequently, we try to explain this phenomenon by comparing the  $S$ and the $D$ wave potentials with or without the recoil corrections. Besides, the $S$\--$D$ wave mixing effect is also considered. We show the effective potentials of the $D_{1}D_{1}$ systems with $I(J^{P})$=$1(0^+)$ and $I(J^{P})$=$1(2^+)$
    in FIG. \ref{fig:75}-\ref{fig:77} and FIG. \ref{fig:78}-\ref{fig:83}, respectively. We show the potentials of the $D_{1}\bar{D}_{1}$ with $I(J^{P})$=$0(0^+)$ and $I(J^{P})$=$0(1^+)$
    in FIG. \ref{fig:84}-\ref{fig:86} and FIG. \ref{fig:87}-\ref{fig:89}, respectively. In all the figures, the black curve represents the
    effective potential without the recoil corrections while the red curve represents the effective potential with the recoil corrections.
    \par 
    For the $D_{1}D_{1}$ system with $I(J^{P})=1(0^{+})$, the $S$ wave potential has a tiny increase around 0.5 fm. However, the $D$ wave potential shows a relatively remarkable decrease around 0.5 fm. Moreover, it is clear to see in FIG.77 that the $S$-$D$ wave mixing effect is more significant after considering the recoil corrections. 
    \par
    For the $D_{1}D_{1}$ system with $I(J^{P})=1(2^{+})$, the $^5D_2$ wave potential shows a remarkable decrease around 0.5 fm. 
    \par
    For the $D_{1}\bar{D}_{1}$ system with $I(J^{P})=0(0^{+})$, the $S$ wave potential and the $S$\--$D$ wave mixing effect are almost unchanged. However, the $D$ wave potential has a relatively remarkable increase. The results are similar for the $D_{1}\bar{D}_{1}$ system with $I(J^{P})=0(1^{+})$.
    \par   
    In general, for the $I=0$ channels of the $D_1 D_1$ and the $D_1 \bar D_1$ systems, the recoil corrections are unfavorable to form the molecular states. But for the $I=1$ channels the recoil corrections are favorable to form the molecular states. For the $D_1 \bar D_1$ system, the $J=0$ and the $J=1$ channels with the same isospin have the similar effective potentials and the numerical results. Compared with the $D_1 D_1$ and the $D_1 \bar D_1$ systems, the $B_1 B_1$ and the $B_1 \bar B_1$ systems are easier to form molecular states and less likely to be influenced by the recoil corrections. For all the systems, the $S$ wave is the main component of the ground state.     
    The great importance of the recoil corrections is shown in this work. If we had ignored this effect, a possible molecular state would be missed. Therefore, researchers should consider the recoil corrections in subsequent works so that they can get more reliable results. 



    \clearpage

    \section{Acknowledgment} \label{sec5}
    
    This work is supported by the Fundamental Research Funds for the Central Universities under Grant No. 2022JBMC037 and the National Natural Science Foundation of China under Grants No. 12105009.

	\clearpage
	
    \section{Appendix}
	\subsection{The representation of polarization vectors}\label{subsecA}
	\begin{table*}[htbp]
		\scriptsize
		\begin{center}
			\caption{\label{tab:22}  The matrix elements $\left\langle f \mid O_{k} \mid i \right\rangle$ for the polarization vector related operators $O_{k}$ in the effective potentials }
			
			\begin{tabular}{ c|ccc}\toprule[1pt]
				\multicolumn{1}{c|}{$\left\langle f \mid O_{k} \mid i \right\rangle$} & \multicolumn{3}{c}{$\left\langle f \mid O_{k} \mid i \right\rangle_{[J]}$}\\
				\midrule[1pt]
				$\left\langle D_{1} D_{1}\left|O_{1}\right| D_{1} D_{1}\right\rangle$ & $\operatorname{Diag}(1,1)_{[0]}$ & $\operatorname{Diag}(1,1)_{[1]}$ & $\operatorname{Diag}(1,1,1)_{[2]}$ \\				
				$\left\langle D_{1} D_{1}\left|O_{2}\right| D_{1} D_{1}\right\rangle$ & $\operatorname{Diag}(2,-1)_{[0]}$ & $\operatorname{Diag}(1,1)_{[1]}$ & $\operatorname{Diag}(-1,2,-1)_{[2]}$ \\
				$\left\langle D_{1} D_{1}\left|O_{3}\right| D_{1} D_{1}\right\rangle$ & $\left(\begin{array}{cc}
					0 & \sqrt{2}\\
					\sqrt{2} & 2
				\end{array}\right)_{[0]}$ & $\left(\begin{array}{cc}
					0 & -\sqrt{2} \\
					-\sqrt{2} & 1
				\end{array}\right)_{[1]} $ & $\left(\begin{array}{ccc}
					0 & \sqrt{\frac{2}{5}} & -\sqrt{\frac{14}{5}} \\
					\sqrt{\frac{2}{5}} & 0 & -\frac{2}{\sqrt{7}} \\
					-\sqrt{\frac{14}{5}} & -\frac{2}{\sqrt{7}} & -\frac{3}{7}
				\end{array}\right)_{[2]}$ \\				
				$\left\langle D_{1} D_{1}\left|O_{4(5)}\right| D_{1} D_{1}\right\rangle$ & $\left(\begin{array}{cc}
					0 & -\sqrt{2} \\
					-\sqrt{2} & 1
				\end{array}\right)_{[0]}$ & $\left(\begin{array}{cc}
					0 & \frac{1}{\sqrt{2} } \\
					\frac{1}{\sqrt{2} } & -\frac{1}{2} 
				\end{array}\right)_{[1]}$ & $\left(\begin{array}{ccc}
					0 & -\sqrt{\frac{2}{5}} & -\sqrt{\frac{7}{10}} \\
					-\sqrt{\frac{2}{5}} & 0 & \frac{2}{\sqrt{7}} \\
					-\sqrt{\frac{7}{10}} & \frac{2}{\sqrt{7}} & -\frac{3}{14}
				\end{array}\right)_{[2]}$ \\
				$\left\langle D_{1} D_{1}\left|O_{6(7)}\right| D_{1} D_{1}\right\rangle$ & $\left(\begin{array}{cc}
					0 & 0 \\
					0 & 3i
				\end{array}\right)_{[0]} $ & $\left(\begin{array}{cc}
					0 & 0 \\
					0 & \frac{3}{2}i 
				\end{array}\right)_{[1]}$ & $\left(\begin{array}{ccc}
					0 & 0 & 0 \\
					0 & 0 & 0 \\
					0 & 0 & \frac{3}{2}i
				\end{array}\right)_{[2]}$ \\

				\bottomrule[1pt]
				
			\end{tabular}
		\end{center}
	\end{table*} 
	$O_1$,$O_2$ $\dots$ $O_7$ represent
	    \begin{eqnarray*}
		O_{1}&=&(\boldsymbol{\epsilon}_1 \cdot \boldsymbol{\epsilon}_3^\dagger )(\boldsymbol{\epsilon}_2 \cdot \boldsymbol{\epsilon}_4^\dagger ),
		\\
		O_{2}&=&(\boldsymbol{\epsilon}_1 \times \boldsymbol{\epsilon}_3^\dagger )\cdot (\boldsymbol{\epsilon}_2 \times \boldsymbol{\epsilon}_4^\dagger ),
		\\
		O_{3}&=&S(\boldsymbol{\hat{r}},\boldsymbol{\epsilon}_1\times \boldsymbol{\epsilon}_3^\dagger,\boldsymbol{\epsilon}_2\times \boldsymbol{\epsilon}_4^\dagger),
		\\   O_{4}&=&S(\boldsymbol{\hat{r}},\boldsymbol{\epsilon}_1,\boldsymbol{\epsilon}_3^\dagger)(\boldsymbol{\epsilon}_2 \cdot \boldsymbol{\epsilon}_4^\dagger),
		\\    O_{5}&=&S(\boldsymbol{\hat{r}},\boldsymbol{\epsilon}_2,\boldsymbol{\epsilon}_4^\dagger)(\boldsymbol{\epsilon}_1 \cdot \boldsymbol{\epsilon}_3^\dagger),
		\\
		O_{6}&=&(\boldsymbol{\epsilon}_2 \cdot \boldsymbol{\epsilon}_4^\dagger)[(\boldsymbol{\epsilon}_1 \times  \boldsymbol{\epsilon}_3^\dagger)\cdot  \boldsymbol{L}],
		\\
		O_{7}&=&(\boldsymbol{\epsilon}_1 \cdot \boldsymbol{\epsilon}_3^\dagger)[(\boldsymbol{\epsilon}_2 \times  \boldsymbol{\epsilon}_4^\dagger)\cdot  \boldsymbol{L}]     
	\end{eqnarray*} 
	with $S(\hat{\boldsymbol{r}}, \boldsymbol{a}, \boldsymbol{b})=3 \frac{(\boldsymbol{a} \cdot \boldsymbol{r})(\boldsymbol{b} \cdot \boldsymbol{r})}{r^{2}}-\boldsymbol{a} \cdot \boldsymbol{b}$, $\boldsymbol{L}=\boldsymbol{r} \times (-i\nabla)$.
	
	\subsection{The Fourier transformation of $\boldsymbol{q}$ and $\boldsymbol{k}$ related terms}
	\begin{eqnarray*}
	-\frac{1}{q^2-m^2+i\epsilon }&=&\frac{m}{4\pi}H_0,\\
	-\frac{q_i}{q^2-m^2+i\epsilon}&=&-i\frac{m^3}{4\pi}H_2 r_i,\\
	-\frac{\boldsymbol{q}}{q^2-m^2+i\epsilon}&=&-i\frac{m^3}{4\pi}H_2 \boldsymbol{r},\\
	-\frac{q_i q_j}{q^2-m^2+i\epsilon}&=& \left(-\frac{m^3}{12\pi}\right)\left[H_3\left(\frac{3r_i r_j}{r^2}-\delta _{ij}  \right)-H_1\delta _{ij}\right],\\
	-\frac{\boldsymbol{q}^2}{q^2-m^2+i\epsilon}&=&\frac{m^3}{4\pi}H_1
	\end{eqnarray*}
	and
	\begin{eqnarray*}
		-\frac{i\boldsymbol{S} \cdot \left(\boldsymbol{q} \times \boldsymbol{k}\right)}{q^2-m^2+i\epsilon }&=&-\frac{m^3}{4\pi}H_2\boldsymbol{S} \cdot \boldsymbol{L},\\
		-\frac{\boldsymbol{k}^2}{q^2-m^2+i\epsilon}&=&-\frac{m}{4\pi}H_0 \nabla ^2+\frac{m^3}{16\pi}H_1-\frac{m^3}{4\pi}H_2 \left(\boldsymbol{r} \cdot \nabla \right)   
	\end{eqnarray*}
	with
	\begin{eqnarray*}
	H_{0}&=&\frac{1}{m r}\left(e^{-m r}-e^{-\Lambda r}\right)-\frac{\Lambda^{2}-m^{2}}{2 m \Lambda} e^{-\Lambda r}, \\
	H_{1}&=&-\frac{1}{m r}\left(e^{-m r}-e^{-\Lambda r}\right)+\Lambda \frac{\Lambda^{2}-m^{2}}{2 m^{2}} e^{-\Lambda r}, \\
	H_{2}&=&-(1+m r) \frac{1}{m^{3} r^{3}} e^{-m r}+(1+\Lambda r) \frac{1}{m^{3} r^{3}} e^{-\Lambda r}+\frac{\Lambda^{2}-m^{2}}{2 m^{2}} \frac{1}{m r} e^{-\Lambda r}, \\
	H_{3}&=&\left(1+\frac{3}{m r}+\frac{3}{m^{2} r^{2}}\right) \frac{1}{m r} e^{-m r}-\left(1+\frac{3}{\Lambda r}+\frac{3}{\Lambda^{2} r^{2}}\right) \frac{\Lambda^{2}}{m^{2}} \frac{1}{m r} e^{-\Lambda r}-\frac{\Lambda^{2}-m^{2}}{2 m^{2}}(1+\Lambda r) \frac{1}{m r} e^{-\Lambda r}.
    \end{eqnarray*}

\end{document}